# Hydrogen in superconductors

## (Review Article)

S.I. Bondarenko, V.P. Timofeev, V.P. Koverya, A.V. Krevsun

B. Verkin Institute for Low Temperature Physics and Engineering of the National Academy of Scinces of Ukraine,

47 Nauky ave., Kharkiv, 61103, Ukraine

E-mail: bondarenko@ilt.kharkov.ua

## Abstract

Information is presented on the state of research on the effect of hydrogen on the superconducting properties of various compounds. The review consists of an introduction, one appendix and four sections: methods for the synthesis of modern hydrogen-containing superconductors, experimental studies of the properties of hydrogen-containing superconductors, on the mechanisms of the influence of hydrogen on superconductivity, problems and prospects of hydrogen-containing superconductors.

**Keywords**: superconductivity, hydrogen, synthesis, pressure, experiment, properties, BCS theory, hydrides, pnictides, chalcogenides, cuprates, deborides, carbon nanotubes, critical temperature, critical current density, critical magnetic field.

## Contents

**Introduction.**

Hydrogen was discovered in the 16th century, and only in the 18th century did the study of its gaseous state begin. Industrial interest in it appeared in the 19th century with the beginning of the production of illuminating gas (a gas mixture of hydrogen and methane) and the manufacture of hydrogen balloons. The increased attention to it in the 20th century is associated with the rapid development of materials science, the nuclear industry, rocketry and environmentally friendly hydrogen energy. One of the new achievements in materials science of this period was the discovery and explanation of the embrittlement of metals caused by their interaction with hydrogen. An important event in 1968 in materials science, directly related to superconductivity, was the theoretical work of N.W. Ashcroft in Physical Review Letters, which indicated the possibility of the existence of high-temperature superconductivity up to 100K in solid hydrogen at very high pressure. This served as a powerful impetus for the development of research and applied work in this direction in the most developed countries of the world. By the beginning of the 21st century, it was experimentally proven that almost all chemical elements become superconducting if they are strongly compressed and the temperature is lowered. The result of these works in the last decade was experimental evidence of the existence of superconductivity under ultra-high pressure at temperatures close to room temperature, although not in pure solid hydrogen, but in a whole series of two-component hydrides (this is described in more detail in Sections 1 and 2 of the review). In addition, a search began for superconductivity in compounds with hydrogen that were more complex in composition at normal pressure.

The review consists of an Introduction and fore sections: 1) Methods for the synthesis of hydrogen with metals and non-metals, 2) Experimental studies and properties of hydrogen-containing superconductors, 3) Physical models explaining the effect of hydrogen on superconductivity,4) Problems and prospects for hydrogen-containing superconductors. and Appendix with physical chemical properties of hydrogen.

The review does not claim to be an exhaustive analysis of all works in the field of hydrogen-containing superconductors. The review includes work mainly from the last approximately 30 years. It was taken into account that there are reviews on superconductors with hydrogen, studied before the advent of high-temperature superconductors (HTSC) in 1986: F.A. Lewis, The Palladium Hydrogen System, 1967, Hydrogen in Metals, ed. G. Alefeld and I. Felkl, 1981, P.V. Geld, R.A. Ryabov, L.P. Mokhracheva, Hydrogen and physical properties of metals and alloys. Transition metal hydrides, 1985.

# 1. Methods for the synthesis of modern hydrogen-containing superconductors.

## 1.1 Synthesis of hydrides having a high critical temperature at high compression pressure.

Table 1 presents hydrides [1-30] in which superconductivity at high pressure is observed. Pressure during the synthesis of hydrides is achieved using diamond anvil cell (DAC). Ammonia borane $NH_3BH_3$ as a source of hydrogen and very thin metal foil located in the DAC are often used in the synthesis of hydrides [6,7,8, 10, 12, 13,14, 16, 17, 18, 19, 20, 21, 22, 23, 24, 25, 27]. Ammonia borane, when heated by laser pulses (up to a temperature of 1600-2400K), dissociates and releases hydrogen (one mole of ammonium borane releases three moles of hydrogen and c-BN – cubic boron nitride epoxy mixture) $NH_3BH_3 \rightarrow 3H_2$+ c-BN. Ammonia borane also serves as a pressure transmitting medium. It should be noted that the pressure in experiments with hydrides is determined either by the shift of the Raman peak for diamond [2,3,6, 8, 9, 13, 16, 17, 18, 23, 24, 25, 27, 28, 29, 30] , or from the $H_2$ vibron scale [9, 11]. Figure 1 shows an image of a cell for $LaH_{10}$ synthesis using ammonia borane in Ref. [7].

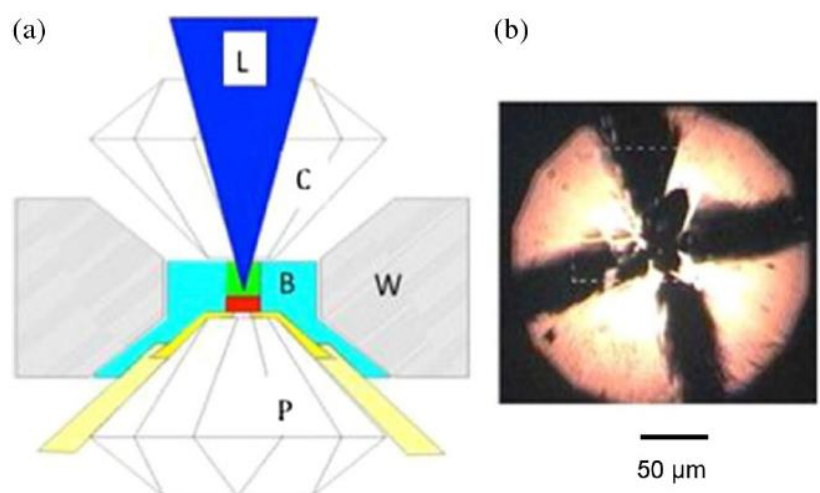

Fig. 1. (a) Schematic of the assembly used for synthesis and subsequent conductivity measurements. The sample chamber consisted of a tungsten outer gasket (W) with an insulating cBN insert (B). The piston diamond (P) was coated with four 1-μm thick Pt electrodes which were pressure-bonded to 25-μm thick Pt electrodes (yellow). The 5-μm thick La sample (red) was placed on the Pt electrodes and packed in with ammonia borane (AB, green). Once the synthesis pressure was reached, single-sided laser heating (L) was used to initiate the dissociation of AB and synthesis of the superhydride. To achieve optimal packing of AB in the gasket hole, we loaded AB with the gasket fixed on the cylinder diamond (C). (b) Optical micrograph of a sample at 178 GPa after laser heating using the above procedure (sample A) [7]. Reprinted from M. Somayazulu, M. Ahart, A. K. Mishra *et al.*, *Phys. Rev. Lett.* **122**, 027001 (2019) with permission of American Physical Society.

In Ref. [2] an $H_3S$ sample was synthesized from gaseous $H_2S$ which was introduced through a capillary into a DAC. At a temperature of 191-213 K, hydrogen sulfide was liquefied and further compressed to the required pressure. At high pressures, $H_2S$ decomposed into $H_3S$ hydride and sulfur.

In Ref. [4] gaseous phosphine was liquefied in DAC at 170 K and then compressed for the synthesis of $PH_3$ hydride.

In Ref. [3] the $H_3S$ hydride was synthesized by heating sulfur with a laser in solid hydrogen at a pressure of 150 GPa, and the reaction $2S+3H_2 \rightarrow 2 H_3S$ was carried out.

In Ref. [15] a C-H-S sample was synthesized by photochemical synthesis. Carbon and sulfur were mixed in a 1:1 ratio, then placed in a DAC and hydrogen gas $H_2$ was introduced. After that, at a pressure of 4 GPa, the sample was heated with a laser with a wavelength of 532 nm at a power of 10-25 mW for several hours.

In Ref. [26] a gas mixture of $H_2/N_2$ (99:1) and lutetium foil in DAC at a pressure of 2 GPa was heated at a temperature of 65 $^0$C to form $LuH_{3-\delta}N_\epsilon$.

In Ref. [5] lanthanum foil (or lanthanum hydride $LaH_3$) and hydrogen in a DAC at high pressures were heated by a laser to 1000-2150K for the synthesis of $LaH_{10}$.

In Ref. [11] an 800 nm thick yttrium film was coated with a 10 nm thick palladium layer and placed in a DAC in a hydrogen environment. Due to this, the yttrium film was transformed into $YH_3$ hydride in 18 hours at a pressure of 4.5 GPa and room temperature. Next, $YH_3$ reacted with hydrogen at a pressure above 130 GPa and heated by a laser to 1800 K, and the $YH_9$ hydride was formed.

In Ref. [9] yttrium superhydride was formed by various methods, either by reacting yttrium foil with liquid hydrogen at 17 GPa and room temperature; either by heating $YH_3$ and hydrogen with a laser at a pressure of 160-175 GPa to 1500 K, or the reaction process took place at even higher pressures and room temperature for a week. The reaction of $YH_3$ with ammonium borane under laser heating was also used.

$BaReH_9$ samples [28] were synthesized through the reduction process of perrhentate in aqueous ethylenediamine solutions.

In Ref. [28, 29] a sulfur plate 8 μm thick was located between two layers of ammonia borane, each 10–15 μm thick, in a DAC. Next, the pressure was increased to 167 GPa and heated with a pulsed laser to 700K to form $H_3S$. Also, a 6-μm-thick $LaH_3$ plate was placed between two layers of ammonium borane, the pressure was increased to 167 GPa and heated to 2000 K to form $LaH_{10}$.

Table 1. Parameters of superconducting hydrides under pressure.

| System | Space group | $T_c$@P | $H_{c1}$, mT | $H_{c2}(0)$, T | $\xi$, nm | $\lambda$, nm | Isotope coefficient, $\alpha$ | $J_c$, $I_c$ | Ref. |
|---|---|---|---|---|---|---|---|---|---|
| $SiH_4$ | $P6_3$ | 17K@96GPa | | | | | | | [1] |
| $H_3S$ | $Im\bar{3}m$ | 203K@155GPa | 30 (exp.) | 60-80 | $\xi_{GL}$=2-2.3 | $\lambda_L$=125 | 0.3 | $10^7 A/cm^2$ (calc.) | [2] |
| $H_3S$ | $Im\bar{3}m$ | 200K@146GPa, $\Delta T_C$=14 K | | | | | | | [3] |
| $PH_3$ | | 103K@207GPa | | | | | | | [4] |
| $LaH_{10}$ | $Fm\bar{3}m$ | 250K@170GPa | | 136 GL | $\xi_{GL}$=1.56-1.86 | | 0.46 | | [5] |
| $LaH_{10}$ | | 250K@165GPa | | 174 GL 223WHH | | | | | [6] |
| $LaH_{10}$ | $Fm\bar{3}m$ | 260K@180-200GPa | | | | | | | [7] |
| $LaH_{10}$ | $Fm\bar{3}m$ $R\bar{3}m$ | 556K@185GPa | | | | | | | [8] |
| $YH_6$ | $Im\bar{3}m$ | 220@183GPa | | 157WHH 107GL | 1.45-1.75 | | 0.39 | | [9] |
| $YH_9$ | $P6_3/mmc$ | 243@201GPa | | 120WHH 92GL | | | 0.5 | | [9] |
| $YH_6$ | $Im\bar{3}m$ | 224@166GPa | | 116 GL 158WHH | $\xi_{BCS}$=2.3 $\xi_{exp}$=1.4-1.7 | $\lambda_L$=93 $\kappa$=40 | 0.4 | $I_c(0)$=1.75A $J_c(0)$=3500A/mm$^2$ @196 GPa | [10] |
| $YH_9$ | | 262@182±8GPaa | | 103 GL | | | 0.48 | | [11] |
| $ThH_9$ | $P6_3/mmc$ | 146K@170GPa | | 38WHH | $\xi_{BCS}$=3 | | | | [12] |
| $ThH_{10}$ | $Fm\bar{3}m$ | 159-161K@174GPa | 24 | 45WHH | $\xi_{BCS}$=2.9 | $\lambda_L$=136 $\kappa$=46 | | $3\times10^8 A/cm^2$ (calc.) | [12] |
| $PrH_9$ | $F\bar{4}3m$ | 9K@145GPa | | | | | | | [13] |
| | $P6_3/mmc$ | | | | | | | | |
| $La_{0.75}Y_{0.25}H_{10}$ | $Fm\bar{3}m$ | 253K@183GPa | | 100 GL 135WHH | $\xi_{BCS}$=1.6 | | | $J_c$(230K)=22 A/mm$^2$ $J_c$(4.2K)= 27700 A/mm$^2$ | [14] |
| $La_{0.7}Y_{0.3}H_6$ | $Im\bar{3}m$ | 237±5K@183GPa | | | | | | | [14] |
| C-H-S | | 287.7±1,2K@267±10GPa | | 62 GL 85WHH | $\xi_{GL}$=2.31 | $\lambda(0)$=3.8 $\kappa$>1.1 | | | [15] |
| $CeH_9$ | $C2/c$ | 100K@130GPa | 13 | | $\xi_{BCS}$=3.4 | 186 $\kappa$=55 | | $1.5\times10^8$ A/cm$^2$ @120GPa | [16] |
| $CeH_9$ | $P6_3/mmc$ | 57K@88GPa | | 24.7GL 33.5WHH | | | 0.49 | | [16] |
| $CeH_{10}$ | $Fm\bar{3}m$ | 115@95GPa | | | | | | | [16] |
| $SnH_{12}$ | $C2/m$ | 70K@200GPa | | 11.2 GL 9.6WHH | $\xi_{GL}$=5.4 | | | | [17] |
| $BaH_{12}$ | $Cmc2_1$ | 20K@140GPa | | 5.1-7GL @150 GPa | $\xi_{BCS}$=6.8 | | 0.47 (calc.) | | [18] |
| $CaH_6$ | $Im\bar{3}m$ | 215K@172GPa | | 203WHH 142GL | 1.27-1.52 | | | | [19] |
| $CaH_6$ | $Im\bar{3}m$ | 210K@160GPa | | 131-196GL 180-268WHH | $\xi_{GL}$=1.1-1.6 | | | | [20] |
| $NdH_9$ | $P6_3/mmc$ | 4.5±0.5K@110GPa | | | | | | | [21] |
| $ZrH_6$ | $Cmc2_1$ | 71K@220GPa, $\Delta T_C$=8K | | | | | | | [22] |
| $HfH_{14}$ | $C2/m$ | 83K@243GPa, $\Delta T_C$=5K | | 31WHH 24GL | $\xi_{GL}$=3.7 | | | | [23] |
| $Lu_4H_{23}$ | $Pm\bar{3}n$ | 71K@218GPa | | 48WHH 36GL | $\xi_{GL}$=3 | | | | [24] |

| | | | | | | | | |
|---|---|---|---|---|---|---|---|---|
| TaH$_3$ | $I\bar{4}3d$ | 30K@197GPa, $\Delta T_C$=5K | | 21WHH 20GL | $\xi_{GL}$=4 | | | [25] |
| LuH$_{3-\delta}$N$_\varepsilon$ | $Fm\bar{3}m$ | 294K@1GPa | | | | | | [26] |
| SbH$_4$ | $P6_3/mmc$ | 116K@184GPa | | 20WHH 16GL | $\xi_{GL}$=4 | | | [27] |
| BaReH$_9$ | $P6_3/mmc$ | 7K@102GPa | | 5 | | | | [28] |
| H$_3$S | $Im\bar{3}m$ | 196K@155GPa | 820@0K | | | $\lambda_L$(0)=22 $\kappa$=12 | $J_c$(100K)=7×10$^6$A/cm$^2$ | [29] |
| LaH$_{10}$ | $Fm\bar{3}m$ | 231K@130GPa | 550@0K | | | $\lambda_L$(0)=30 $\kappa$=20 | $J_c$(100K)=7×10$^6$A/cm$^2$ | [29] |
| H$_3$S | $Im\bar{3}m$ | 195K@155GPa | 360@10K | | | $\lambda_L$(10)=37 $\kappa$(10)=20 | $J_c$(30K)=7.1×10$^6$A/cm$^2$ | [30] |
| LaH$_{10}$ | $C2/m$ | 200K@120GPa | | | | | | [30] |

## 1.2 Synthesis of hydrogen-containing compounds having superconductivity at normal ambient pressure.

### 1.2.1 Electrochemical implantation of hydrogen ions

In Ref. [31] hydrogen was implanted non-uniformly at room temperature into the interlayer spaces of single crystals of compounds of families 11 ($FeSe_{0.93}S_{0.07}$ and $FeS$) and 122 ($BaFe_2As_2$) using the electrolysis of residual water contained in the ionic liquid. A platinum negative electrode was attached to the surface of the sample using silver paste (Fig. 2). A plastic container with an ionic liquid was located between the platinum electrodes. Two types of ionic liquid were used (DEME-TFSI ($C_{10}H_{20}F_6N_2O_5S_2$), FMIM-$BF_4$ ($C_9H_{17}BF_4N_2$)), each giving the same critical temperature increasing. At a voltage of 3V electrolysis of water occurs, hydrogen ions move to the negative electrode and oxygen ions move to the positive electrode. Thus, the surface layer of the sample is saturated with hydrogen. Hydrogen enters the interstices. The process lasts 6 days. In this case, hydrogen penetrates to a depth of about 1-3 microns.

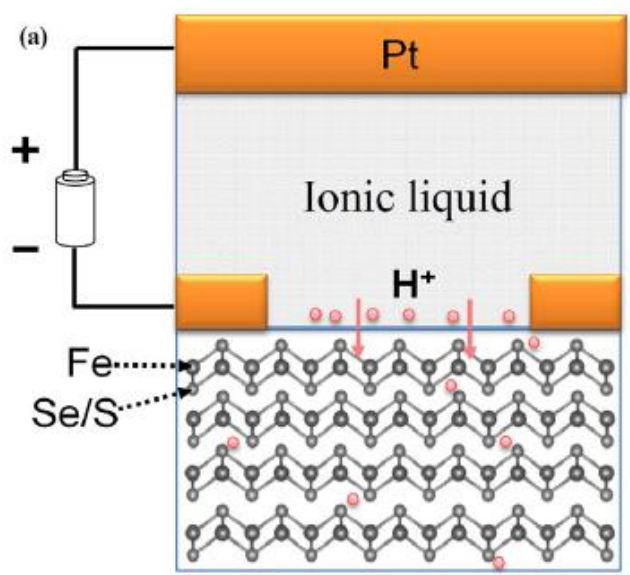

Fig. 2. The configuration for protonation. Two parallel Pt electrodes are placed in the ionic liquid with a distance of 15 mm, applied with a 3 V voltage difference. The sample is attached to the negative electrode [31]. Reprinted from Y. Cui, G.Zhang, H. Li *et al.*, *Science Bulletin* **63**, 11 (2018) with permission of Elsevier.

The paper [32] is devoted to the uniform implantation of hydrogen into layered compounds using the electrolysis of residual water contained in the EMIM-$BF_4$ ($C_6H_{11}BF_4N_2$) ionic liquid. The process of hydrogen implantation occurred at an elevated temperature of 350 K and lasted 12 days. During this time, the residual water in the ionic liquid was completely decomposed into hydrogen and oxygen ions (Fig. 3).

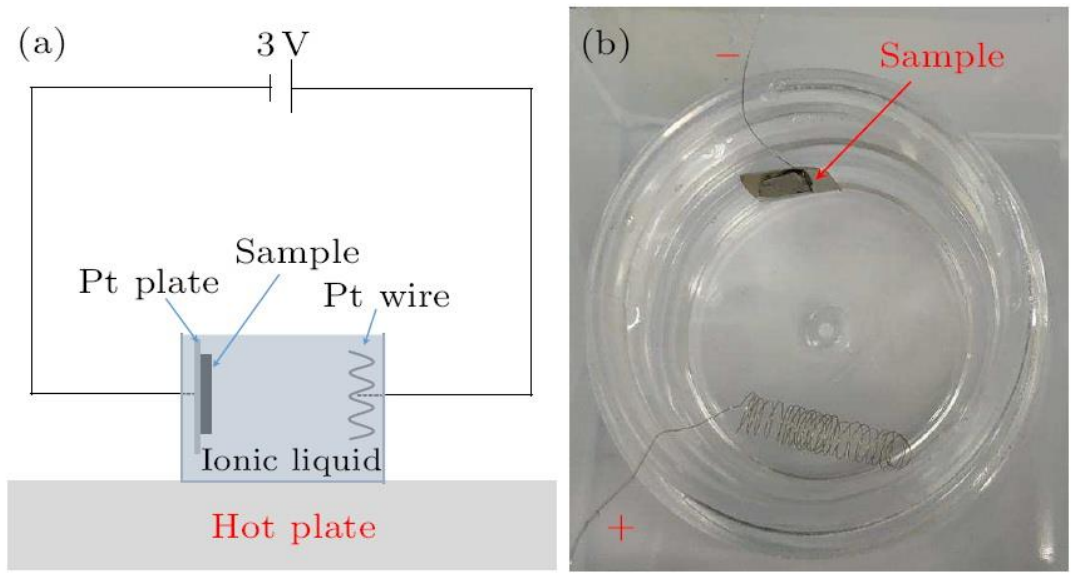

Fig. 3. (a) An illustration of the protonation setup. Platinum electrodes are placed in a container filled with the ionic liquid. The gating voltage is set to be about 3.0 V. The ionic liquid is heated up to 350K by a hot plate. (b) A picture of the positive and negative platinum electrodes, with the sample attached on the negative electrode [32]. Reprinted from Y. Cui, Z. Hu, J.-S. Zhang *et al.*, *Chinese Physics Letters* **36** (7), 077401 (2019).

In Ref. [33] a container with an EMIM-BF$_4$ (1-ethyl-3-methylimidazolium tetrafluoroborate) ionic liquid is used, the electrodes are made of platinum, a FeSe sample (15 μm thick) is attached to the negative electrode using silver paste, the voltage on the electrodes is 3 V, the liquid heated to 330 K – these are optimal conditions. The distance between the electrodes is 15 mm. In this process, as in Ref. [31, 32], electrolysis of the residual water contained in the ionic liquid occurs, oxygen ions move to the positive electrode and hydrogen ions move to the negative electrode (incorporate into FeSe). The process lasts up to 20 days (in this case, the critical temperature of FeSe is reached up to 44 K).

### 1.2.2 Hydrogen thermal diffusion

In this method [34, 35], FeTe$_{0.65}$Se$_{0.35}$ single crystals were saturated with hydrogen by heating the samples in a hydrogen environment at a pressure of 5 atm. and temperatures of 20, 100, 150, 180, 200 and 250 $^0$C for 10-90 hours in a reaction chamber made of stainless steel. At temperatures above 200 $^0$C, dissociation of hydrogen molecules occurs on the catalytically active centers of the iron group metals, i.e. at these temperatures the sample is saturated with atomic hydrogen.

### 1.2.3 Hydrogen ion bombardment

The authors of Ref. [36] implanted molecular and ionized hydrogen into $FeSe_{0.88}$. The intercalation scheme for the FeSe compound in an atmosphere of molecular hydrogen is implemented in a closed, sealed volume with a special device that allows us to regulate the sample temperature from room temperature to ~ 800 °C with an error of ± 2 °C (Fig. 4). Intercalation with hydrogen ions ($H^+$) was carried out using a special ion source (Fig. 4) under the action of a cathode discharge. The current strength in the ion beam was ~ 0.5 – 0.6 mA, the accelerating potential difference was from 500 to +4000 V. The residual pressure in the chamber before the injection of hydrogen was ~ $10^{-4}$ Pa, and when the source is operating, due to the presence of hydrogen, the pressure increases to $10^{-2}$ – $10^{-1}$ Pa. Hydrogen is ionized by primary electrons when voltage is applied to the anode. Subsequently, the discharge is supported by secondary electrons formed during the ionization of the gas. Protons formed during hydrogen ionization were implanted directly into the crystal lattice of the intercalated $FeSe_{0.88}$ sample. Their kinetic energy significantly exceeds the binding energy of a solid body. The concentration of the implanted impurity can reach large values (~10%), and the process of its implantation is possible at sample surface temperatures close to room temperature.

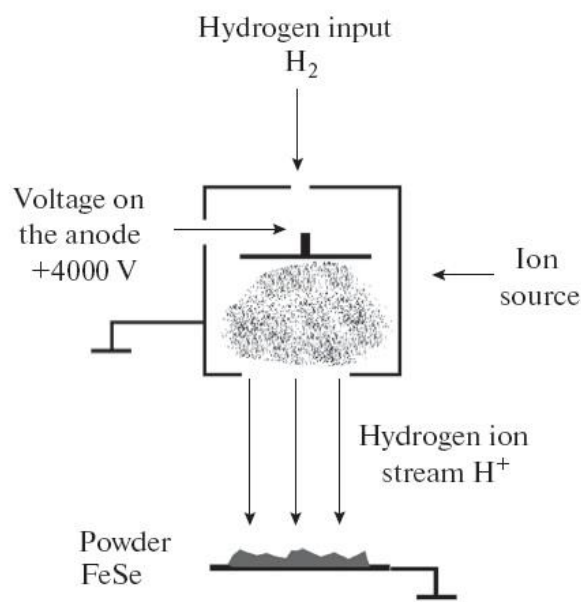

Fig. 4. The ion source circuit [36]. Reprinted from G. S. Burkhanov, S. A. Lachenkov, M. A. Kononov *et al.*, *Inorganic Materials: Applied Research* **8**(5), 759 (2017).

## 2. Experimental studies of the properties of hydrogen-containing superconductors.

### 2.1 Properties of hydrides under pressure

Hydrogen has the smallest atomic mass, therefore high phonon frequencies and high critical temperatures should be expected in hydrogen compounds [37]. However, this requires high pressures created using diamond anvil cells. Here we encounter two obstacles – the size of the samples and the possibility of their application.

Table 1 summarizes the main parameters for 21 superconducting hydrides under pressure, which have been confirmed experimentally [1-30]. This table provides data regarding the structure, the highest critical temperature $T_c$ at a given pressure $P$, the first critical magnetic field $H_{c1}$, the second critical magnetic field $H_{c2}$, coherence length $\xi$, magnetic field penetration depth $\lambda$, Ginzburg-Landau parameter ($\kappa$), isotope coefficient $\alpha$ and critical current density $I_c$ (or critical current $J_c$) for silicon hydride [1], sulfur hydride [2, 3, 29, 30], phosphorus hydride [4], lanthanum hydride [5-8, 29, 30], yttrium hydride [9 -11], thorium hydride [12], praseodymium hydride [13], ternary lanthanum and yttrium hydride [14], carbon-hydrogen-sulfur system [15], cerium hydride [16], tin hydride [17], barium hydride [18], calcium hydride [19-20], neodymium hydride [21], zirconium hydride [22], hafnium hydride [23], lutetium hydride [24], tantalum hydride [25], nitrogen-doped lutetium hydride [26], antimony hydride [27], $BaReH_9$ hydride [28 ]. Pressures from 1 GPa [26] to 280 GPa [15] are required to observe superconductivity in hydrides. The lowest critical temperature is observed for neodymium hydride (5K), and the highest for lutetium hydride doped with nitrogen $LuH_{3-\delta}N_\varepsilon$ (294K). For $LaH_{10}$, the value $T_c$=556K [8] is observed only after thermal cycling (initial value 294K@180GPa). The highest value of the second critical magnetic field (about 200 T at zero temperature) is observed for $LaH_{10}$ [6] and $CaH_6$ [20]. The dimensions of the samples range from a few microns to tens of microns with a thickness from 1 micron to several microns (the exact cross-sectional dimensions are difficult to determine). The small size of the samples and the presence of diamond anvil cells makes it difficult to carry out measurements using SQUID-magnetometers. Isotope effect data indicate a standard pairing mechanism [38, 39].

As a rule, the dependence of $R(T)$ at various pressures and external magnetic field values (up to 5-16 T) is studied by means the four-probe (with the exception of Ref. [3]) van der Pauw method [40, 41]. The value of $H_{c2}(0)$ using the Ginzburg-Landau (GL) [42] or Werthamer-Helfand-Hohenberg (WHH) formula [43] can be estimated from experimental data. Gold-coated tantalum [2, 10, 12, 14] or platinum [20, 25, 26] are used as electrodes. The critical temperature decreasing in an external magnetic field is observed in experiments (Fig. 5).

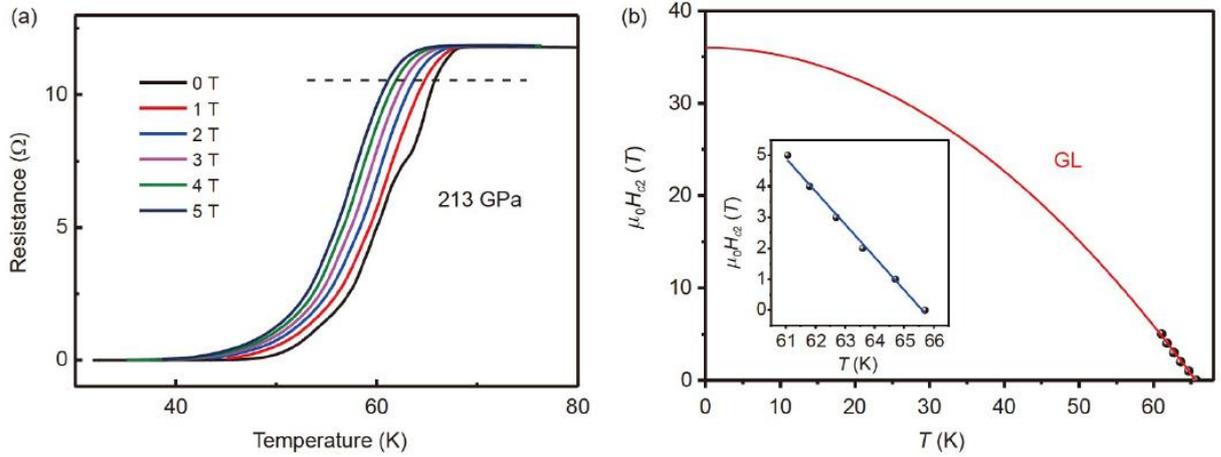

Fig. 5 (Color online) The superconducting parameters. (a) Temperature dependence of electric resistance measured at 213 GPa in different magnetic fields; (b) the upper critical magnetic field $\mu_0 H_{c2}(T)$ with $T_c^{90\%}$ being adopted. The red line is from the GL fitting. The inset shows the linear fitting results [24]. Reprinted from Z. Li, X. He, C. Zhang *et al.*, *Sci. China-Phys. Mech. Astron.* **66**, 267411 (2023).

Replacing hydrogen with deuterium leads to a noticeable (up to 70K) critical temperature decreasing in accordance with the BCS model in Ref. [2, 5, 9, 10, 11, 16]. Fig. 6, Fig. 7 and Fig. 8 shows the results for YH$_6$, YH$_9$ and LaH$_{10}$.

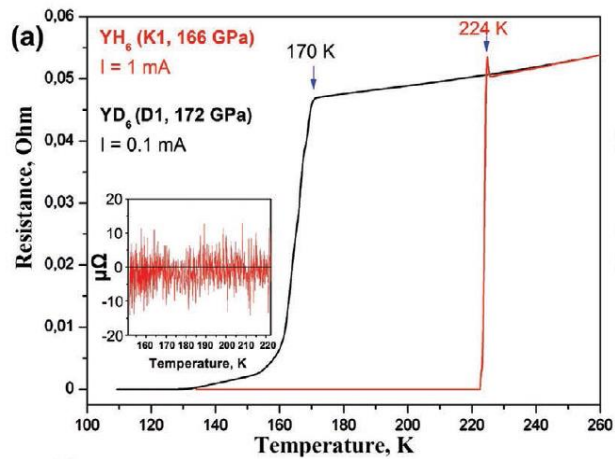

Fig. 6. Superconducting transitions in the Im-3m-YH$_6$: temperature dependence of the electrical resistance $R(T)$ in the YH$_6$ (DAC K1) and YD$_6$ (DAC D1). Inset: the resistance drops to zero after cooling below $T_c$. [10]. Reprinted from I. A. Troyan, D. V. Semenok, A. G. Kvashnin *et al.*, *Adv. Mater.* **2021**, 2006832 (2021) with permission of John Wiley and Sons.

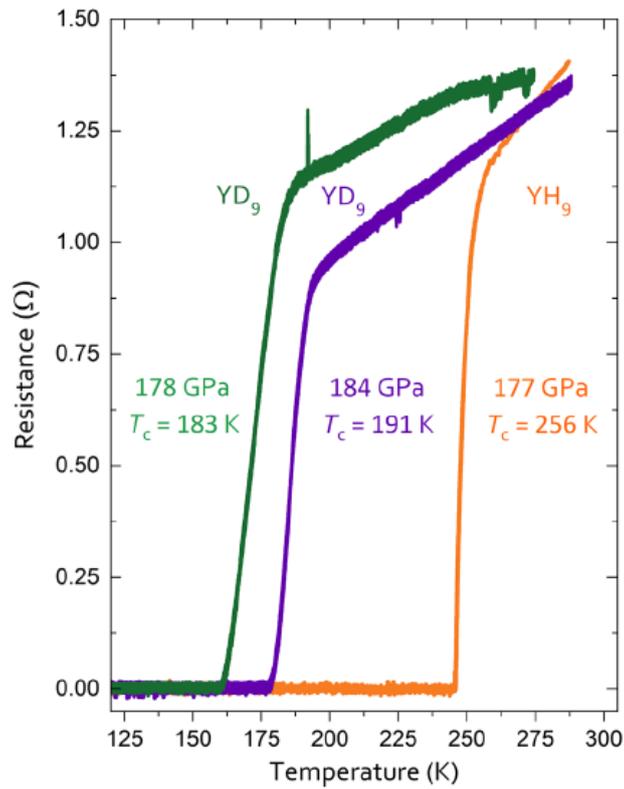

Fig. 7. The substitution of hydrogen with deuterium noticeably affects the value of $T_c$, which shifts to 183 K at 177 GPa. The calculated isotope coefficient at 177 GPa with $T_c$=256 K for YH$_{9\pm x}$ (orange curve) and $T_c$=183 K for the YD$_{9\pm x}$ (green curve) sample is 0.48. By comparing the transition temperatures at around 183 GPa, we obtained α=0.46. Both values are in very good agreement with the Bardeen-Cooper-Schrieffer value of α=0.5 for conventional superconductivity [11]. Reprinted from E. Snider, N. Gammon, R. McBride *et al.*, *Phys. Rev. Lett.* **126**, 117003 (2021) with permission of American Physical Society.

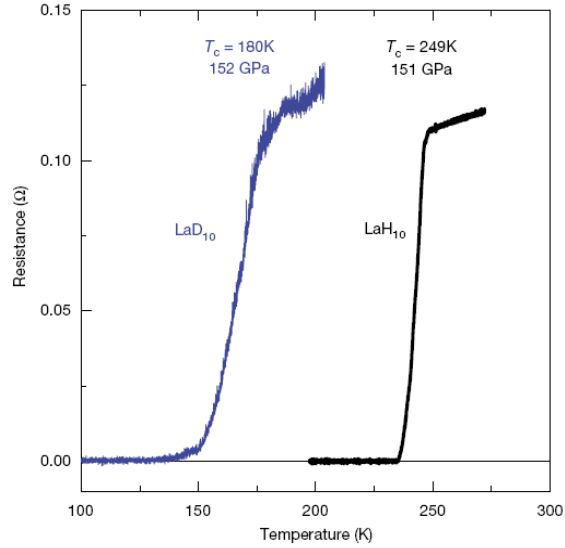

Fig. 8. The isotope effect. The superconductive transition shifts to markedly lower temperatures after hydrogen is replaced by deuterium in samples with the same fcc crystal structure. The black curve corresponds to LaH$_{10}$ (sample 1) and the blue curve corresponds to LaD$_{10}$ (sample 17) [5]. Reprinted from A. P. Drozdov, P. P. Kong, V. S. Minkov *et al.*, *Nature* **569**, 528 (2019).

The dependences of the critical current on temperature and current-voltage characteristics near the transition for YH$_6$ are presented in Fig. 9 [10]. Calculated dependences of the critical current density on the external magnetic field in the low temperature region and comparison with the dependences $J_c(H)$ for known industrial superconductors NbTi, Nb$_3$Sn, YBCO, BSCCO, MgB$_2$ are also given.

Figure 10 shows the dependences of the critical current on temperature and on the external magnetic field for La$_{0.75}$Y$_{0.25}$H$_{10}$ [14].

Figure 11 shows data on magnetization measurements in two modes of the H$_3$S sample [2, 86]. The curve is combined with the resistive transition curve.

Figure 12 presents magnetic susceptibility data for LuH$_{3-\delta}$N$_\varepsilon$ [26].

Figure 13 presents data on measuring the specific heat capacity for LuH$_{3-\delta}$N$_\varepsilon$ samples with a diameter of 60-100 μm [26]. An alternating current calorimetric method was used to measure the specific heat capacity [45].

It should be noted that the dependence of the critical temperature of a hydride on the pressure applied to it often exhibits a maximum with a subsequent critical temperature decreasing with increasing pressure [2, 3, 5, 9, 11, 16, 19, 26] for hydrides H$_3$S, LaH$_{10}$, YH$_9$, CeH$_9$, CaH$_6$, LuH$_{3-\delta}$N$_\varepsilon$. As an example Fig. 14 shows the $T_c(P)$ dependence for sulfur hydride [2, 46, 86]. As pressure increases, the critical temperature decreasing is explained by decreasing of the electron-phonon interaction constant due to anharmonic hardening of soft phonon modes. As the pressure decreases, the critical temperature decreasing is associated with distortion of the crystal structure and decomposition of the compound [9, 47].

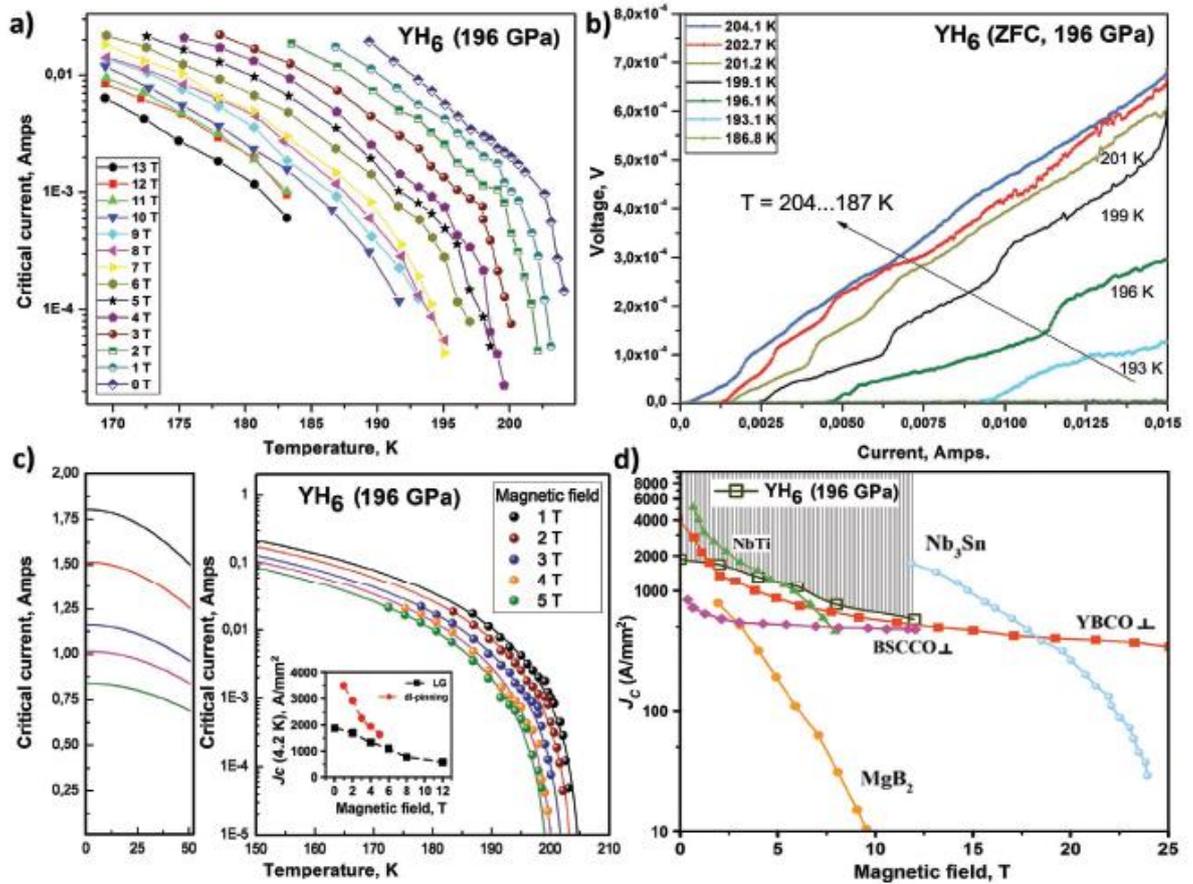

Fig. 9. Dependence of the critical current on temperature and external magnetic fields (0–13 T) in $Im\bar{3}m$-YH$_6$ at 196 GPa. a) The critical current at different magnetic fields near $T_c$ (defined below 50% resistance drop). b) The voltage–current characteristics of the YH$_6$ sample near $T_c$. c) Extrapolation of the temperature dependence of the critical current using the single vortex model [44] $J_c = J_{c0}(1 - T/T_c)^{5/2}(1 + T/T_c)^{-1/2}$; inset: dependencies of the critical current density at 4.2 K on the magnetic field. d) Critical current densities $J_c$ of various industrial superconducting wires and of YH$_6$ (shaded area) at 4.2 K. The lower bound of the critical current density of YH$_6$ was calculated assuming the maximum possible cross-section of the sample of 10×50 μm$^2$ [10]. Reprinted from I. A. Troyan, D. V. Semenok, A. G. Kvashnin *et al.*, *Adv. Mater.* **2021**, 2006832 (2021) with permission of John Wiley and Sons.

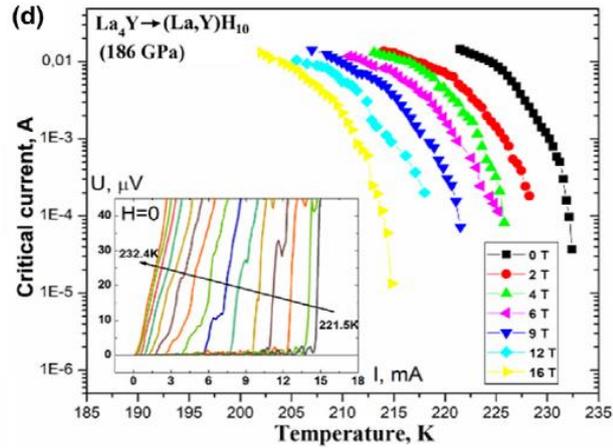

Fig. 10. Dependence of the critical current on the temperature and external magnetic field (0–16 T). The critical currents were measured near $T_c$. Inset: current–voltage characteristic near a superconducting transition [14]. Reprinted from D. V. Semenok, I. Troyan, A. G. Ivanova *et al.*, *Matls. Today* **48**, 18 (2021) with permission of Elsevier.

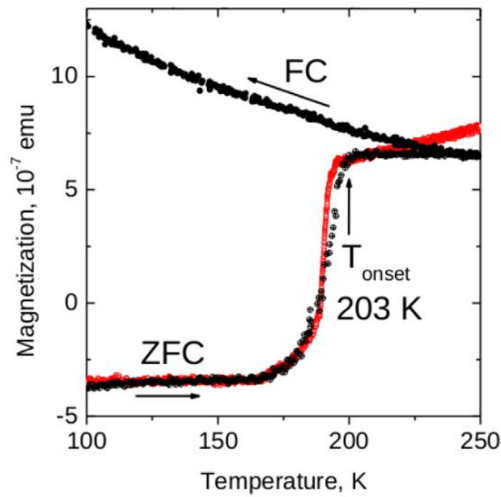

Fig. 11. Temperature dependence of the magnetization of sulfur hydride at 155 GPa in zero-field cooled (ZFC) and 20 Oe field cooled (FC) modes (black circles). The onset temperature is $T_c$ = 203 K. For comparison, the superconducting step obtained for sulfur hydride from electrical measurements at 145 GPa is shown by red circles. [86]. Reprinted from J. A. Flores-Livas, *et al.*, *Phys. Rep.* **856**, 1 (2020).

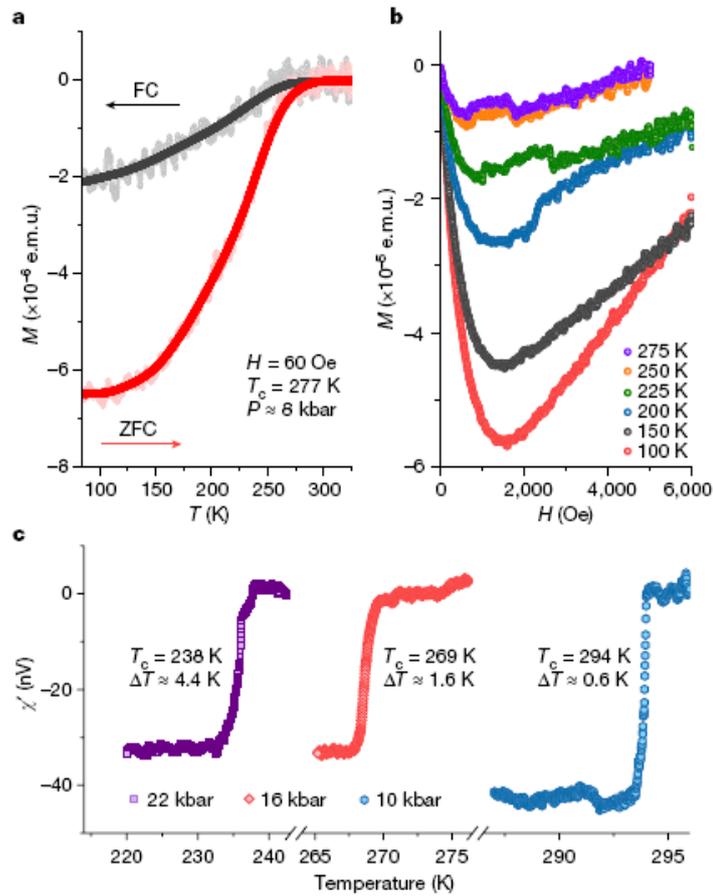

Fig. 12. **Magnetic susceptibility. a,** Magnetic susceptibility ($\chi = M/H$, in which $M$ is magnetization and $H$ is magnetic field) as a function of temperature ($T$) under conditions of zero field cooling (ZFC) and field cooling (FC) at a d.c. field of 60 Oe. **b,** $M$–$H$ curves recorded close to zero field. **c,** a.c. susceptibility ($\chi$) in nanovolts versus temperature at select pressures, showing marked diamagnetic shielding of the superconducting transition for pressures of 10–22 kbar [26]. Reprinted from N. Dasenbrock-Gammon, E. Snider, R. McBride *et al.*, *Nature* **615**, 244 (2023).

The superconducting transition shifts rapidly under pressure to lower temperatures. $T_c$ is determined from the temperature at the onset of the transition. A cubic or quadratic fit of the background signal has been subtracted from the data. We have applied a ten-point adjacent average smoothing for all d.c. magnetization data.

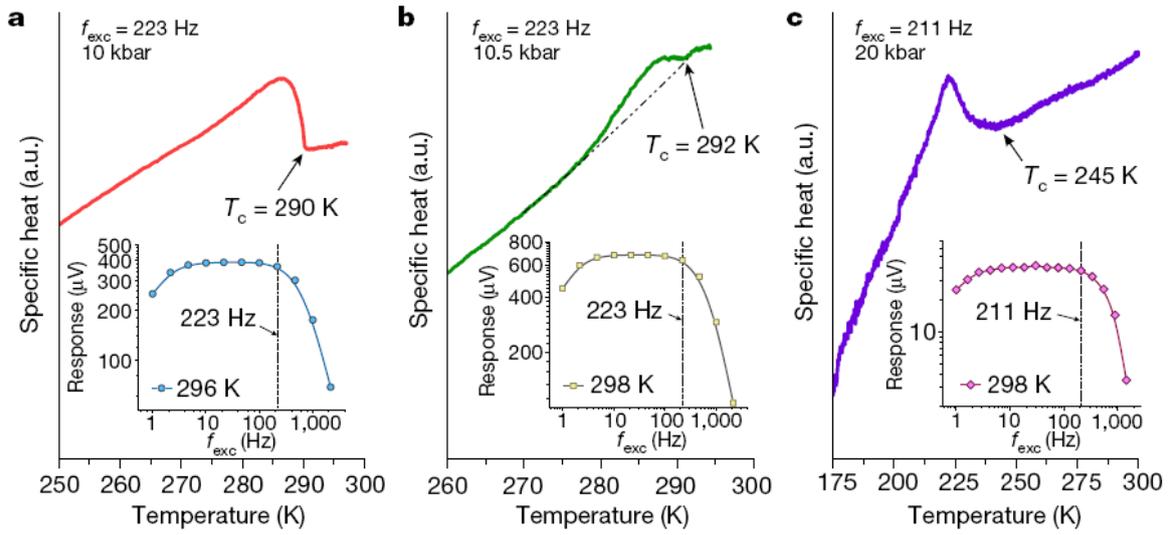

Fig. 13. Specific-heat-capacity measurement on the superconducting lutetium–nitrogen–hydrogen system. **a**–**c**, Specific heat capacity of nitrogen-doped lutetium hydride at 10 kbar (**a**), 10.5 kbar (**b**) and 20 kbar (**c**), showing the superconducting transition as high as 292 K at 10.5 kbar in **b** [26]. Reprinted from N. Dasenbrock-Gammon, E. Snider, R. McBride *et al.*, *Nature* **615**, 244 (2023).

The drive frequency ($f_{exc}$) and frequency sweeps of each measurement are depicted in the insets. The strength of the heat-capacity anomaly associated with superconductivity varied owing to volume fraction as shown in **c**. The dashed line is a guide to the eye to distinguish the trend of the heat-capacity anomaly before and after the transition.

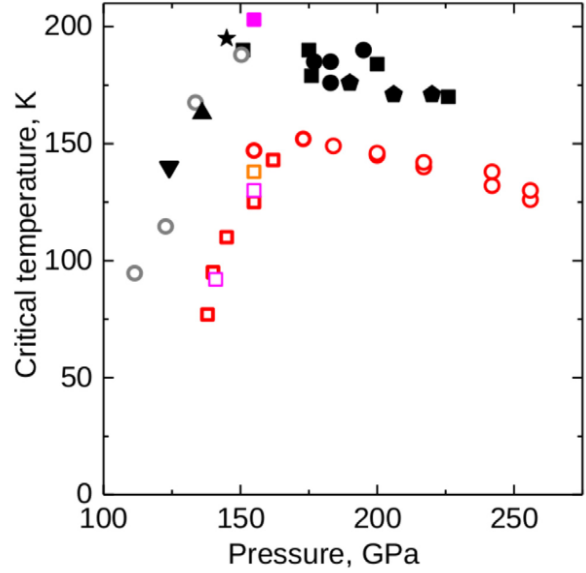

Fig. 14. Summary of critical temperatures of superconductivity upon pressure, solid shapes (black and pink) shows for sulfur hydride ($H_xS$), and open shapes (except gray) represent sulfur deuteride ($D_xS$). Gray represent a different batch. Shown only data on annealed samples. The highest measured $T_c$ is 203 K [86]. Reprinted from J. A. Flores-Livas, *et al.*, *Phys. Rep.* **856**, 1 (2020).

Sulfur hydride is the most studied of all hydrides synthesized at high pressures. The energy gap value of $2\Delta=76$ meV was determined at a temperature of 50 K by means optical spectrometry for an $H_3S$ sample with a diameter of about 80 μm [48]. This value turned out to be higher than the calculated value $2\Delta=73$ meV [49] at the same temperature. In addition, experiments to study the dependence of the magnetization of an $H_3S$ sample with a thickness of 2.8 μm and a diameter of 85 μm (demagnetization correction $(1-N)^{-1} = 8.6$) and a $LaH_{10}$ sample with a thickness of 1.9 μm and a diameter of 70 μm (demagnetization correction $(1-N)^{-1} = 13.5$) on temperature and magnetic field [29, 50 ] were carried out. The penetration field values $H_P(0K)=96$ mT for $H_3S$ and $H_P(0K)=41$ mT for $LaH_{10}$ are obtained from these experiments. Taking into account the relationship

$$H_{c1} = \frac{H_P}{1-N}, \quad (1)$$

the value $H_{c1}(0K) = 820$ mT for $H_3S$ and 550 mT for $LaH_{10}$ was determined. Based on the Bean model [51, 52], the critical current density can be determined using the formula for the magnetic moment of the sample ($m$):

$$m = \frac{\pi}{3} J_c h \left(\frac{d}{2}\right)^3, \quad (2)$$

where $J_c$ is the critical current density, $h$ is the sample thickness, $d$ is the sample diameter. The estimation from formula (2) in [29] gave the value $J_c(100K)=7\times10^6$ A/cm$^2$ for sulfur and lanthanum hydride, which is comparable to the critical current density for HTSC wires [53].

Measurements of the trapped magnetic moment [30] allowed us to most accurately determine the magnitude of the penetration field in the ZFC mode in the H$_3$S sample. It amounted to $H_P(10K)=42\pm3$ mT (Fig. 15), taking into account the demagnetizing factor $H_{c1}(10K)=0.36$ T.

Figure 16 shows the dependence of the trapped magnetic moment on the external magnetic field. They correspond to Bean's model. Based on the value of the trapped magnetic moment at 30K $m_{trap}{}^S(30K)=1.6\times10^{-8}$ Am$^2$, the authors of Ref. [30] using formula (2) determined the value of the critical current density $J_c(30K)=7.1\times10^6$ A/cm$^2$. At a temperature of 0K, extrapolation of experimental data gives $J_c(0K)=7.31\times10^6$ A/cm$^2$. This value is almost two orders of magnitude lower than the critical current density at a temperature of 4.2 K for single crystals of iron-based superconductors [54, 55].

Also, the authors of Ref. [30] estimated the depairing current density $J_d=4\times10^9$ A/cm$^2$ using the formula from Ref. [56]:

$$J_d = \frac{\Phi_0}{3\sqrt{3\pi}\mu_0} \frac{1}{\lambda_L^2\xi} \ . \qquad (3)$$

Figure 17 shows the dependences of the trapped magnetic moment $m_{trap}$ on time $t$. From these dependencies, the flow creep rate for H$_3$S was calculated using the formula

$$S = \frac{1}{m_{trap,0}} \frac{dm_{trap}}{d\ln t} \ . \qquad (4)$$

It amounted to $S=0.002$ at a temperature of 165K and $S=0.005$K at a temperature of 185K, which is typical for type II superconductors and MgB$_2$ [57, 58].

Estimation of the Ginzburg-Levanyuk number (it gives an idea of the scale of creep and melting of vortices in a superconductor) for H$_3$S in Ref. [30] gives the value $Gi=7\times10^{-6}$ according to the formula

$$Gi = \frac{1}{2}\left(\frac{2\pi\mu_0 k_B T_c \lambda_L^2}{\Phi_0^2\xi}\right)^2 \ , \qquad (5)$$

where $k_B$ is Boltzmann's constant. This value is lower than for cuprate ($10^{-2}$) and iron-based HTSC ($10^{-5}$-$10^{-2}$) [57, 59], and is comparable in value for LTSC ($10^{-9}$-$10^{-6}$) and MgB$_2$ ($10^{-6}$). All this indicates strong pinning in sulfur hydride.

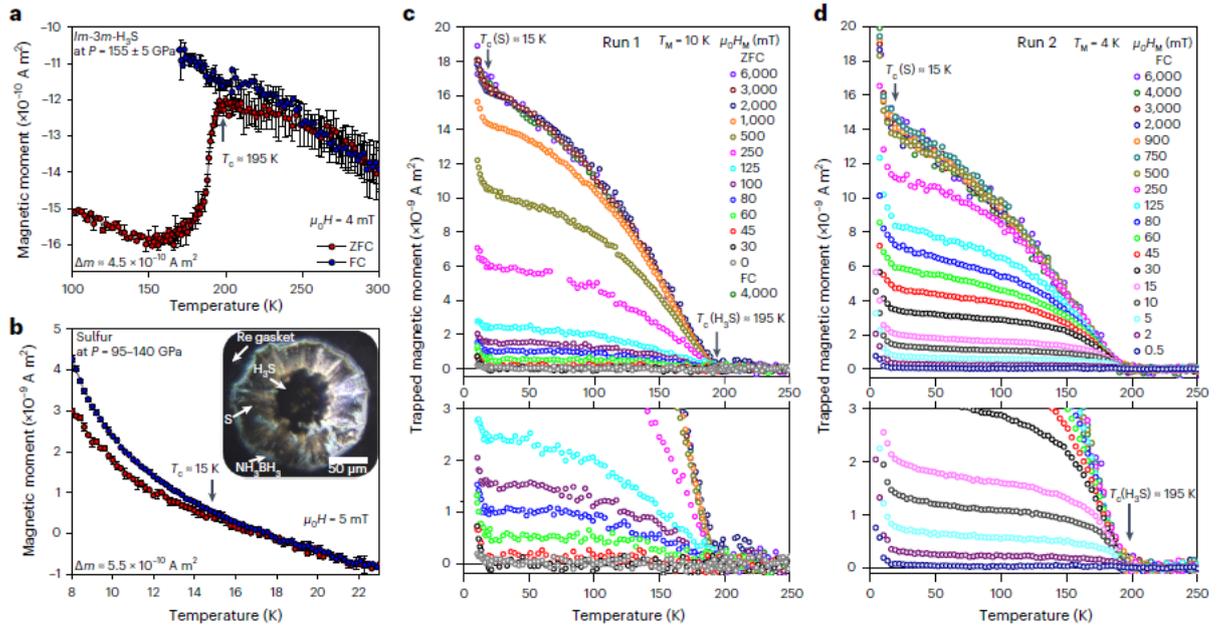

Fig. 15. Magnetic measurements of sample containing $H_3S$ and elemental sulfur. **a**, ZFC and FC $m(T)$ data of $Im$-$3m$-$H_3S$ at $155 \pm 5$ GPa. **b**, ZFC and FC $m(T)$ data of sulfur at 95–140 GPa. The error bars represent the standard deviation from multiple measurements. The inset shows a photograph of the sample and the spatial distribution of $H_3S$ and S. **c,d**, Temperature dependence of a trapped magnetic moment at zero field generated under ZFC (run 1; **c**) and FC (run 2; **d**) conditions. The open circles of different colours correspond to the temperature dependence of the trapped flux created at different magnetic fields $\mu_0 H_M$ (0–6 T). The lower panels reveal the beginning of the penetration of flux lines into the $H_3S$ sample above $\mu_0 H_M = 45.0$ mT (run 1) and already at $\mu_0 H_M = 0.5$ mT (run 2) [30]. Reprinted from V. S. Minkov, V. Ksenofontov, S. L. Bud'ko *et al.*, *Nature Physics* **19**, 1293 (2023).

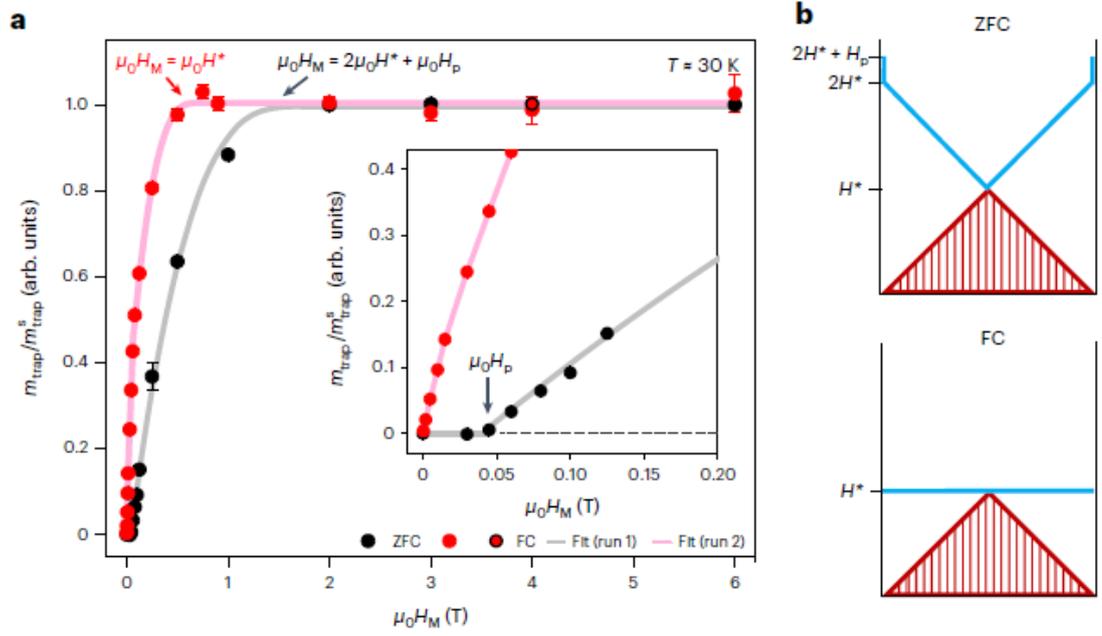

Fig. 16. Trapped magnetic moment in *Im-3m*-H$_3$S sample at 30 K. **a**, Dependence of a trapped magnetic moment at zero field on magnetization field $\mu_0 H_M$ measured in the ZFC and FC modes (runs 1 and 2). The circles correspond to the experimental data, and the magenta and grey curves are guides for the eye. The trapped flux was created in the ZFC (black circles) and FC (red circle) modes at several applied magnetic fields $H_M$. The inset with the enlarged plot shows the entry of the magnetic field into the sample at low $H_M$. The error bars represent uncertainties in the estimation of values of $m_{trap}/m^s_{trap}$. **b**, Profile of magnetic field in the disc-shaped sample in applied magnetic field $H_M = 2H^* + H_p$ (ZFC mode) and $H_M = H^*$ (FC mode) (area below the blue lines) and after removing the applied magnetic field (hatched red area) [30]. Reprinted from V. S. Minkov, V. Ksenofontov, S. L. Bud'ko *et al.*, *Nature Physics* **19**, 1293 (2023).

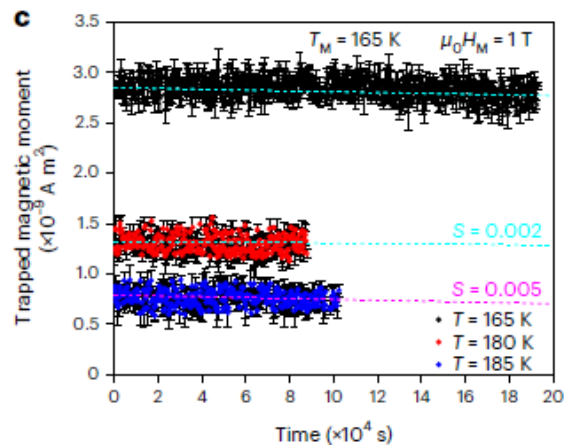

Fig. 17. Vortex pinning in H$_3$S. Creep of the trapped flux in H$_3$S at several temperatures near $T_c$ (the trapped flux was generated at $\mu_0 H_M = 1$ T and $T_M = 165$ K). The error bars represent the standard deviation from multiple measurements [30]. Reprinted from V. S. Minkov, V. Ksenofontov, S. L. Bud'ko *et al.*, *Nature Physics* **19**, 1293 (2023).

Let's touch on some calculations of the critical temperature for hydrides that exhibit superconducting properties at high pressure. As a rule, calculations were obtained before the synthesis and experimental discovery of superconductivity in the corresponding hydrides. Table 2 shows the calculated values of the critical temperature ($T_c$) at a given pressure ($P$), electron-phonon interaction constants ($\lambda$), values of the Coulomb pseudopotential ($\mu^*$), information on the crystal structure and calculation method. It should be noted that these calculations do not take into account anharmonic effects and anisotropy of the superconducting gap. Anharmonic effects can lead to a decreasing of the critical temperature by 20-25K and the electron-phonon interaction constant by 20-25%, as well as an increasing of the logarithmic frequency by 40-50%. Calculations were carried out by numerically solving the Eliashberg equations [60] (E in the Method column), using the Allen-Dines formula [61, 62] (A.-D. in the Method column) and using density functional theory for superconductors (SCDFT) [63, 64]. In the Allen-Dines expression, the critical temperature

$$T_c = \frac{\omega_{\log}}{1.2} \exp\left( \frac{-1.04(1+\lambda)}{\lambda - \mu^*(1+0.62\lambda)} \right), \qquad (6)$$

$$\lambda = 2\int_0^\infty \frac{\alpha^2 F(\omega)}{\omega} d\omega, \qquad (7)$$

$$\omega_{\log} = \exp\left[ \frac{2}{\lambda} \int \frac{d\omega}{\omega} \alpha^2 F(\omega) \ln(\omega) \right], \qquad (8)$$

where $\alpha^2 F(\omega)$ is the Eliashberg spectral phonon function, $\omega$ is the frequency, $\omega_{\log}$ is the logarithmic frequency.

As we see from a comparison of the experimental values of the critical temperature (Table 1) and the calculated values of the critical temperature (Table 2), good agreement between experiment and calculations is observed in some cases for hydrides of sulfur, zirconium, praseodymium, hafnium, silicon, calcium, yttrium ($YH_9$), cerium ($CeH_9$). At the same time, the experiment observed 25-30K lower $T_c$ compared to the calculation for $YH_6$, 25-35K lower for $LaH_{10}$, 60-80K lower for $ThH_{10}$, 20-30K lower for $BaH_{12}$, 10K lower for $SbH_4$, 10-25K lower for $SnH_{12}$, 40-50K lower for $CeH_{10}$. This can be explained, as noted above, by the fact that the calculations do not take into account the anisotropy of the superconducting gap and anharmonicity [47].

More information on calculations of the critical temperature of hydrides can be found in articles and reviews [47, 85, 86, 87, 88, 89].

We will also touch upon issues of crystal structure. Hydrides $H_3S$, $PH_3$, $SiH_4$, $SbH_4$, $SnH_n$ are covalent, and hydrides $CaH_6$ (ionic hydride), $LaH_{10}$, $Lu_4H_{23}$ are clathrate, $BaH_{12}$ is a molecular hydride. Figure 18 shows the crystal structures for hexagonal $P6_3$-$SiH_4$, $Im$-$3m$-$H_3S$, hexagonal $P6_3/mmc$-$NdH_9$, cubic $Fm$-$3m$-$LaH_{10}$, cubic $I$-$43d$-$TaH_3$, monoclinic $C_2/m$-$SnH_{12}$, cubic $Im$-$3m$-$CaH_6$, cubic $Pm$-$3n$-

Lu$_4$H$_{23}$, pseudocubic *Cmc2$_1$*-ZrH$_6$. They were initially obtained by computer calculations and modeling and confirmed experimentally by X-ray diffraction. It should be noted that the X-ray diffraction method cannot accurately determine the position of hydrogen atoms in the hydride lattice due to the small X-ray scattering factor of hydrogen.

Table 2. Calculated critical temperatures for hydrides

| System | *P* (GPa) | λ | μ* | *T$_c$*(K) | Method | Ref. | Space group |
|--------|-----------|---|-----|-----------|--------|------|-------------|
| SiH$_4$ | 190 | 0.58 | 0.13 | 16.5 | A.-D. | [65] | *Pbcn* |
| SiH$_4$ | 70-150 | 1.2-0.7 | 0.1 | 75-20 | A.-D. | [66] | *Cmca* |
| H$_2$S | 160 | 1.2 | 0.13 | 80 | A.-D. | [67] | *Cmca* |
| H$_3$S | 200 | 2.19 | 0.13-0.1 | 191-204 | A.-D. | [68] | *Im-3m* |
| PH$_2$ | 200 | 1.04 | 0.18-0.1 | 49.2-75.6 | A.-D. | [69] | *C2/m* |
| PH$_2$ | 200 | 1.13 | 0.18-0.1 | 48.0-70.4 | A.-D. | [70] | *I4/mmm* |
| PH$_3$ | 200 | 1.45 | 0.13 | 83 | A.-D. | [71] | *C2/m* |
| PH$_3$ | 200 | 1.05 | | 55 | SCDFT | [72] | *C2/m* |
| SbH$_4$ | 150 | 1.26 | 0.13-0.1 | 95-106 | A.-D. | [73] | *P6$_3$/mmc* |
| CaH$_6$ | 150 | 2.69 | 0.13-0.1 | 220-235 | E | [74] | *Im-3m* |
| YH$_6$ | 120 | 2.93 | 0.13-0.1 | 251-264 | E | [75, 76] | *Im-3m* |
| YH$_9$ | 150 | 4.42 | 0.13-0.1 | 253-276 | E | [75] | *P6$_3$/mmc* |
| LaH$_{10}$ | 200 | 2.28 | 0.1 | 288 | E | [75] | *Fm-3m* |
| LaH$_{10}$ | 210 | 3.41 | 0.13-0.1 | 274-286 | E | [77] | *Fm-3m* |
| LaH$_{10}$ | 200 | 3.57 | 0.13-0.1 | 238 | E | [78] | *C2/m* |
| LaH$_{10}$ | 200 | 3.57 | 0.13-0.1 | 229-245 | A.-D. | [78] | *C2/m* |
| CeH$_{10}$ | 200 | 0.7 | 0.13-0.1 | 50-55 | E | [75] | *Fm-3m* |
| CeH$_{10}$ | 94 | 1.952 | 0.13-0.1 | 156.66-168.13 | A.-D. | [79] | *Fm-3m* |
| CeH$_9$ | 95 | 1.596 | 0.13-0.1 | 130.78-142.55 | A.-D. | [79] | *F-43m* |
| CeH$_9$ | 200 | 2.30 | 0.15-0.1 | 105-117 | A.-D. | [80] | *P6$_3$/mmc* |
| CeH$_9$ | 100 | 1.48 | 0.15-0.1 | 63-75 | A.-D. | [80] | *C2/m* |
| ZrH$_6$ | 215 | 0.92 | 0.13 | 70 | E | [81] | *Cmc2$_1$* |
| HfH$_{14}$ | 300 | 0.93 | 0.1 | 76 | A.-D. | [82] | *C2/m* |
| TaH$_3$ | 80 | | | 23 | A.-D. | [83] | *I-43d* |
| ThH$_{10}$ | 100 | 2.5 | 0.15-0.1 | 220-241 | E | [84] | *Fm-3m* |
| SnH$_{12}$ | 250 | 1.25 | 0.13-0.1 | 83-93 | A.-D. | [85] | *C2/m* |
| BaH$_{12}$ | 150 | 1.02 | 0.15-0.1 | 39-53 | A.-D. | [18] | *Cmc2$_1$* |
| PrH$_9$ | 120 | 0.48 | 0.1 | 8.4 | A.-D. | [13] | *P6$_3$/mmc* |
| LaYH$_{20}$ | 180 | 3.87 | 0.15-0.1 | 232-266 | A.-D. | [14] | *R-3m* |
| LaYH$_{12}$ | 180 | 2.82 | 0.15-0.1 | 176-203 | A.-D. | [14] | *Pm-3m* |

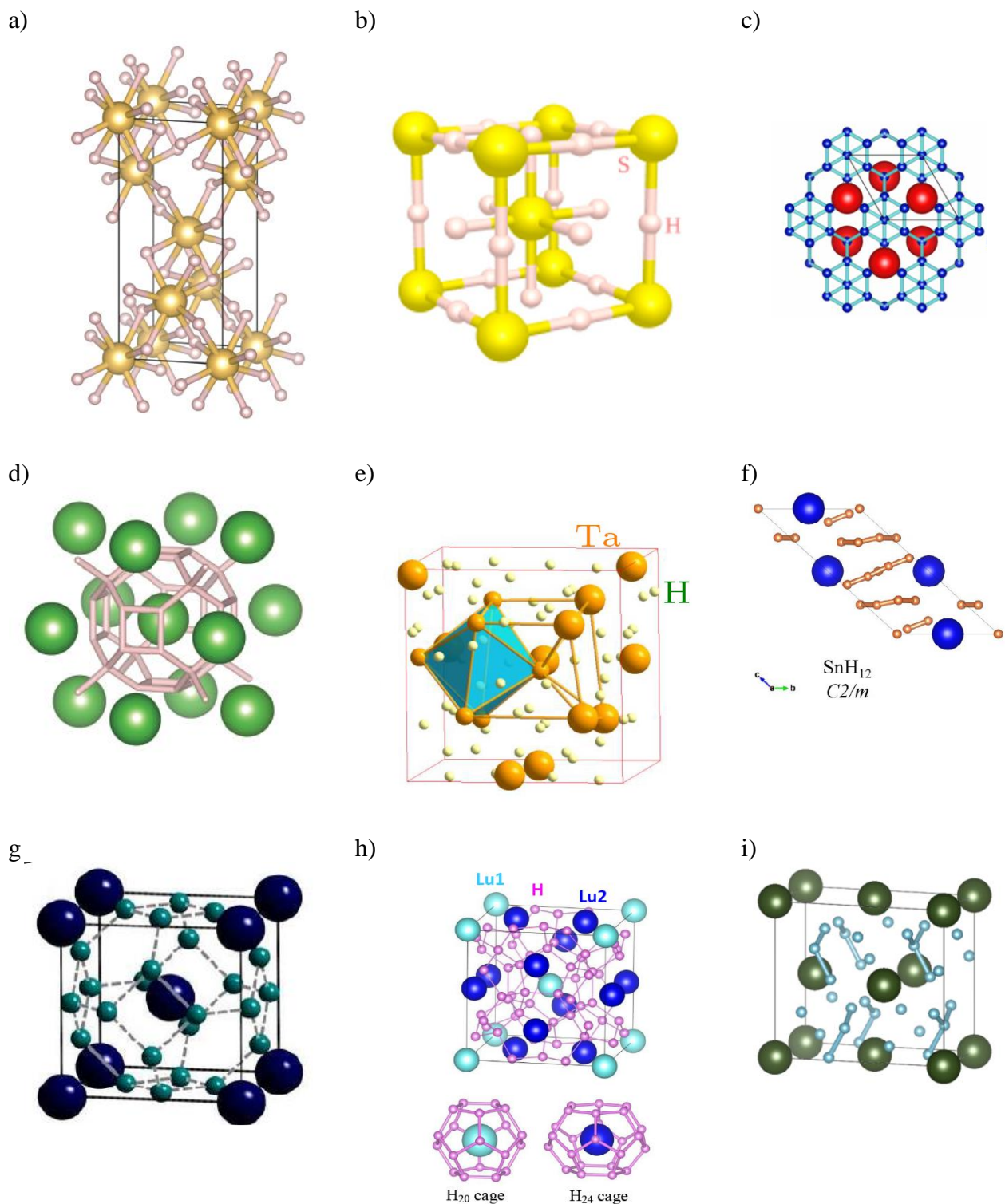

a)

b)

S

H

c)

d)

e)

Ta

H

f)

SnH₁₂

*C2/m*

g)

h)

Lu1    H    Lu2

H₂₀ cage    H₂₄ cage

i)

Fig. 18. Crystal structure of SiH₄[85] at 100 GPa (a) reprinted from C. J. Pickard, I. Errea, M. I. Eremets, *Annu. Rev. Condens. Matter Phys.* **11**, 57(2020); SH₃[86] at 200 GPa (b) reprinted from J. A. Flores-Livas, *et al.*, *Phys. Rep.* **856**, 1 (2020); NdH₉ [21] at 120 GPa (c) reprinted from D. Zhou, D. V. Semenok, H. Xie *et al.*, *J. Am. Chem. Soc.***142**, 2803 (2020) with permission of American Chemical Society; LaH₁₀ [85] at 200 GPa (d) reprinted from C. J. Pickard, I. Errea, M. I. Eremets, *Annu. Rev. Condens. Matter Phys.* **11**, 57(2020); TaH₃ [ 25] at 195 GPa (e) reprinted

from X. He, C. L. Zhang, Z. W. Li *et al.*, *Chinese Physics Letters* **40**, 057404 (2023); $SnH_{12}$ [17] at 250 GPa (f) reprinted from F. Hong, P.F. Shan, L.X. Yang *et al.*, *Materials Today Physics* **22**, 100596 (2022) with permission of Elsevier; $CaH_6$ [19] at 200 GPa (g) reprinted from L. Ma, K. Wang, Y. Xie *et al.*, *Phys. Rev. Lett.* **128**, 167001 (2022) with permission of American Physical Society; $Lu_4H_{23}$ [24] at 185 GPa (h) reprinted from Z. Li, X. He, C. Zhang *et al.*, *Sci. China-Phys. Mech. Astron.* **66**, 267411 (2023); $ZrH_6$ [80] at 160 GPa ( i) reprinted from K. Abe, *Physical Review B* **98**, 134103 (2018) with permission of American Physical Society.

If we consider metal hydrides $REH_X$ (RE is a metal), then in the case of $REH_6$, the metal atom RE is located in the center of the hydrogen cage $H_{24}$ of 24 hydrogen atoms [74] and this metal atom is an electron donor. The hydrogen atoms are weakly bonded covalently to each other within the $H_{24}$ cage. In the case of $REH_9$ hydride, the RE metal atom is located at the center of the $H_{29}$ hydrogen cage of 29 hydrogen atoms. If we are dealing with the $REH_{10}$ hydride, then the RE metal atom is located in the center of the $H_{32}$ hydrogen cage of 32 hydrogen atoms (Fig. 19).

The structure of the hydrides $PrH_9$, $CeH_9$, $YH_9$, $ThH_9$ is similar to the structure of $NdH_9$. The structure of $YH_{10}$, $ThH_{10}$ is similar to $LaH_{10}$, and the structure of $YH_6$ is similar to the structure of $CaH_6$.

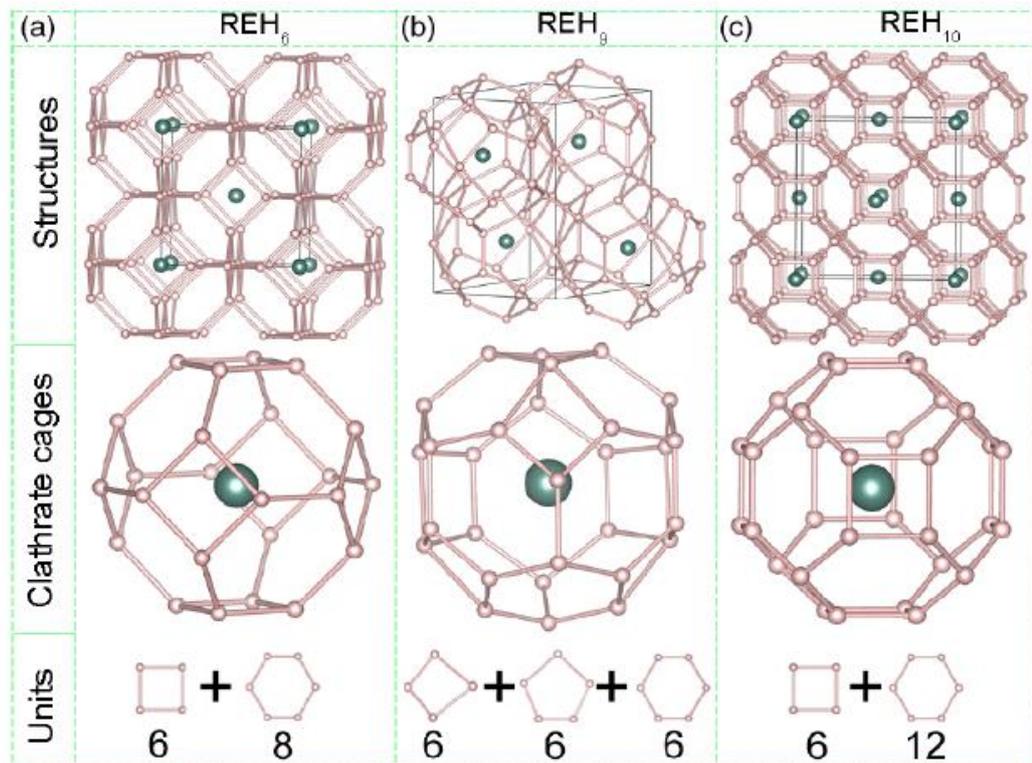

Fig. 19. Clathrate structures of $REH_6$ (a), $REH_9$ (b), and $REH_{10}$ (c). The small and large spheres represent H and RE atoms, respectively. The middle panel depicts the RE-centered $H_{24}$, $H_{29}$, and $H_{32}$ cages of $REH_6$, $REH_9$, and $REH_{10}$, respectively. Each $H_{24}$ or $H_{32}$ cage with $O_h$ or $D_{4h}$ symmetry contains six squares and eight hexagons or six squares and twelve hexagons. One $H_{29}$ cage consists of six irregular squares, six pentagons, and six hexagons [74]. Reprinted from F. Peng, Y. Sun, C. J. Pickard *et al.*, *Phys. Rev. Lett.* **119**, 107001 (2017) with permission of American Physical Society.

Among the important structural parameters of hydrides, one can highlight the minimum (nearest) distance between hydrogen atoms. At high pressures it is in the range $d_{min}$(H-H) = 1-2 Å for different hydrides and decreases with increasing pressure. Figure 20 from Ref. [79] shows the dependence of the nearest distance between hydrogen atoms on pressure for the hydrides $CeH_9$ [79], $FeH_3$ [90], $FeH_5$ [91], $AlH_3$ [92], $H_3S$ [68], $LaH_{10}$ [93], and atomic hydrogen [91]. Note that despite the successful synthesis of iron hydrides, for example $FeH_5$, there is no superconductivity in them. As for aluminum hydride, even if it is superconducting, its critical temperature is below 4K [47]. Uranium hydrides have also been successfully synthesized, but have a very low critical temperature of less than 4K [47].

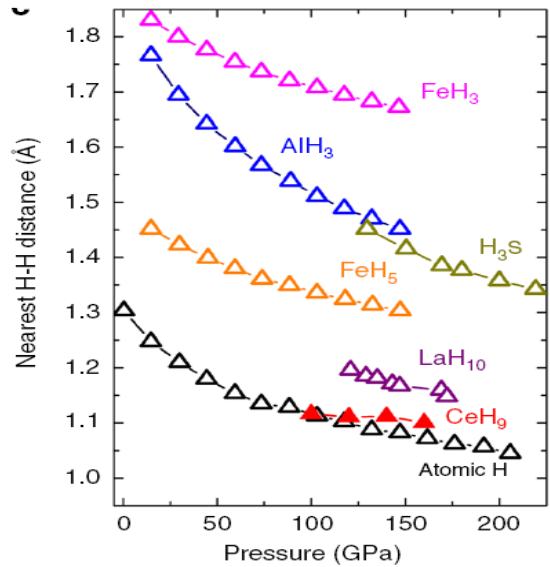

Fig. 20. Comparison of the pressure dependence of the nearest H–H distances for CeH$_9$, FeH$_3$, FeH$_5$ , AlH$_3$ , LaH$_{10}$ , H$_3$S  and atomic H. Magenta, blue, dark yellow, orange, purple, and black open triangle symbol line represents FeH$_3$, AlH$_3$, H$_3$S, FeH$_5$, LaH$_{10}$, and atomic H, respectively. Red solid triangle symbol line represents CeH$_9$ [79]. Reprinted from N. P. Salke, M. M. DavariEsfahani, Y. Zhang *et al.*, *Nature Communications* 10, 4453 (2019).

## 2.2. Properties of hydrogen-containing superconductors at normal atmospheric pressure.

In accordance with the Bardeen, Cooper, Schrieffer (BCS) theory of superconductivity (SC), the effect of hydrogen on superconductivity can be caused by the peculiarity of the interaction of phonons at the optical frequency with the electronic system of the superconductor while maintaining the isotopic effect, in which heavier interstitial atoms reduce critical temperature $T_c$, and lighter ones should increase $T_c$ [94]. The discovery of a strong inverse isotope effect for PdH$_x$ and PdD$_x$ clearly contradicts the original version of the BCS theory. An even larger inverse isotope effect was found for tritium in Pd. The picture became significantly more complicated after reports of experiments on the hydrogenation of superconducting cuprates [95], magnesium diboride MgB$_2$ [96], carbon-based [97] and iron-based superconductors (IBSC, pnictides and chalcogenides) [98, 99].

### 2.2.1 Properties of iron-based pnictides and chalcogenides.

The discovery of a new class of superconductors based on iron compounds by Japanese researchers in 2008 revived research in the field of SC and HTSC materials. Iron-pnictide superconductors have a relatively higher critical temperature than conventional low-temperature superconductors [100]. Along with many similarities to high-temperature cuprates, the proximity of antiferromagnetism to superconductivity in these semimetallic materials has attracted much attention.

**Iron pnictides**. Hydrogen plays an important role in the synthesis of IBSC materials. In [101], a synthesis method was proposed for inducing superconductivity with hydrogen in iron pnictides of type 1111. The compounds CaFeAsF$_{1-x}$H$_x$ ($x$ = 0–1.0) and SmFeAsO$_{1-x}$H$_x$ ($x$ = 0–0.47) were synthesized using high hydrogen pressure [101]. Figure 21 shows the resistive ($\rho$-$T$) dependences of the samples (a), the crystal structure of the compound and temperature influence (d), and the phase ($x$-$T$) diagram (e).

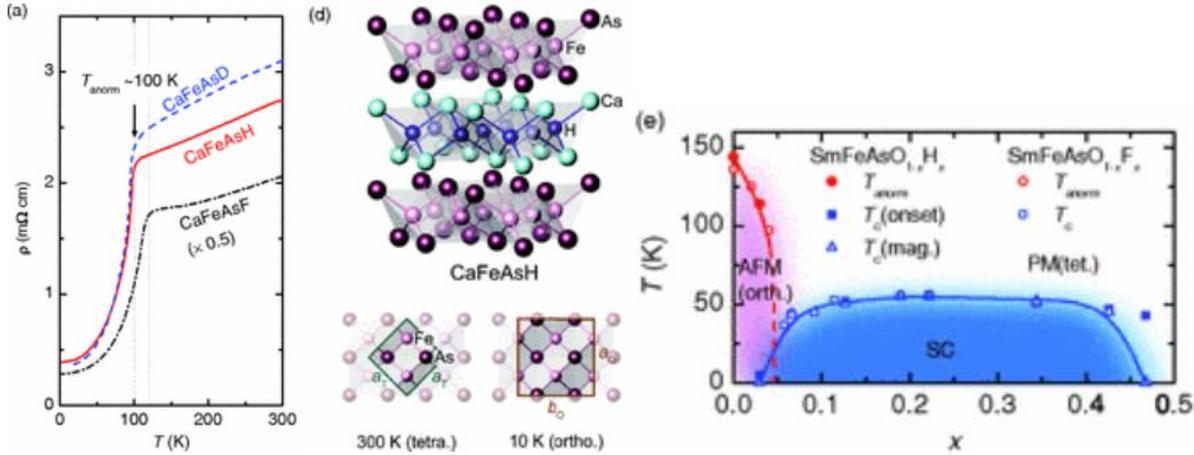

*Fig. 21.* Data from work [94]: (a) ρ-T profile of CaFeAsH and CaFeAsD compared with CaFeAsF; (d) Crystal structure of CaFeAsH at 300 K (tetragonal) and 10 K (orthorhombic); (e) $x$-$T$ diagram of SmFeAsO$_{1-x}$H$_x$ superimposed by that of SmFeAsO$_{1-x}$F$_x$ [101] Reprinted from T. Hanna, Y. Muraba, S. Matsuishi *et al.*, *Phys Rev* B84, 024521 (2011) with permission of American Physical Society.

Neutron diffraction and density functional theory calculations showed that hydrogens are introduced in the form of H⁻ ions. The resulting CaFeAsF$_{1-x}$H$_x$ is not superconducting, while SmFeAsO$_{1-x}$H$_x$ is a superconductor with an optimal temperature $T_c$ = 55 K at $x \sim$ 0.2. It has been found that up to 40% of O²⁻ ions can be replaced by H⁻ ions, with electrons entering the FeAs layer to maintain neutrality (O²⁻ = H⁻ + e⁻). When $x$ exceeded 0.2, the critical temperature $T_c$ decreased, which corresponded to the region of electron overdoping.

In [102], a heavy lanthanide-like superconductor of type 1111 ErFeAsO$_{1-x}$H$_x$ with $T_c$ = 44.5 K was synthesized by doping with hydrogen. A method for stabilizing a superconductor based on ErFeAsO with the lowest lattice constants in the series $Ln$FeAsO$_{1-y}$ ($Ln$ = lanthanide) by doping with hydrogen was demonstrated. Polycrystalline samples were synthesized by heating pellets of the nominal composition ErFeAsO$_{1-y}$ (*1−y*=0.75–0.95), sandwiched between pellets of the composition LaFeAsO$_{0.8}$H$_{0.8}$, at 1100 °C under a pressure of 5.0–5.5 GPa. The sample with lattice constant $a$ = 3.8219 Å and $c$ = 8.2807 Å exhibits the highest critical superconductivity temperatures 44.5 K and 41.0 K, determined from the temperature dependence of resistivity and susceptibility, respectively. The phase diagram of the *Ln*- dependence of $T_c$ in superconductors based on *Ln*FeAsO is discussed in publication [102].

In [103], the authors tried to explain the nature of the increase in the critical temperature $T_c$, which is caused by hydrogen doping. They concluded that in 1111 LaFeAsOH$_x$ type superconductors, the most stable arrangement of hydrogen atoms is located near the positions of Fe ions, which makes the crystal structure and electronic interactions more suitable for the occurrence of superconductivity effects.

Paper [31] reports a method for protonation of Fe-based superconductors (structures of types 122 and 11) at room temperature using an ionic liquid. Figure 22 shows the temperature dependences of the magnetization $\chi$ of the H$_x$-BaFe$_2$As$_2$ sample. The original BaFe$_2$As$_2$ sample was not superconducting before treatment with hydrogen. After protonation, the critical temperature was determined by the authors of the cited article from the discrepancy between the temperature dependences of the magnetic susceptibility obtained in the FC and ZFC modes and was about 20 K. The same $T_c$ is characteristic of Ba(Fe$_{(1-x)}$Co)$_2$As$_2$ upon electron doping with hydrogen ($x$ =0.07).

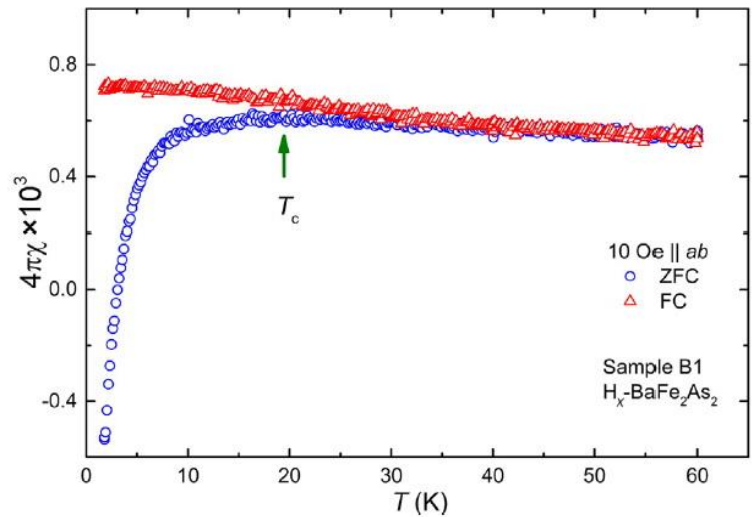

*Fig. 22.* DC magnetization $\chi$ on protonated BaFe$_2$As$_2$(Sample B1) measured under ZFC and FC conditions. $T_c$ marks the onset temperature of superconductivity by the drop of ZFC $\chi$ upon cooling and the deviation between the FC and the ZFC data [31]. Reprinted from Y. Cui, G.Zhang, H. Li *et al.*, *Science Bulletin* 63, 11 (2018) with permission of Elsevier.

The maximum critical temperature $T_c$= 55 K was observed for SmFeAsO$_{0.8}$H$_{0.2}$, which suggests that the replacement of hydrogen (H$^-$) by an anion (O$^-$) supplies additional electrons to the FeAs layer, which increases the density of carrier states near the Fermi surface (FS).

**Iron chalcogenides** include primarily FeSe and Fe-Te-Se compositions [104]. They attract the attention of researchers for several reasons: FeSe has a simple crystal structure, monomolecular FeSe films have the highest critical temperature of the superconducting transition (~100 K) among IBSCs [105], and the calculated critical magnetic fields have high values.

When introducing hydrogen into FeTe$_{0.65}$Se$_{0.35}$ single crystals by thermal diffusion from the gas phase [34, 106, 35] at a temperature of 200 $^0$C and a hydrogen pressure of 5 atm during 10 houres, dissociation of hydrogen molecules occurs at the catalytically active centers of iron group metals. In this case, the symmetry of the crystal lattice decreases from tetragonal to orthorhombic. The unit cell volume decreases by 15% at a hydrogen saturation temperature of 250 $^0$C.

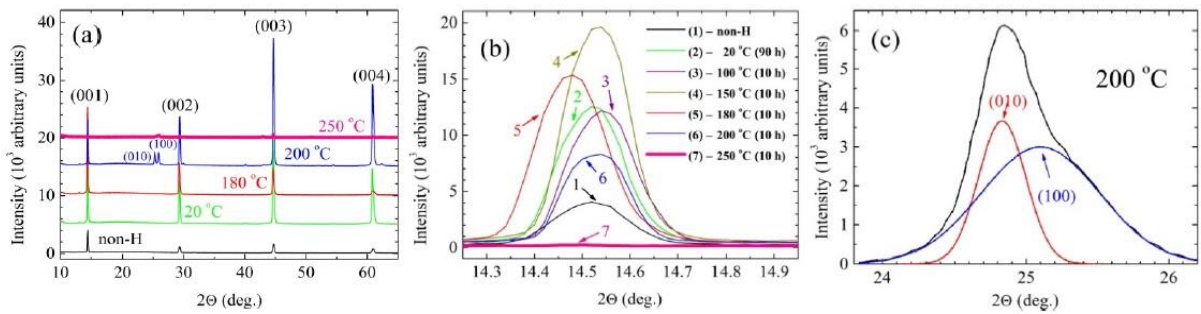

*Fig. 23.* (**a**) X-ray diffraction patterns for FeTe$_{0.65}$Se$_{0.35}$ crystals obtained for the initial state after prolonged exposure to air and after exposing the crystal to hydrogen at a pressure of $5 \times 10^5$ Pa at a temperature of 20 $^0$C for 90 h, 180 $^0$C for 10 h, 200 $^0$C for 10 h, and 250 $^0$C for 10 h. All of the diffractograms have been shifted vertically, by $5 \times 10^3$ between each pattern, for clarity. (**b**) Changes in the structural characteristics of the first diffraction line (001) after the hydrogenation: (1) initial state after prolonged exposure to air, (2) after exposing the crystal to hydrogen at a pressure of $5 \times 10^5$ Pa at temperature of 20 $^0$C for 90 h, (3) – exposing at 100 $^0$C for 10 h, (4) exposing at 150 $^0$C for 10 h, (5) exposing at 180 $^0$C for 10 h, (6) exposing at 200 $^0$C for 10 h, (7) exposing at 250 $^0$C for 10 h. (**c**) Formation of a qualitatively new diffraction pattern, displaying a rather intense asymmetric maximum at reflection angles $2\theta \sim 24$–$26^0$, appearing after exposing the crystal to hydrogen at temperature of 200 $^0$C. Black line is a sum of red (peak 010) and blue (peak 100) lines [34].

If another method of introducing hydrogen (ion source) is used [36, 107, 108], then with an increase in the time of treatment of the FeSe$_{0.88}$ sample with ionized hydrogen (protons), the lattice parameter *a* almost did not change, the crystal lattice parameter *c* decreased, also by 15 % the weight of the sample decreased, which is apparently due to the formation of volatile selenium hydride (Table 3 [36]).

Table 3. The changes in hexagonal lattice periods of FeSe$_y$ superconducting compound depending on the hydrogen ion exposure time [36]. Reprinted from G. S. Burkhanov, S. A. Lachenkov, M. A. Kononov *et al.*, *Inorganic Materials: Applied Research* 8(5), 759 (2017).

| Sample number FeSe$_y$ | Exposure time $t$, min | Lattice periods | | |
|---|---|---|---|---|
| | | $a$, Å | $b$, Å | $c$, Å |
| 1 | 0 | 3.7718(12) | 3.7718(12) | 5.523(2) |
| 2 | 5 | 3.7700(12) | 3.7700(12) | 5.521(2) |
| 3 | 10 | 3.7721(7) | 3.7721(7) | 5.5199(12) |
| 4 | 30 | 3.768(6) | 3.768(6) | 5.511(8) |

In Fig.24 are shown the results of magnetization and NMR measurements for the H$_x$-FeSe$_{0.93}$S$_{0.07}$ sample [31]. Figure 24(a) shows the dependences of the magnetic susceptibility in the FC and ZFC modes in a field of 10 Oersted. A drop in magnetic susceptibility is visible at two temperatures: 42.5 K and 9 K. Temperature $T_{c2}$ corresponds to the transition of the upper layer of the sample with hydrogen to the superconducting state, and $T_{c1}$ corresponds to the transition of the entire sample to the superconducting state, i.e. that part in which there is no introduced hydrogen. Based on the fact that above 9 K the susceptibility drops by 1% of the total diamagnetism, the thickness of the layer with incorporated hydrogen is about 2.8 μm. Figure 24(b) shows the magnetic susceptibility data of the sample in an alternating field. They are obtained based on the shift of the resonant frequency of the NMR circuit. There are four features here. The values of $T_{c1}$ and $T_{c2}$ coincide with the values from Fig. 24a, and the temperature $T_s$ corresponds to the structural transition of the intermediate phase at $T_{c3}$= 25.5 K. Figure 24(c) shows proton NMR spectra at different temperatures.

As the temperature decreases, a broadening of the maximum is observed. Figure 24(d) shows the temperature dependence of the Knight shift. Below 40 K, a rapid decrease in the dependence is observed, indicating superconductivity, which does not contradict the data on magnetic susceptibility. In Fig. 24(e) is shown the dependence of the full width at the half-amplitude level (FWHM) on temperature, and the inset shows the dependence of the inverse square of the magnetic field penetration depth in the *ab* plane on temperature. A rapid increase in the full width at half maximum of the absorption signal peak below 40 K (Fig. 24f) indicates the presence of vortices in the superconductor. The penetration depth of the magnetic field for the sample before treatment with hydrogen is about 0.4 μm, and after treatment is 0.25 μm at a temperature of 2 K. The density of superconducting carriers is proportional to the inverse square of the magnetic field penetration depth. A slight increase in the carrier density for the hydrogenated sample is observed.

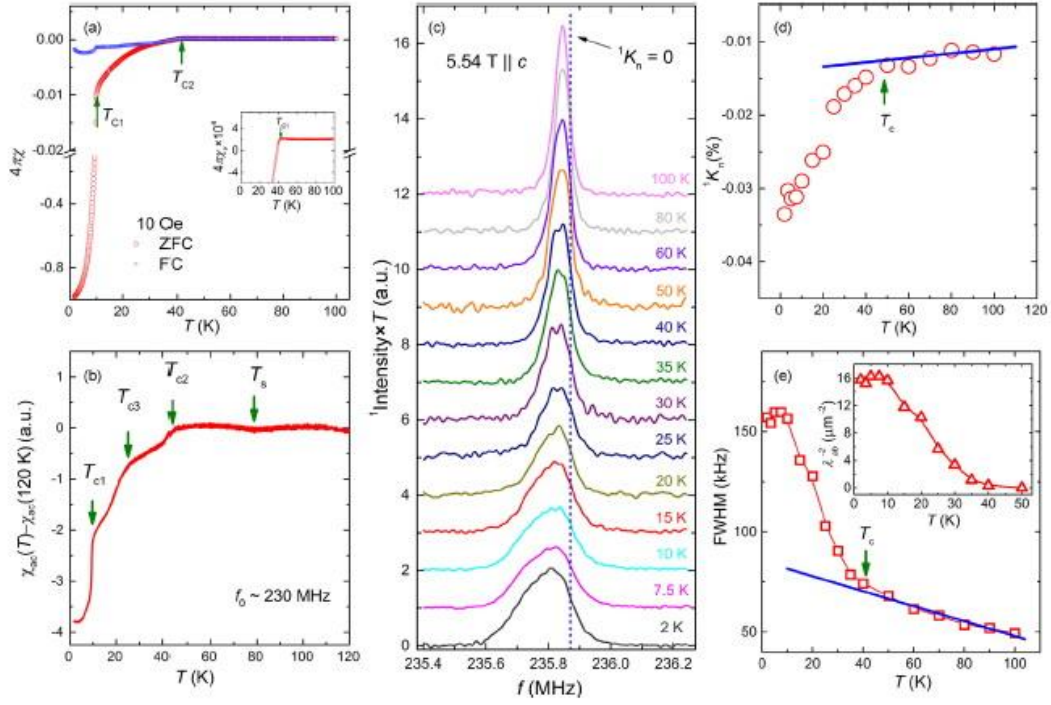

*Fig.24.* Magnetization and NMR measurements on $H_x$-FeSe$_{0.93}$S$_{0.07}$ (Sample S1). (a) The magnetic susceptibility of the sample $\chi_v$ measured under zero-field cooled (ZFC) and field-cooled (FC) conditions with a field of 10 Oe. Two superconducting transitions are shown by the sharp decrease of $\chi_v$ with the transition temperature marked as $T_{c1}$ ($\approx$ 9 K) and $T_{c2}$ ($\approx$ 42.5 K). (b) The ac susceptibility measured by the NMR coil with Sample S1 inside, which shows the structure transition and three superconducting transitions by kinked features. (c) The $^1$H NMR spectra at different temperatures under a fixed field of 5.54 T applied along the c-axis. Data at different temperatures are offset for clarity. (d) The Knight shift $^1K_n$ as a function of temperature, with $T_c$ marking the onset temperature of superconductivity. (e) The full-width-at half-maximum (FWHM) of the NMR spectra. Inset: the $\lambda^{-2}_{ab}$ as a function of temperature, where $\lambda_{ab}$ is the in-plane penetration depth [31]. Reprinted from Y. Cui, G.Zhang, H. Li *et al.*, *Science Bulletin* 63, 11 (2018) with permission of Elsevier.

In Fig. 25 [31] are shown the results of NMR measurements of the $H_x$-FeS sample. In the temperature dependence of the magnetic susceptibility of the sample in an alternating field (Fig. 25a), we see three features. The critical temperature $T_{c2}$ is about 18 K, $T_{c1}$ is about 5 K. The feature at $T_{c2}$ was not observed on the sample not treated with hydrogen. The nature of $T^*$ is not clear to the authors of the article. In Fig. 25b is shown proton NMR spectra at different temperatures. When the sample is cooled, the resonant maximum of the absorption signal broadens and shifts towards lower frequencies. In Fig. 25c is shown the temperature dependence of the Knight shift, and Fig.25d shows the dependence of the full width of the resonance maximum of the absorption signal at half amplitude (FWHM) on temperature. The Knight shift data does not contradict the magnetic susceptibility data. The increase in FWHM in the dependence (fig. 25d) below 20 K is associated with vortices. In Fig. 25f is shown the temperature dependence of the spin-lattice relaxation rate. Features at $T_{c1}$ and $T_{c2}$ are associated with the opening of superconducting gaps. The absence of low-energy spin fluctuations above $T_{c2}$ indicates significant electron doping via hydrogen insertion.

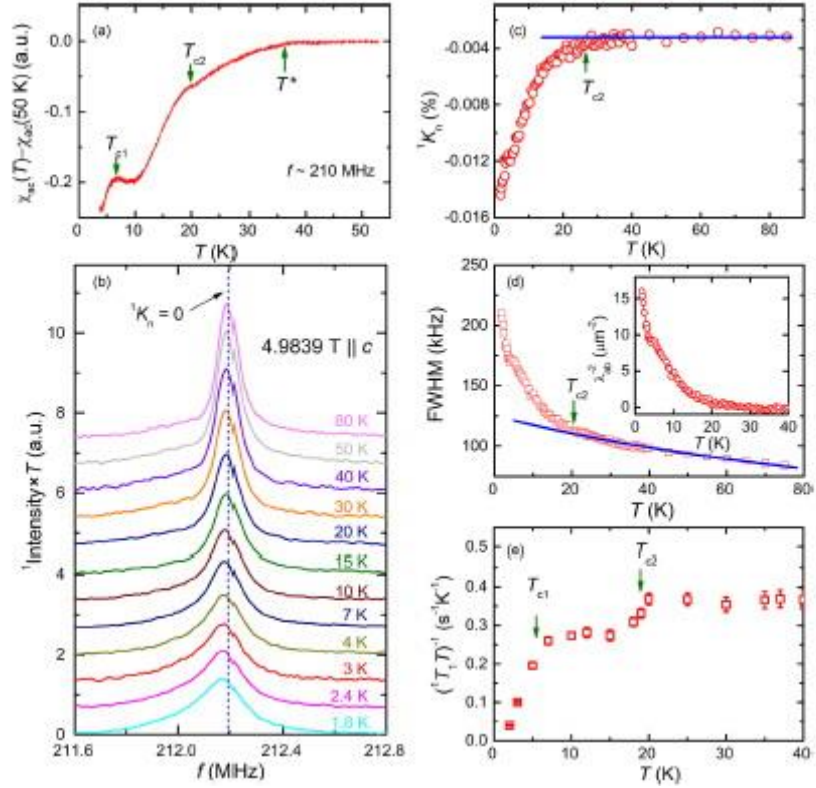

*Fig. 25.* NMR study on H$_x$-FeS (Sample S2). (a) The ac susceptibility $\chi_{ac}(T)$ as a function of temperature, where $T_{c1}$ and $T_{c2}$ mark the onset temperatures of superconductivity by the sharp drop of $\chi_{ac}$. The $T^*$ at 37 K suggests an additional phase-transition like behavior by a weak decrease of $\chi_{ac}$ upon cooling. (b) The NMR spectra measured under a constant field of 4.9839 T applied along the c-axis. (c) The Knight shift $^1K_n$ as a function of temperature. A sudden drop of $^1K_n$ at $T_{c2} \sim 18$ K indicates the onset of a superconducting transition. (d) The FWHM of the spectra as a function of temperature. The high temperature data are fit with a Curie-Weiss function with a deviation to data below 18 K, also suggests a superconducting transition at $T_{c2}$. Inset: the $\lambda^{-2}_{ab}$ as a function of temperature derived from FWHM, where the Curie-Weiss contribution is subtracted. (e) The spin-lattice relaxation rate $1/^1T_1T$ as a function of temperature, where $T_{c1}$ and $T_{c2}$ marks the double superconducting transitions by the drop of $1/^1T_1T$ [31]. Reprinted from Y. Cui, G.Zhang, H. Li *et al.*, *Science Bulletin* 63, 11 (2018) with permission of Elsevier.

In the dependence (Fig. 26) of DC-magnetic susceptibility on temperature [32] in two cooling modes of a FeSe single crystal sample with dimensions $5 \times 5 \times 1$ mm$^3$, a sharp decrease is observed at 41 K. This corresponds to the temperature of the onset of the transition to the superconducting state, while one HTSC phase. The critical temperature of the sample before its treatment with hydrogen was 9 K.

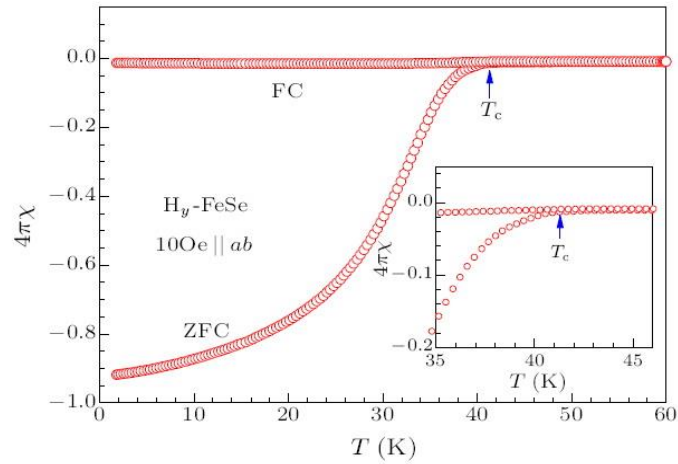

*Fig. 26.* The dependence of the magnetic susceptibility on temperature in two cooling modes of the FeSe sample [32]. Reprinted from Y. Cui, Z. Hu, J.-S. Zhang *et al.*, *Chinese Physics Letters* 36 (7), 077401 (2019).

A sharp decrease is observed at a temperature of 41 K, which corresponds to the temperature at which the transition to the superconducting state begins. One HTSC phase is observed, instead of two phases at 25 and 42.5 K as in [31]. A 100% share of the superconducting phase is also observed. In the dependence of resistance on temperature (Fig. 27b), the onset of the superconducting transition is observed at a higher temperature value of 43.5 K than from the magnetic susceptibility data in Fig. 27a.

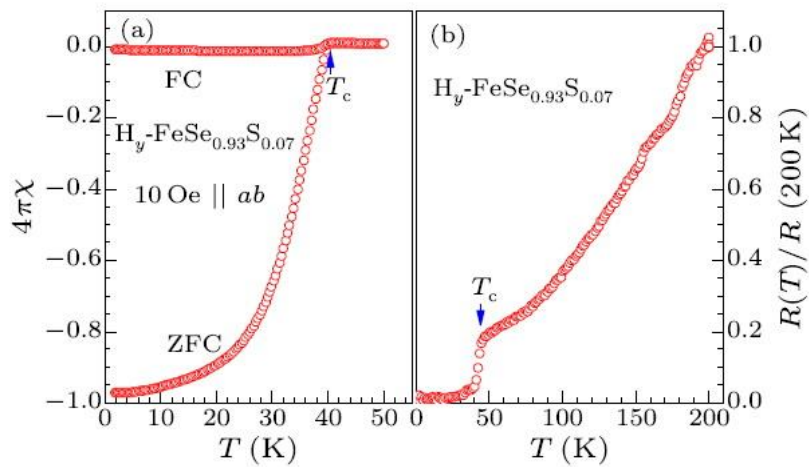

*Fig.27.* (a) The dc susceptibility of a $H_y$-FeSe$_{0.93}$S$_{0.07}$ single crystal as a function of temperature, measured under ZFC and FC conditions. (b) The resistance of the crystal as a function of temperature. The arrows mark the onset temperature of superconductivity [32]. Reprinted from Y. Cui, Z. Hu, J.-S. Zhang *et al.*, *Chinese Physics Letters* 36 (7), 077401 (2019).

In [33], FeSe single crystals with a size of 0.8 mm × 0.5 mm × 15 μm were grown using the CVT (chemical vapor transport method) method. Depending on the time of hydrogen introduction, superconducting phases were obtained at 10K, 25K, 34K, 44K. A homogeneous sample with $T_c$=44 K (resistive measurements) was formed in 20 days. Bulk superconductivity is confirmed by magnetic measurements. At a temperature of 5 K and zero magnetic field, the critical current density ($J_c$) was calculated using the Bean model. It was about $10^6$ A/cm$^2$ (Fig. 28). Since the amount of introduced hydrogen (the $x$ value in the H$_x$-FeSe formula) could not be clearly determined, the time of introduction or removal of hydrogen in days at a temperature of 200 K is used to estimate the degree of saturation with hydrogen [33]. From a sample with $T_c$ = 44 K, hydrogen at a temperature of 200 K it is completely released in 40 days.

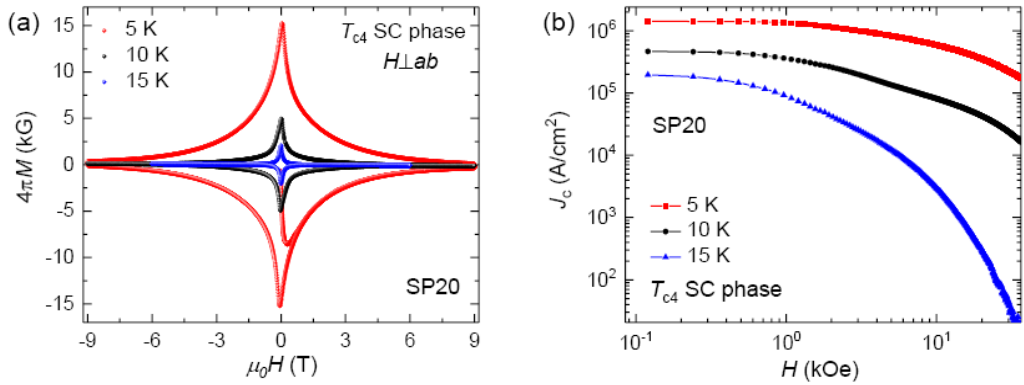

*Fig. 28.* (a) Magnetization hysteresis loops (MHLs) of the optimal $T_{c4}$ SC phase (SP20) at $T$ = 5 K, 10 K, and 15 K for $H \perp ab$. (b) Corresponding magnetic field dependence of critical current densities, $J_c$, derived from the Bean model [33]. Reprinted from Y. Meng, X. Xing, X. Yi *et al.*, *Phys. Rev. B* 105, 134506 (2022) with permission of American Physical Society.

The authors of [33] analyzed the exponent $\alpha$ of the dependence $\rho(T)/\rho_{200K} = \rho_0 + AT^{\alpha}$ based on experimental data (Fig. 29 and 30). Non-Fermiliquid behavior with $\alpha \approx 1$ is observed inside the nematic phase. If the nematic phase is completely suppressed, $\alpha$ increases to ~1.5 and remains unchanged. Then $\alpha$ reaches a value of 2, indicating an evolution from non-Fermi-liquid to Fermi-liquid charge transfer as the time of saturation of the FeSe sample with hydrogen increases. In this case, the effective concentration of holes is suppressed and electrons dominate.

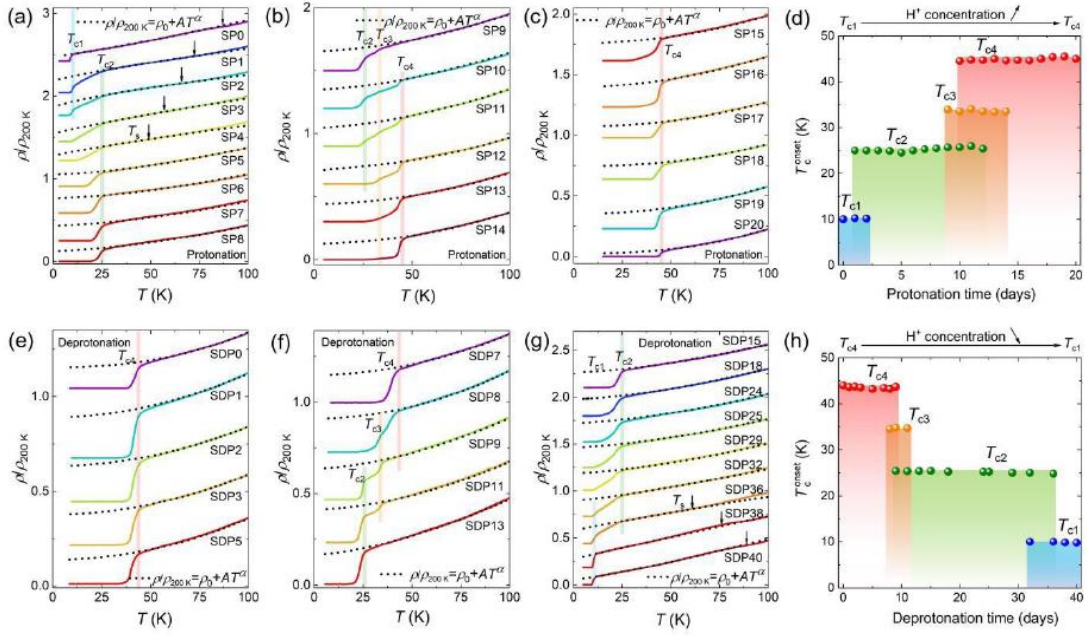

*Fig. 29.* (a–c) Temperature dependence of the normalized resistivity, $\rho(T)/\rho_{200K}$, for different $H_x$-FeSe single crystals during the protonation process with different protonation times. (e–g) $\rho(T)/\rho_{200K}$ curves for $H_x$-FeSe single crystals that have been protonated for 20 days (SP20, shown in (c)) in the deprotonation process with different deprotonation time. The curves have been vertically shifted for clarity. Black arrows indicate the nematic transition at $T_s$. Thick vertical lines are guides for the eye and highlight onset SC transition temperatures, $T_{cn}(n = 1, 2, 3,$ and $4)$, for different SC phases. $T_s$ and $T_{cn}(n = 1, 2, 3,$ and $4)$ are respectively defined as the dip and peak positions in the temperature derivative of $\rho/\rho_{200 K}$, as shown in [33]. Black dotted lines represent fits to the formula $\rho(T)/\rho_{200K} = \rho_0 + AT^\alpha$ (see text for details). (d) and (h) summarize the evolution of $T_c$ with (d) protonation and (h) deprotonation time [33]. Reprinted from Y. Meng, X. Xing, X. Yi *et al.*, *Phys. Rev. B* **105**, 134506 (2022) with permission of American Physical Society.

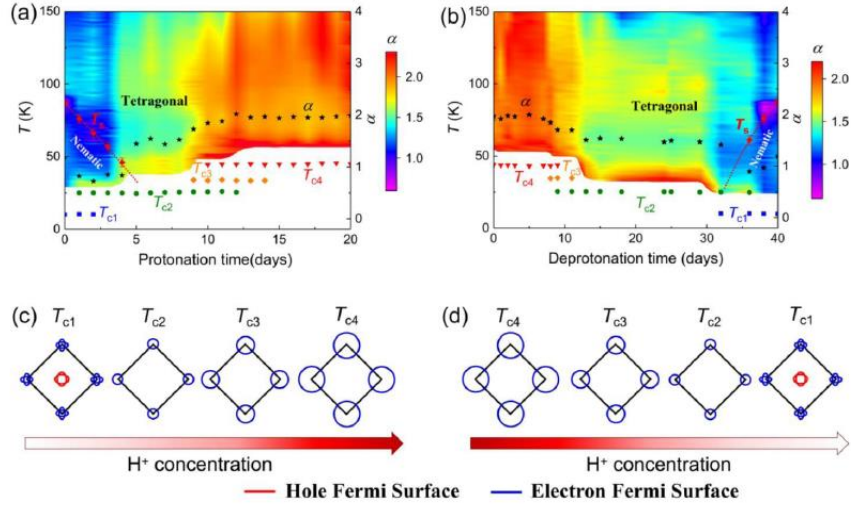

*Fig. 30.* (a and b) Phase diagrams of $H_x$-FeSe single crystals derived from the (a) protonation and (b) deprotonation processes. Color maps represent the temperature dependence of the exponent, α, extracted from $d\ln(\rho - \rho_0)/d\ln T$. (c and d) Schematics describing the evolution of FS topology with (c) protonation and (d) deprotonation. Blue and red lines represent the electron and hole FeSe, respectively [33]. Reprinted from Y. Meng, X. Xing, X. Yi *et al.*, *Phys. Rev. B* 105, 134506 (2022) with permission of American Physical Society.

After thermal diffusion of hydrogen into FeTe$_{0.65}$Se$_{0.35}$ monocrystals, the temperature dependence of the magnetic susceptibility (Fig. 31), taken in the ZFC mode [35], shows that the temperature of the onset of the transition for the initial sample ($T_c^{bulk}$ = 13 K) is lower than that of the sample saturated with hydrogen at a temperature of 200 $^0$C by 1K ($T_c^{bulk}$= 14K). For a sample annealed at a temperature of 250 $^0$C in a hydrogen environment for 10 hours, the transition onset temperature decreases to 9 K, which is associated with partial amorphization of the sample.

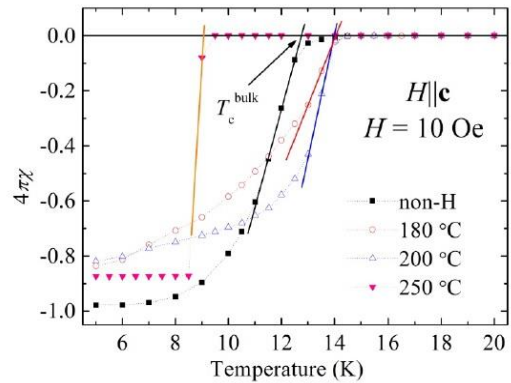

*Fig.31.* Temperature dependence of dc magnetic susceptibility, recorded in zero field cooling mode in $H$ = 10 Oe parallel to the $c$-axis, for the as-grown single crystal of FeTe$_{0.65}$Se$_{0.35}$ and after hydrogenation at 180, 200, and 250 $^0$C [35].

Figure 32 shows the magnetization loops for the initial and hydrogen-treated FeTeSe samples, from which the critical current densities were calculated using the Bean model [35]. An increase in the critical current density up to ~30 times in a magnetic field of 7T for a sample processed at 200 $^0$C compared to the original one is associated with the emergence of additional pinning centers due to significant mechanical stresses that arise during the restructuring of the FeTe$_{0.65}$Se$_{0.35}$ crystal lattice after saturation hydrogen.

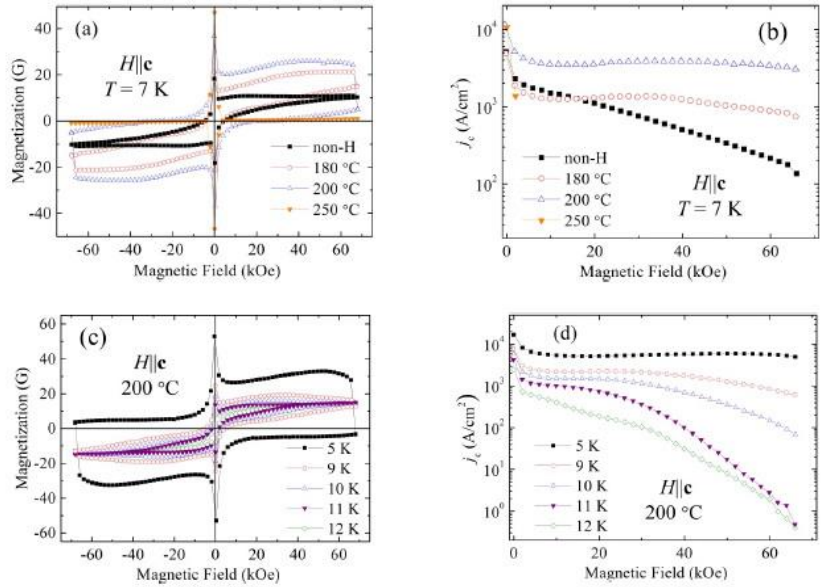

*Fig.32*. (a) Magnetization hysteresis loops recorded at 7 K for pristine FeTe$_{0.65}$Se$_{0.35}$ crystal and for the crystals hydrogenated at 180, 200, and 250 $^0$C. (b) Field dependence of the critical current density,$j_c$, at 7 K for the FeTe$_{0.65}$Se$_{0.35}$ pristine single crystal hydrogenated at 180, 200, and 250 $^0$C. (c) Hysteresis loops recorded in the temperature range from 5 to 12 K for the crystal hydrogenated at 200 $^0$C. (d) Comparison of field dependence of $j_c$, recorded at various temperatures in the temperature range from 5 to 12 K, for the crystal hydrogenated at 200 $^0$C [35].

The increase in the critical temperature for the FeTe$_{0.65}$Se$_{0.35}$ sample treated with hydrogen by thermal diffusion at 200 $^0$C compared to the initial one is also confirmed by EPR studies [35] (Fig. 33c). In this case, the maximum absorption intensity of microwave radiation at the frequency of the EPR spectrometer (9.4 GHz) occurs at zero magnetic field.

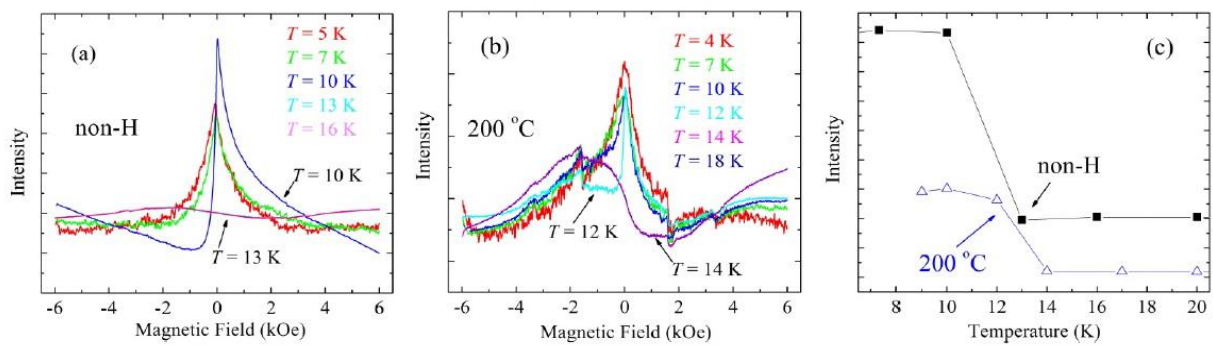

*Fig.33.* (a,b) Wide asymmetric lines observed in the temperature range of 4–18 K, which change significantly with decreasing temperature from 13 to 10 K (a), and from 14 to 12 K for the crystal hydrogenated at 200 $^0$C (b). (c) Temperature dependence of the integrated intensity of the absorption line in the region close to the bulk superconducting transition at about 13 K for both the as-grown and hydrogenated at 200 $^0$C single crystals of FeTe$_{0.65}$Se$_{0.35}$[35].

The introduction of accelerated hydrogen protons into the FeSe$_{0.88}$ compound carried out in [36] did not change its $T_c^{bulk}$. The introduction of protons (H+) using an ion source led to an increase in the mid-transition temperature by 1 K and a decrease in the transition width (the authors claim that the sample becomes more homogeneous after hydrogen treatment).

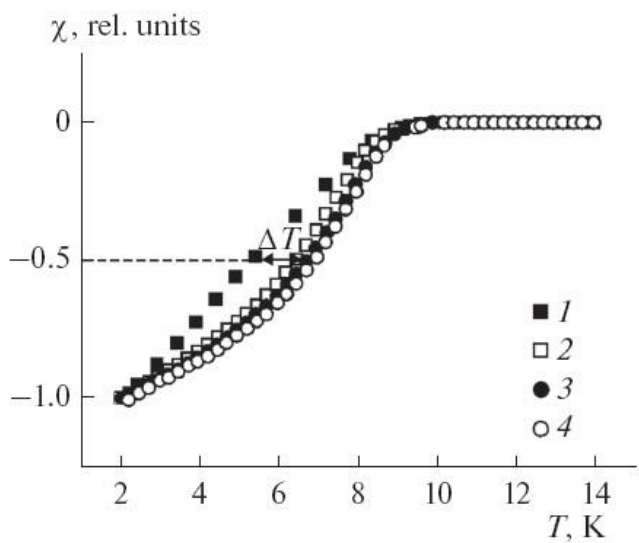

*Fig. 34.* Magnetic susceptibility as function of temperature (superconducting transition curves) for FeSe$_y$ samples ($y$ ~0.88) at different times of hydrogen plasma exposure, min: (*1*) initial sample; (*2*) 5, (*3*) 10, (*4*) 30. $\Delta T$ is the shift of the middle of the superconducting transition [36]. Reprinted from G. S. Burkhanov, S. A. Lachenkov, M. A. Kononov *et al.*, *Inorganic Materials: Applied Research* 8(5), 759 (2017).

Iron selenide (Li,Fe)OHFeSe [109] intercalated with (Li,Fe)OH molecules has a strongly layered structure similar to quasi-two-dimensional (2D) bismuth-based cuprate superconductors and exhibits both high-temperature ($T_c$) and topological superconductivity. However, the question of the dimension of its superconductivity has not yet been fully studied. Features of quasi-2D superconductivity, including high ($\gamma = 151$) anisotropy and associated quasi-2D vortices, are also reported to be found for (Li,Fe)OHFeSe based on systematic electrical transport and magnetization experiments and simulations. Thus, a new vortex phase diagram has been established for (Li,Fe)OHFeSe, which outlines the emerging quasi-two-dimensional vortex-liquid state and the subsequent vortex-solid dimensional transition from pancake-shaped vortices to a three-dimensional state with decreasing temperature and magnetic field. In addition, it is found that all quasi-2D characteristics discovered for the high-temperature iron selenide superconductor are very similar to those described for the high-temperature bismuth cuprate (BiCaCuO) superconductors.

The intercalated iron selenide system (Li,Fe)OHFeSe is a convenient platform for studying the influence of two-dimensionality on superconductivity due to its prominent quasi-2D nature. and availability of samples. Weak interlayer hydrogen bonds and large interlayer distance $d{\sim}9.3$ Å are characteristic features of high-temperature ($T_c{\sim}42$ K) (Li,Fe)OHFeSe superconductors in comparison with the prototype low-temperature bulk superconductors. ($\sim$8.5 K) FeSe with a significantly smaller $d{\sim}5.5$ Å (Fig. 35a). Accordingly, (Li,Fe)OHFeSe has a quasi-two-dimensional electronic structure similar to a two-dimensional FeSe monolayer on a SrTiO$_3$ substrate. In addition, the development in recent years of methods for hydrothermal synthesis of single crystals and epitaxial films of (Li,Fe)OHFeSe makes it possible to measure their properties on high-quality samples.

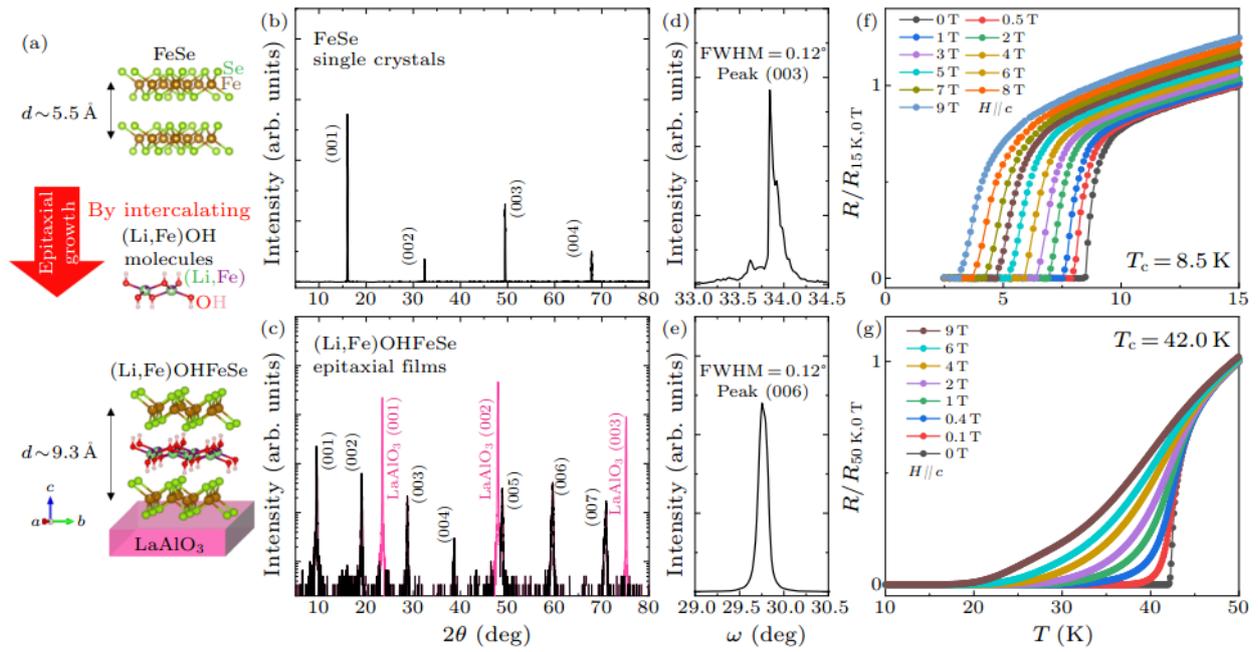

*Fig. 35.* The crystal structure, XRD characterization, and superconducting transition behavior of FeSe and (Li,Fe)OHFeSe. (a) Schematic illustrations of the crystal structures of FeSe single crystals and (Li,Fe)OHFeSe epitaxial films grown on LaAlO₃ substrates. The strongly layered structure of (Li,Fe)OHFeSe ($d$∼9.3Å ) is achievedby intercalating (Li,Fe)OH molecules into the bulk FeSe structure ($d$∼5.5Å ). (b) and (c) The (00l) XRD patterns of the samples. (d) and (e) X-ray rocking curves for the (003) peak of FeSe and (006) peak of (Li,Fe)OHFeSe, respectively. (f) and (g) Temperature dependence of the normalized resistance near $T_c$ under $c$-axis fields up to 9 T for FeSe single crystals ($T_c$= 8.5 K) and (Li,Fe)OHFeSe epitaxy films ($T_c$= 42.0 K), respectively [109]. Reprinted from Dong Li, Yue Liu, Zouyouwei Lu *et al.*, *Chinese Phys.Lett.* 39, 127402 (2022).

In [110], a study of the critical current density $J_c$ and vortex properties of H⁺-intercalated H$_x$-FeSe single crystals was presented. The $J_c$ value for the H$_x$-FeSe single crystal is significantly increased, exceeding $1.3 \times 10^6$ A/cm² at 4 K, which is more than two orders of magnitude higher than $1.1 \times 10^4$ A/cm² of the original FeSe. The vortex pinning mechanism of H$_x$-FeSe is found to involve surface pinning, which is different from the dominant strong point pinning in pure FeSe. Moreover, systematic study of the vortex phase transition and its underlying mechanism provides rich information about the vortex phase diagram of H$_x$-FeSe single crystal. The results confirm that H⁺ intercalation in FeSe not only increases $T_c$, but also significantly increases the $J_c$ value, which is necessary for practical applications (Figs. 36, 37).

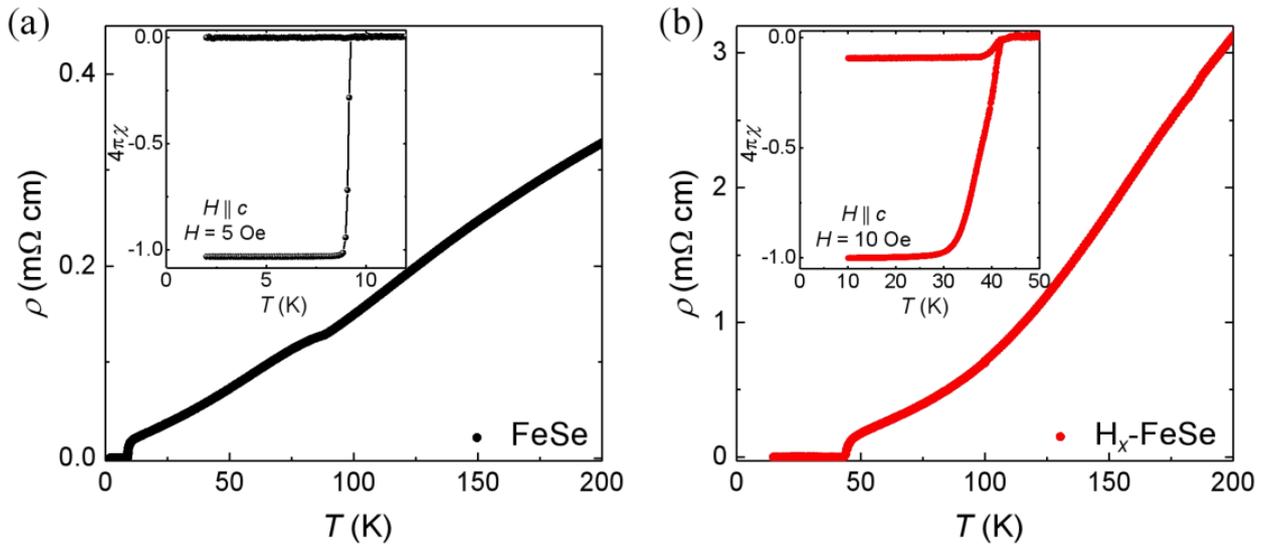

*Fig. 36.* (a) Temperature dependence of the electrical resistivity of FeSe single crystal. The inset shows the temperature dependences of ZFC and FC magnetizations for $H\|c$ under 5 Oe. (b)Temperature dependence of the electrical resistivity of $H_x$-FeSe single crystal. The inset shows the temperature dependences of ZFC and FC magnetizations for $H\|c$ under 10 Oe [110]. Reprinted from Yan Meng, Wei Wei, Xiangzhuo Xing *et al.*, *Supercond. Sci. Technol.* 35, 075012 (2022).

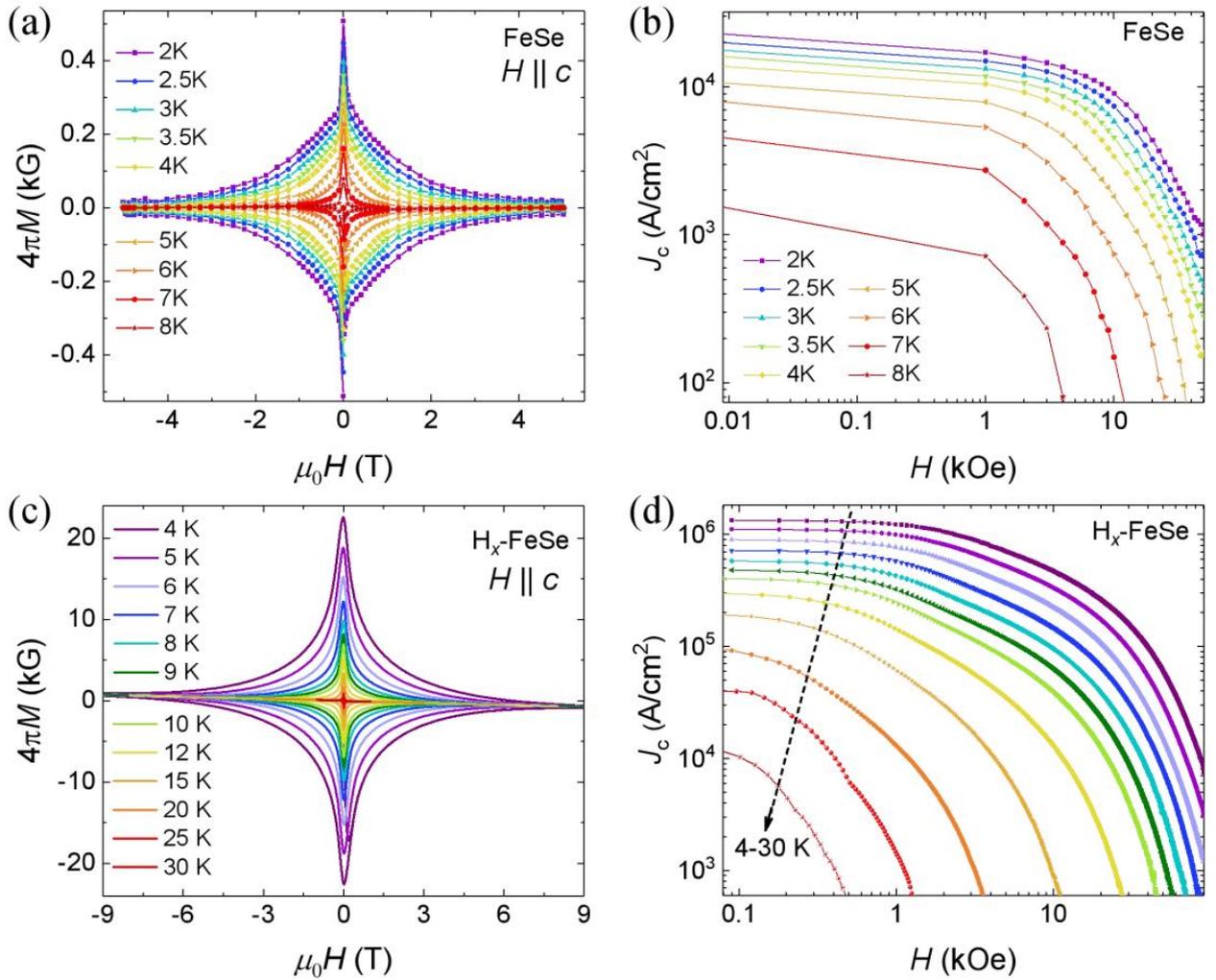

*Fig. 37.* MHLs (magnetization loops) at different temperatures for $H\|c$ and corresponding magnetic field dependence of $J_c$, derived from the Bean model ranging from 2 to 8K for FeSe single crystal (a and b), and 4−30K for optimal $H_x$-FeSe single crystal (c and d) [110]. Reprinted from Yan Meng, Wei Wei, Xiangzhuo Xing *et al.*, *Supercond. Sci. Technol.* 35, 075012 (2022).

The effect of transition metal doping (TMs = Mn, Co, Ni and Cu) on the critical parameters of superconductivity ($T_c$, $H_{c2}$ and $J_c$) in iron selenide (Li,Fe)OHFeSe films has been studied. Samples are grown using matrix hydrothermal epitaxy. The results show that TM doping with Mn and Co can be achieved more easily than others using the matrix hydrothermal epitaxy (MHE) method, which was developed for film growth. Among HMs, there is a lighter inclusion of Mn and Co elements adjacent to Fe into the crystal lattice. It is assumed that doped TMs occupy mainly the iron positions of the intercalated (Li,Fe)OH layers, rather than the superconducting FeSe layers. We found that the critical current density $J_c$ can be increased much more by Mn doping than other TMs, while the critical temperature $T_c$ is weakly dependent on TM doping [111]. However, through systematic experiments, it was found that among all HMs, Mn dopant can increase $J_c$ in a field (9 T, 10 K) to a value reaching 0.51

MA/cm$^2$ at an optimal doping level of ~ 12%. Due to the high reactivity of alkali metals and the easy formation of impurity phase, the superconducting transition temperature ($T_c$) of FeSe intercalated with alkali metals is usually limited to 45 K. To avoid the formation of impurities and improve $T_c$, a more chemically inert organic ion (non-reactive alkali metals) was intercalated into the FeSe single crystal. By intercalating FeSe singlecrystal with organic tetrabutylammonium ion (TBA$^+$) by electrochemical intercalation method, a new FeSe-based superconductor, namely (TBA)$_{0.3}$FeSe, with $T_c$ 50 K, which has the highest critical temperature among bulk FeSe-based superconducting materials, is synthesized. The product intercalated with organic ions consists of an alternate stacking of a FeSe monolayer and an organic molecule. The superconductivity of (TBA)$_{0.3}$FeSe is confirmed by both magnetic susceptibility and transport measurements. The chemically inert organic ion is expected to play a key role in increasing $T_c$, avoiding the formation of impurities and disorder in the FeSe plane as much as possible. We also suggest that intercalated TBA$^+$FeSe with a well-defined shape and higher $T_c$ represents a good platform for further studies in bulk measurements [112].

The simple crystal structure, large external pressure effect and the highest superconducting transition temperature (up to ~100K) in the FeSe/SrTiO$_3$ monolayer interface make FeSe an interesting system in iron-based superconductors. At normal ambient pressure, the key to improving the $T_c$ of bulk FeSe-based superconductors is electron doping of the FeSe plane to form an intercalated structure or charge transfer interface.

By intercalating a bulk FeSe single crystal with an organic tetrabutylammonium ion (TBA$^+$), a new FeSe-based superconductor, namely (TBA)$_{0.3}$FeSe, with $T_c^{onset} \geq 50$ K was synthesized [106]. This is one of the highest critical temperatures of bulk FeSe-based superconducting materials at normal atmospheric pressure.

(TBA)$_{0.3}$FeSe is synthesized through an electrochemical intercalation process using FeSe single crystal as the starting material. First, the FeSe single crystal is weighed on a microgram balance (AX 26). Secondly, the suspended FeSe single crystal is fixed on an indium wire, which is used as a positive electrode. The negative electrode consists of a silver piece. The electrolyte is prepared by dissolving 6 g TBAB (Aladdin, AR, 99.0%) in 20 ml DMF (Innochem, 99.9%, extra dry with molecular screening, water content less than 50 ppm). Finally, the above electrodes are inserted into the electrolyte and a direct constant current (20–30 μA) is applied through the electrolyzer. As current flows through the electrolyzer, the negative electrode loses electrons and the positive electrode gains electrons.

The structure of (TBA)$_{0.3}$FeSe consists of alternately stacked monolayers of FeSe and an organic molecule. In this case, there are 16 hydrogen atoms per FeSe molecule The superconductivity of (TBA)$_{0.3}$FeSe is confirmed by measurements of both magnetic susceptibility and resistivity of the samples.

Fig. 38(a) shows the optical images of the FeSe single crystal and the intercalated product. The morphology of the intercalated FeSe maintains a well-defined shape, like that of a FeSe single crystal. The area of the *ab* plane remains virtually unchanged; only an increase is observed along the *c* axis, which indicates the intercalation of the TBA$^+$ ion. Note that such an electrochemical intercalation process can be carried out on a very large FeSe single crystal with a FeSe mass of more than 10 mg, which is useful for volumetric measurements.

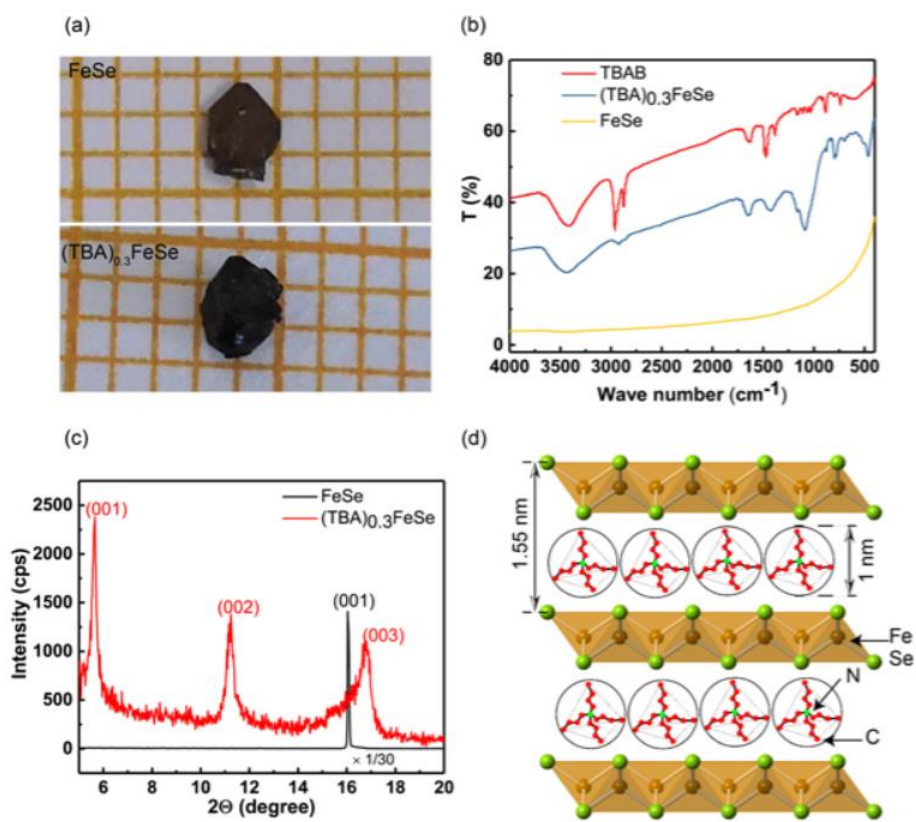

*Fig. 38*. Composition and structure characterization of the organic ion TBA$^+$ intercalated FeSe sample. (a) The optical image of FeSe single crystal and (TBA)$_{0.3}$FeSe; (b) FTIR of FeSe, TBAB and (TBA)$_{0.3}$FeSe; (c) x-ray diffraction pattern of (TBA)$_{0.3}$FeSe; (d) the structure model of (TBA)$_{0.3}$FeSe with hydrogen atom neglected [112]. Reprinted from M.Z. Shi, N. Z. Wang, B. Lei *et al.*, *New J. Phys.* 20, 123007 (2018).

The magnetic susceptibility and transport measurements (TBA)$_{0.3}$FeSe are shown in Figure 39. Figure 39(a) shows the temperature dependence of the magnetic susceptibility (*M-T* curve) in a 5 Oe magnetic field applied parallel to the crystallographic *c*-axis (TBA)$_{0.3}$FeSe. The *M–T* curve exhibits a sharp transition at 48 K (inset in Fig. 39a), indicating the onset of superconductivity at $T_c$ = 48 K. In addition, the transition is absent in the region of 8.9 K, which indicates that the intercalation process is full and even. The difference between the ZFC and FC curves in Figure 39(a) is very small and practically does not demonstrate the effect of flux pinning. Such results indicate that the impurities and defects in the FeSe plane should be very small and the intercalation process uniform, since the small difference means that the fraction of the sample volume in which the magnetic flux is fixed due to defects or impurities is small. Fig.39(b) shows the magnetic susceptibility versus magnetic field ($M$–$H$ curve) at 15 K. The $M$–$H$ curve shows a typical magnetic hysteresis profile of type II superconductors, and the lower critical field $H_{c1}$ is about 890 Oe.

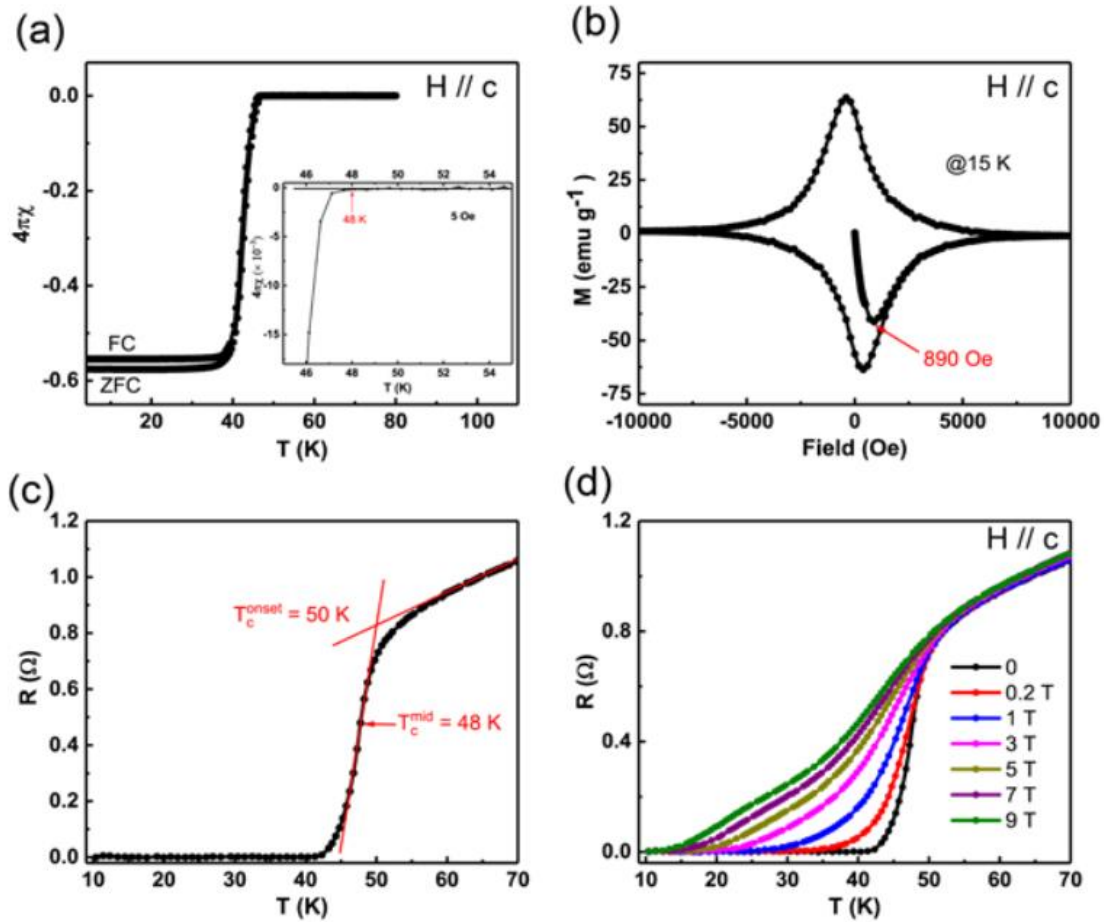

*Fig.39.* Magnetic susceptibility and transport measurement of (TBA)$_{0.3}$FeSe. (a) The $M$–$T$ curve of (TBA)$_{0.3}$FeSe at 5 Oe with demagnetization effect corrected; (b) the $M$–$H$ curve of (TBA)$_{0.3}$FeSe at 15 K; (c) the resistance–temperature curve ($R$–$T$ curve) of (TBA)$_{0.3}$FeSe; (d) the $R$–$T$ curve of (TBA)$_{0.3}$FeSe under different magnetic field [112]. Reprinted from M.Z. Shi, N. Z. Wang, B. Lei *et al.*, *New J. Phys.* 20, 123007 (2018).

It should be noted that the temperature at which the resistance deviates from the linear behavior is more than 55 K, which is slightly higher than the value observed from the results of transport measurements carried out on a monolayer FeSe film grown on undoped SrTiO$_3$ (54.5 K).

The results of measuring the resistance of the $(TBA)_{0.3}$FeSe sample under the influence of external pressure (0–2.46 GPa) are presented in Fig.40. Under external pressure $(TBA)_{0.3}$FeSe, the critical temperature of the onset of the phase transition $T_c^{onset}$ gradually decreases to 29.3 K at 2.46 GPa with an initial 50 K at ambient pressure, with a negative pressure effect at $dT_c/dP$ = -8.4 K $GPa^{-1}$.

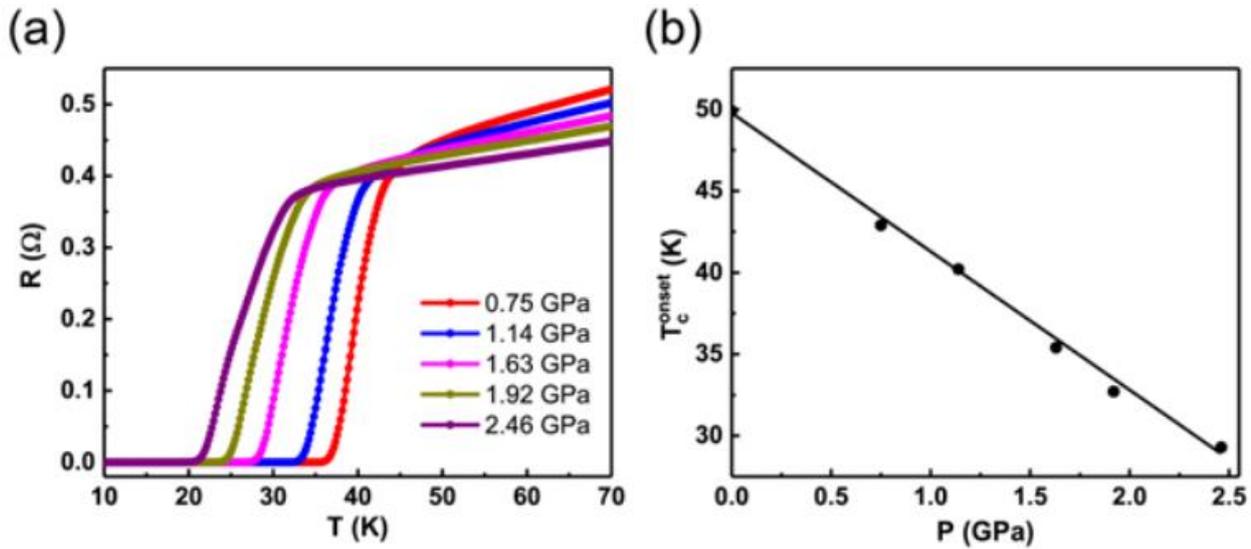

*Fig. 40.* The resistance of the $(TBA)_{0.3}$FeSe sample under the influence of external pressure (0–2.46 GPa) [112]. Reprinted from M.Z. Shi, N. Z. Wang, B. Lei *et al.*, *New J. Phys.* 20, 123007 (2018).

The conclusion made by the authors of [112] based on the last cited cycle of works. It has been shown that the $T_c$ of FeSe can be improved from 8.9 to 50 K by intercalating FeSe with the $TBA^+$ ion. The increase in $T_c$ should be associated with the transfer of electron charge to the FeSe plane. The intercalated sample retains a well-defined morphology and allows volumetric measurements. It is suggested that intercalated $TBA^+$FeSe represents a good platform for further studies of the effect of hydrogen on superconductivity at normal ambient pressure.

### 2.2.2. Properties of high-temperature cuprates.

High-temperature oxide superconductors (HTSC), like many other compounds with a relatively short order parameter coherence length, demonstrate a strong dependence of their critical parameters ($T_c$, $J_c$, $H_c$) on crystal structure defects [113]. Various planar boundaries (interblock boundaries, twinning [111], [114] and the presence of impurities after doping procedures) have a great influence. One of the tasks of studying HTSCs is to improve their superconducting properties by modifying the electronic subsystem of copper oxide when doped with electrons or holes. In particular, such doping can occur as a result of intercalation of hydrogen atoms.

Pure ceramic materials, like unalloyed copper oxides, are generally insulators. After doping, copper oxides can become bad metals in the normal state and high-temperature superconductors when the temperature decreases (see Fig. 41 for the phenomenological phase diagram of electron- and hole-doped high-temperature superconductors $Nd_{2-x}Ce_xCuO_4$ and $La_{2-x}Sr_xCuO_4$) [113].

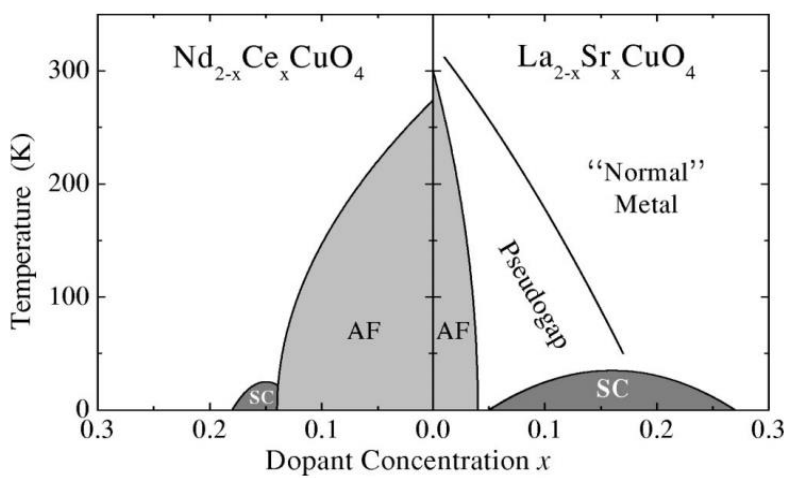

*Fig. 41.* Phase diagram of n - and p -type doped superconductors, showing superconductivity (SC), antiferromagnetic (AF), pseudogap, and normal-metal regions [113]. Reprinted from A. Damascelli, Z. Hussin, and Z.-X. Shen, *Rev. Mod. Phys.* 75, 473 (2003) with permission of American Physical Society.

In particular, significant changes in the crystal lattice parameters and charge distribution in the copper-oxygen subsystem were detected by X-ray phase analysis and nuclear quadrupole resonance (NQR) methods. Discrete changes in interplanar distances along the *c* axis and Cu NQR frequencies with a smooth change in the degrees of hydration and oxidation indicate the formation of hydrocuprate in the matrix of the starting material. The effect of hydrogenation at $T = 150°C$ and $200°C$ on the superconducting properties of highly textured $YBa_2Cu_3O_y$ ceramics with a reduced oxygen content was studied in [115]. The oxidation of the hydrogenated sample at $T = 400°C$ results in the noticeable growth (compared to the initial state) of critical current density $j_c$ and pinning force $F_p$ in both magnetic-field directions. The values of the first critical current like-wise exceed the initial values, especially in the direction of an external field applied perpendicular *c*-axis ($\perp c$). In this case, anisotropy of the material increases due to the growth of the irreversibility field ($B_{irr}$) in the direction of an external field applied $\perp c$ , and the maximum in the field dependence of the pinning force shifts toward the range of higher fields. This suggests that, during hydrogenation, extra planar defects arise. As after hydration, this is due to the formation of planar defects during low-temperature annealing. In addition, during the hydrogenation process, partial reduction of copper occurs with the formation of microinclusions of $Cu_2O$ and other chemical decomposition products, which are additional centers for the attachment of magnetic vortices.

It was shown in [115] that hydrogen is capable of being incorporated into the structure of $YBa_2Cu_3O_{6.96}$. It has been established that its presence leads to structural changes, in many ways similar to the changes that occur during the absorption of water. The interaction of $YBa_2Cu_3O_{6.96}$ with hydrogen does not lead to degradation of grain boundaries, in contrast to treatment in a humid atmosphere. This may be a significant advantage of hydrogenation (saturation with hydrogen) over hydration (saturation with an aqueous base). In this work, we investigated the effect of low-temperature annealing in a hydrogen atmosphere and reduction annealing at $T = 930°C$ on the critical characteristics of $YBa_2Cu_3O_{6.96}$ and $YBa_2Cu_3O_{6.3}$ ceramics with a strong texture. The purpose of the study is to determine the field dependences of critical characteristics and optimize the composition and annealing conditions to improve the current-conducting ability of cuprates that are promising for practical use. Low-temperature treatment was carried out at $T = 150$ and $200 °C$ in a hydrogen atmosphere for 1–20 hours. The samples were processed in a Sieverts-type apparatus filled with hydrogen obtained from the decomposition of $LaNi_5H_x$ Before filling them with hydrogen, the operating chamber was evacuated to a pressure of $\sim 10^{-2}$ mm Hg. Hydrogen absorption was recorded by changes in pressure in the chamber and by the gravimetric method. Reductive annealing after hydrogenation was carried out at $T = 930 °C$, followed by oxidation in an oxygen atmosphere at $T = 400 °C$ (24 h) to an oxygen index of $\sim 7$.

Magnetization measurements were carried out in pulsed magnetic fields at temperature $T = 77$ K. The pulse duration was $\sim 7.5$ ms. The maximum amplitude of the magnetic induction pulse was 35 Tesla. The hysteresis loops were recorded in a field applied both parallel to the $c$ axis ($\|c$) and perpendicular to it ($\perp c$). The samples were cut from one section of a highly textured ceramic block with a uniform macrostructure and had dimensions of $\sim 2.4 \times 2.5 \times 1.5$ mm$^3$. The critical current density was calculated using the modified Bean formula $j_c = 20 \ \Delta M/a(1 - a/3b)$, where $\Delta M$ is the width of the magnetization loop, $a$ and $b$ are the dimensions of the rectangular sample ($a < b$, cm). The vortex pinning force $F_p$ was calculated using the equation $F_p = j_c \times B$. The first critical field was determined by the deviations of the initial section of the curve $\Delta M = f(B)$ from the linear course. To determine $B_{c1}$, hysteresis loops with a maximum amplitude of the magnetic induction pulse $B \sim 3$ T were used. It was shown that during the hydrogenation of $YBa_2Cu_3O_{6.96}$ samples at $T = 200 °C$, partial reduction of copper occurs in them with the formation of $Cu_2O$. The appearance of this phase was observed both visually and according to X-ray diffraction analysis (Fig. 42, shown in the inset). After annealing, a mass loss of the sample (0.12%) was recorded. Partial chemical decomposition of the $YBa_2Cu_3O_{6.96}$ compound upon interaction with hydrogen is also observed during optical examination of the single crystal.

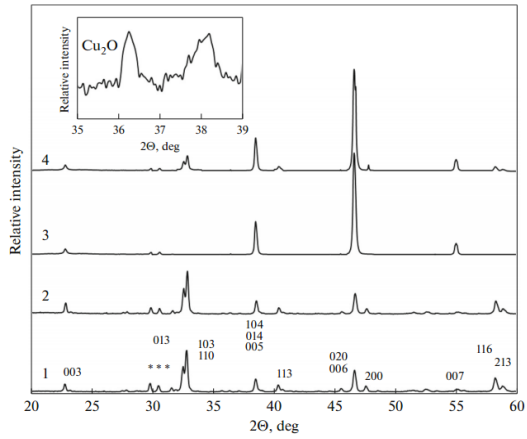

*Fig.42.* Diffraction patterns of YBa$_2$Cu$_3$O$_{6.96}$: (*1*) initial state (powder pattern, Y$_2$BaCuO$_5$ phase is shown by asterisks); lattice parameters (Å): *a* = 3.819(3)Å, *b* = 3.889(7)Å, *c* = 11.696(4)Å; (*2*) after 1-h hydrogenation at *T* = 200 °C (powder pattern): *a* =3.837(2)Å, *b* = 3.895(1)Å, *c* = 11.674(5)Å. Inset: section in the angular range 2Θ = 35°–39°; (*3*) initial state (X-ray diffraction patternis taken from the plane ( *ab* ); (*4* ) after 5-h hydrogenation at *T* = 150 °C and 7-h recovery at *T* = 930 °C [115]. Reprinted from I.B. Bobylev, E.G. Gerasimov, N.A. Zyuzeva *et al.*, *Phys. Metals Metallogr.* 118 (10), 954 (2017).

The field dependences $j_c = f(B)$ show that the results obtained after hydrogenation of YBa$_2$Cu$_3$O$_{6.96}$ and reduction annealing for 2 h are similar to previous results after hydration [116]. The absorption of hydrogen, like the absorption of water, leads to an increase in $j_c$ and $B_{c1}$ compared to the initial state, mainly in an external magnetic field ǁ *c* (Fig. 43, curves 4).

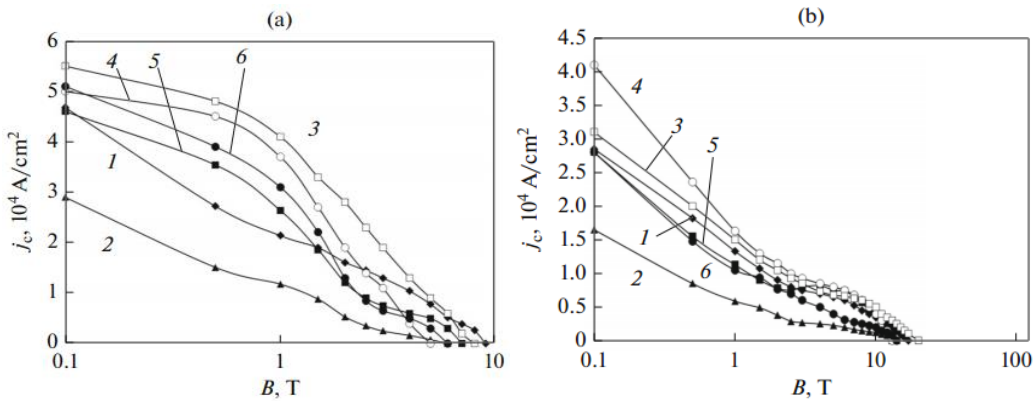

*Fig. 43.* Dependences $j_c = f(B)$ at *T* = 77 K in a magnetic field applied (a) ǁ*c* and (b) ⊥*c*: (*1*) YBa$_2$Cu$_3$O$_{6.96}$, initial state; (*2*) after 5-h hydrogenation at *T* = 150°C; (*3*) after hydrogenation and oxidation; (*4*) after hydrogenation and recovery (930°C, 2 h); (*5*) after 4-h hydrogenation at *T* =150°C and oxidation; (*6*) after hydrogenation and recovery (930°C, 7 h) [115]. Reprinted from I.B. Bobylev, E.G. Gerasimov, N.A. Zyuzeva *et al.*, *Phys. Metals Metallogr.* 118 (10), 954 (2017).

After a 2-hour recovery annealing, an increase in the pinning force is observed in fields up to ~12 T ($\perp c$) and in moderate fields in the $\parallel c$ direction (Fig. 43, curves 4). After a longer recovery treatment (4–7 hours) on the dependences $F_p = f(B)$ in a field directed $\parallel c$, a maximum appears at $B = 5$ T, which indicates the presence of two types of pinning centers in the material (Fig. 44, curves 5, 6).

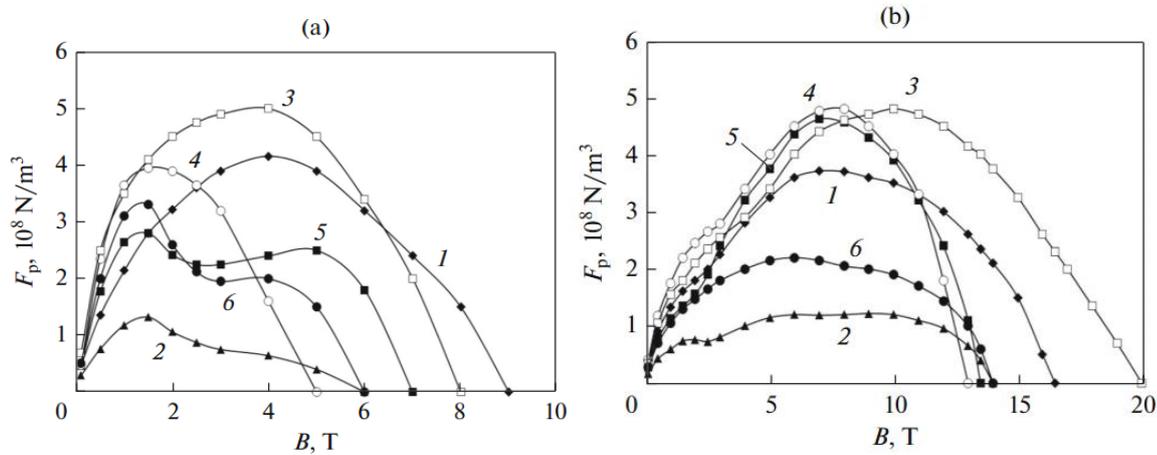

*Fig.44.* Dependences of $F_p = f(B)$ at $T = 77$ K in a magnetic field applied (a) $\parallel c$ and (b) $\perp c$: (1) YBa$_2$Cu$_3$O$_{6.96}$, initial state; (2) after 5-h hydrogenation at $T = 150$ °C; (3) after hydrogenation and oxidation; (4) after hydrogenation and recovery (930 °C, 2 h); (5) after 4-h hydrogenation at $T = 150$°C and oxidation; (6) after hydrogenation and recovery (930 °C, 7 h) [115]. Reprinted from I.B. Bobylev, E.G. Gerasimov, N.A. Zyuzeva *et al.*, *Phys. Metals Metallogr.* 118 (10), 954 (2017).

Thus, after hydrogenation at $T = 150$ °C and a short duration of reduction annealing (930 °C), the critical parameters of YBa$_2$Cu$_3$O$_{6.96}$ increase slightly compared to the initial state. During the hydrogenation process, along with the introduction of hydrogen into the structure of YBa$_2$Cu$_3$O$_{6.96}$, partial reduction of copper is observed. Additional maxima appear in the field dependences of the pinning force; the maxima indicate the formation of new pinning centers, which may represent highly dispersed inclusions of products of the reduction process. Depending on the processing conditions, the growth of $j_c$ and $B_{c1}$ can be observed predominantly in the applied magnetic field $\perp c$, which is associated with the formation of plane defects that carry out correlated pinning. With the formation of microparticles that carry out random pinning of magnetic vortices, the growth of $j_c$ and $B_{c1}$ occurs in both directions of the external field.

Cuprate superconductors (YBa$_2$Cu$_3$O$_7$) strongly absorb hydrogen [117]. This work describes the preparation of the compound H$_x$YBa$_2$Cu$_3$O$_7$ by direct reaction with hydrogen gas. A sample of 0.5g of powdered oxide was introduced into a Pyrex glass reactor of known volume and outgassed at room or slightly elevated temperature (400K) for a period of 30 min. After out-gassing, a known quantity of H$_2$ was introduced to give a pressure of 650 mm Hg. In order to initiate the absorption reaction the reactor was briefly heated to 415 K after which the temperature was reduced to 385 K and kept constant. After a short period of time, which varied from sample to sample, apparently depending on its refractory character, the reaction began. The uptake of hydrogen by the sample was monitored by tracking the hydrogen pressure as a function of time. The rate of hydrogen uptake is a function of temperature and pressure, and in most cases, the reaction was complete in a few hours, although in order to obtain a composition of x=5.9 several days were required. The expansion of the initial orthorhombic unit cell, determined by X-ray diffraction, indicates that a hydrogen solid solution is formed up to x~0.2. The solid solution phase exhibits a superconducting transition at ~94 K, which is typically somewhat higher than that of the parent cuprate. At a higher $H_2$ content, the volumetric homogeneity of the crystal is disrupted, and regions with phase rich in hydrogen are formed. The hydrogen-rich phases, which are believed to be interstitial hydrides, are not superconducting at 4 K or higher. Hydrogen-enriched phases are stable at room temperature and can be nuclei of pinning centers, which is important for increasing the critical current density in these compounds. X-ray patterns were obtained using a step-scan goniometer and Cu$K\alpha$ radiation. The lattice parameters of the solid solution phase were determined by the least squares method for the positions of 20 peaks in the back reflection region, i.e. scattering angles $2\theta$ from 120° to 160°.

The critical superconductivity temperatures $T_c$ were determined using the induction coil method, in which voltage changes across a gradiometric (back-to-back) pair of coils are measured as a function of temperature. This method is essentially the same as the susceptibility method, except that it only measures the relative change in susceptibility rather than the absolute change. Since the transition width for thin powders is relatively large, which is primarily due to the dependence of the magnetic field penetration depth on temperature, the value of $T_c$ was determined as shown in Fig. 45. The relative volume fraction of superconductivity of hydrogen-containing samples was determined by measuring the change in inductive voltage per unit weight of the sample, expressed in $\mu$V/g, and dividing it by the same parameter of the original oxide. The error in this determination is large (+ or - 10%) due to the small weight of the sample and the unknown influence of particle size and separated phases.

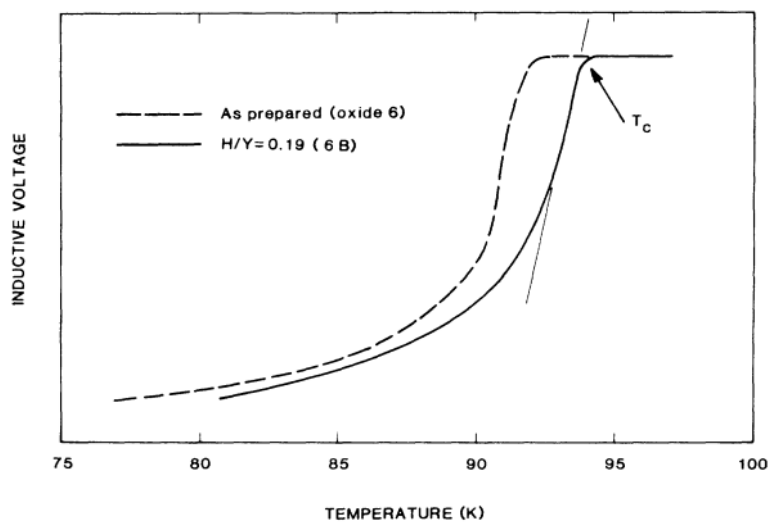

*Fig. 45.* Inductive voltage *vs* absolute temperature. Dashed line represents original oxide (sample 6); solid line represents solid solution phase, $x = 0.19$ (sample 6B) [117]. Reprinted from J. J. Reilly, M. Suenaga, J. R. Johnson *et al.*, *Phys. Rev. B* 36, 5694(R) (1987) with permission of American Physical Society.

Ref. [117] also contains a table of superconducting transition temperatures and hydrogen content in all hydrogenated samples and the corresponding starting oxides. It follows from the table that one sample was prepared containing deuterium, and the limiting $T_c$ for the $H_x YBa_2Cu_3O_7$ system is about 94 K, regardless of the $T_c$ of the original oxide. However, despite a slight increase in $T_c$, in most samples there was a progressive decrease in the strength of the induced voltage signal as the H content increased above $x = 0.19$ at $x = 2.7$ the signal was very weak, and at $x = 4$ the superconducting transition was not observed until up to 4 K.

From preliminary equilibrium measurements of absorption pressure and composition, it is clear that $H_x YBa_2Cu_3O_7$ is quite stable. The authors also obtained X-ray diffraction patterns for each initial oxide sample and most hydrogen compounds. In particular, line broadening effects typical of metal hydride phases obtained at low temperatures were observed in the diffraction patterns of samples with high hydrogen concentrations. A series of samples were prepared to determine the effect of hydrogen concentration on the volume of lattice cells. The lattice parameters of expanded cells are given in Table. 4 [117]; The graph of the increase in cell volume with increasing $x$ is shown in Fig. 46.

Table 4. Expansion of orthorhombic cell of $H_x YBa_2Cu_3O_7$ [117]. Reprinted from J. J. Reilly, M. Suenaga, J. R. Johnson *et al.*, *Phys. Rev. B* 36, 5694(R) (1987) with permission of American Physical Society.

| H content | Lattice parameters | | | Cell volume |
| $x$ | $a$ (Å) | $b$ (Å) | $c$ (Å) | (Å³) |
|---|---|---|---|---|
| 0 | 3.8198(5) | 3.8860(4) | 11.6828(5) | 173.42(6) |
| 0.13 | 3.8215(7) | 3.8866(5) | 11.6882(9) | 173.60(7) |
| 0.19 | 3.8248(8) | 3.8860(5) | 11.6907(7) | 173.76(6) |
| 0.38 | 3.8246(7) | 3.8861(4) | 11.6900(6) | 173.75(6) |
| 0.49 | 3.8260(9) | 3.8861(4) | 11.6937(9) | 173.86(8) |

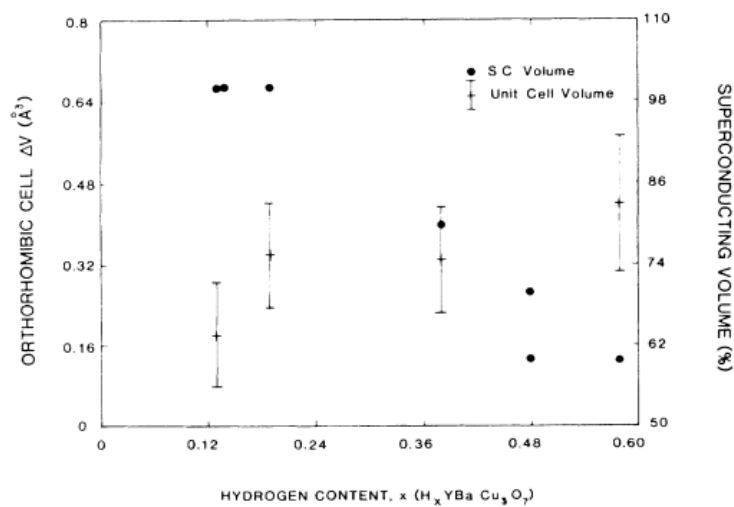

*Fig. 46.* Volume increase of orthorhombic cell and relative superconducting fraction *vs* H content *x*. + with error bars represents volume increase. •: % superconducting (+ or -10%) [117]. Reprinted from J. J. Reilly, M. Suenaga, J. R. Johnson *et al.*, *Phys. Rev. B* 36, 5694(R) (1987) with permission of American Physical Society.

The expansion of the orthorhombic unit cell indicates that hydrogen at room temperature dissolves in the oxide phase to a final solubility value of $x = 0.19$. This defines the approximate phase boundary between the solid solution and the hydrogen-rich phase that forms at higher H contents. This behavior is similar to the behavior of metal-hydrogen systems in which H occupies an interstitial position in the crystal lattice. Crystallographic data (Table 4 [117]) also allow us to make some assumptions about the location of H in the solid solution phase. The $c$ and, to a lesser extent, $a$ axes increase; the $b$ axis does not change. This observation suggests that hydrogen occupies sites that leave intact Cu-O chains along the $b$ axis, which are presumably required for the superconducting transition to occur. Thus, it seems reasonable to assume that in the solid solution region, H occupies an oxygen vacancy on the $c$ or $a$ axis, or both, and coordinates predominantly with Cu. In general, the behavior of this system seems similar to those systems of metal and hydrogen, where H occupies an interstitial position coordinated by the metal atom. Thus, we will conventionally call the crystalline hydrogen-rich phases interstitial hydrides, and the solid solution phase, according to convention, the alpha (α) phase.

The slight increase in $T_c$ in the α phase may be due to a slightly more favorable configuration of the electronic band structure and/or random elastic distortion of the unit cell caused by the introduction of an H atom. In contrast, none of the hydride phases was superconducting, and in the two-phase region (α- phase + hydride) the relative volume fraction of superconductivity depends on the volume fraction of the α phase in the mixture.

Review article [118] was motivated by an analysis of the influence of hydrogen on the temperature of transition to the superconducting state, on the physical properties during hydrogenation, on the diffusion of hydrogen in superconductors for a general idea of the behavior and role of hydrogen in HTSC. Instead of changes in composition due to substitution, much attention has been paid to hydrogen as an interstitial site, since hydrogen atoms trapped in interstitial sites of HTSCs also affect their structure as well as physical properties. HTSCs readily react with hydrogen gas or can be implanted with protons. In general, it is believed that hydrogen atoms/ions occupy some interstitial positions in HTSC due to its small size, forming either dilute solid solutions or hydrogen-concentrated phase(s), i.e. hydride. It is also possible that decomposition or amorphization of HTSC occurs at higher hydrogen concentrations, as observed in hydrogen-absorbing materials. The lattice parameters of $H_xYBa_2Cu_3O_{6.91}$ and $H_xGdBa_2Cu_3O_{6.89}$ (bulk samples) were measured as a function of $x$. The lattice constants $c$ and $b$ decrease, while the lattice constant $a$ increases with increasing $x$ up to $x = 0.8$, as shown in Fig. 47; a further increase in $a$ is observed with increasing $x$, while the lattice constant $c$ remains almost constant until $x = 1.8$, expanding the unit-cell volume.

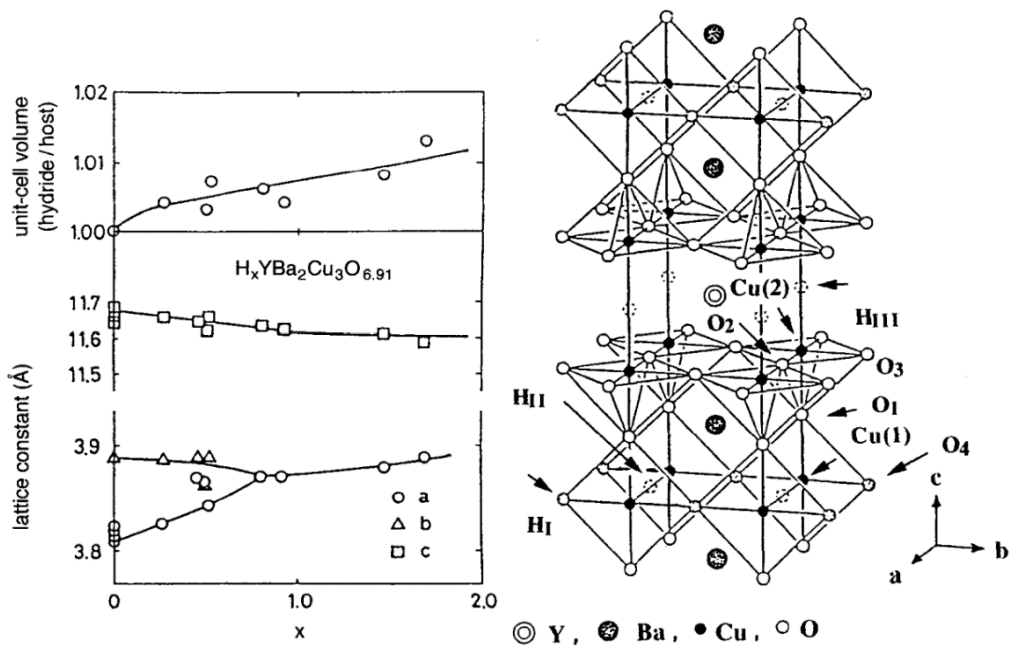

*Fig. 47.* Crystal structure of orthorhombic $YBa_2Cu_3O_{7-\delta}$. The possible occupation sites for hydrogen atom $H_I$, $H_{II}$, and $H_{III}$ are indicated along with the Cu(1), Cu(2), $O_1$, $O_2$, $O_3$, and $O_4$ sites; note that the site $O_5$ (1/2,0,0) in the Cu(1) chain is not shown but corresponds to the site $H_{II}$, and that the apical oxygen site is labelled as $O_1$ instead of $O_4$ for our consistent argument. The sites $H_I$ and $H_{II}$ correspond to the occupied oxygen and vacant positions, respectively [118]. Reprinted from T. Hirata, *Phys.stat.sol.* (a) 156, 227 (1996) with permission of John Wiley and Sons.

Previously, researchers discovered [115] that for the $YBa_2Cu_3O_{6-\delta}$ compound ($\delta = 1$; $T_c = 91$ K; bulk samples), an increase in $T_c$ to 93.5 K is observed after hydrogenation, The solid solution of hydrogen (up to $x \approx 0.2$) has $T_c = 94$ K, which is higher than that of the original oxide. In Fig. 48 it is shown a typical example of $\rho(T)$ for $H_xYBa_2Cu_3O_{7-\delta}$ thin films. Obviously (d$\rho$/d$T$) increases with $x$. There is no significant change in the residual resistance extrapolated to $T = 0$ K. This change in resistance can be explained by a decrease in the concentration of charge carriers.

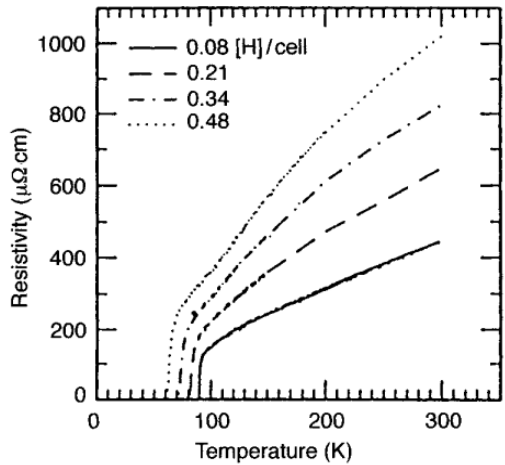

*Fig. 48.* Temperature dependence of the electrical resistivity $\rho$ for $H_xYBa_2Cu_3O_{7-\delta}$ thin films with different *x* values. Note that the transition to the superconducting state shifts to lower temperature but remains sharp with increasing *x*; $\rho$ increases with *x* as well. Reproduced from [118]. Reprinted from T. Hirata, *Phys.stat.sol.* (a) 156, 227 (1996) with permission of John Wiley and Sons.

In Fig. 49 the dependence of the susceptibility $\chi$ on temperature and hydrogen concentration are shown for $YBa_2Cu_3O_{6.9}$. Ceramic samples of $YBa_2Cu_3O_{6.9}$ were prepared by heating the stoichiometric mixture of $Y_2O_3$, BaCO and CuO powder at about 900°C for 12h. The reacted samples were heated again in air at 950°C for 48 h. Hydrogenation was performed by exposing the powdered samples to hydrogen gas of 0.9 bar at 230°C. The samples were then sealed in a Pyrex ampoule under helium exchange gas. It can be seen that the volume fraction of superconductivity decreases due to the formation of a dielectric phase during hydrogenation.

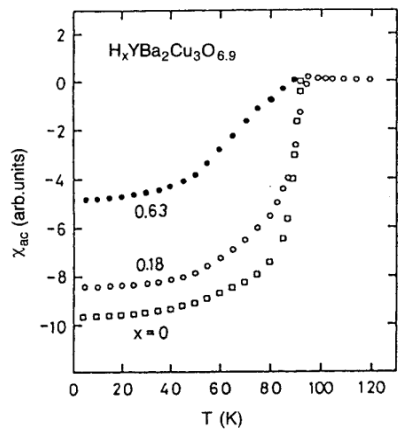

*Fig. 49.* Magnetic ac susceptibility $\chi$ as a function of temperature and hydrogen concentration for $H_xYBa_2Cu_3O_{6.9}$ [118]. Reprinted from T. Hirata, *Phys.stat.sol.* (a) 156, 227 (1996) with permission of John Wiley and Sons.

The electrical, magnetic and physicochemical properties of yttrium-barium hydrocuprate ($H_2YBa_2Cu_3O_7$) and its oxidized form ($H_2YBa_2Cu_3O_{7.3}$) are largely determined by the poorly studied behavior of hydrogen ions in the lattice. Using the methods of isotope exchange and inelastic neutron scattering, the results of studies of the crystal chemical state and mobility of protons intercalated in YBCO are presented in [119]. Based on their behavior in the process of inelastic neutron scattering, a significant part of protons can be considered mechanically free particles that do not have a chemical bond with oxygen ions.

Currently, cuprates are considered one of the most studied families of complex materials, and more than several hundred thousand scientific papers are devoted to them. Despite efforts, the main problems have not yet been resolved. The microscopic mechanism governing superconductivity is unknown, and the origin and nature of the various phases are generally not yet understood.

The origin of the effect of dissolved hydrogen on $T_c$ can be judged from the effect of hydrogen on the crystal structure. In Fig. 50, given in review [94], the expansion values of the crystal lattice of the $H_xYB_2Cu_3O_7$ cell is shown depending on the hydrogen content $x$. The lattice parameters $c$ and, to a lesser extent, $a$ increase, but the $b$ parameter remains almost unchanged with increasing hydrogen content, indicating that hydrogen takes up space, leaving chains of Cu(2)-O layers unbroken along the $b$ axis, which were thought to be important for the occurrence of superconductivity. Hydrogen inhibits the oxygen vacancy on the $c$ or $a$ axis or both, and is especially correlated with Cu in the solid solution region.

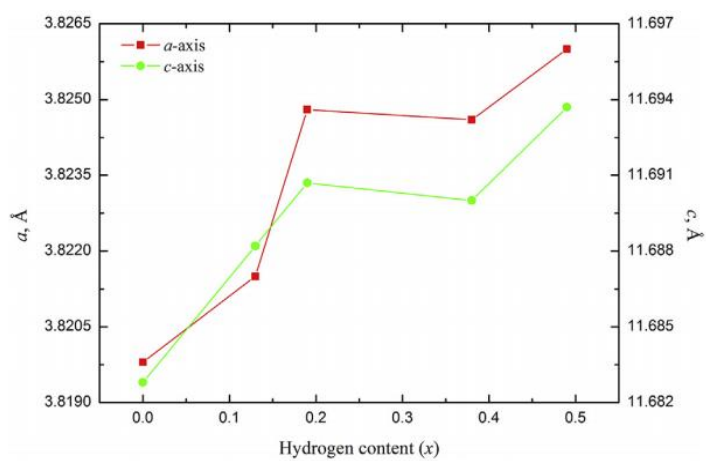

Fig. 50. The dependence of expansion values of the crystal lattice of the $H_xYB_2Cu_3O_7$ cell on the hydrogen content $x$ [94]. Reprinted from H.M. Syed, C.J. Webb, E. MacA. Gray, *Progress in Solid State Chemistry* 44, 20 (2016) with permission of Elsevier.

Subsequent studies using infrared spectroscopy and X-ray absorption spectroscopy confirmed that H is not associated with O atoms, but is located in interstices near Cu sites and forms Cu-H bonds.

In general, the reaction of this system is similar to those systems of metals and hydrogen in which H occupies an interstitial position coordinated with the metal atom, although no explanation has been proposed for the effect of H on $T_c$. It is noted that neutron diffraction on deuterium-modified $YB_2Cu_3O_7$ could provide direct evidence for the position of D and the opportunity to explore possible effects of H isotopes and further theoretical mechanisms by which hydrogen influences $T_c$ [120].

### 2.2.3. Properties of the magnesium-boron compound and carbon nanotubes.

Superconductivity in $MgB_2$ with a critical temperature $T_c = 39$ K was reported in 2001 in [121, 122]. Review [123] provides some information on the effect of hydrogen on this superconductor.

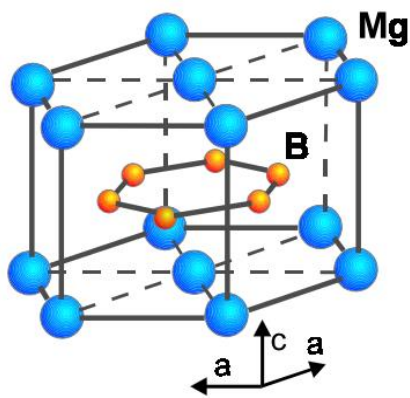

*Fig.51.* The structure of $MgB_2$ containing graphite-type B layers separated by hexagonal close-packed layers of Mg [123]. Reprinted from C. Buzea and T. Yamashita, *Supercond., Sci. & Technol.* 14, R115 (2001) with permission of IOP Publishing.

$MgB_2$ has a simple $A_lB_2$-type hexagonal structure (space group P6/mmm), common among borides. The structure of $MgB_2$ is shown in Fig. 51. It contains graphite-type boron layers separated by hexagonal close-packed magnesium layers. Magnesium atoms are arranged in a lattice in the form of hexagons and give up their electrons to boron atoms. Like graphite, $MgB_2$ has an anisotropic lattice in which the distance between boron planes is much greater than the distance between boron atoms in the plane. Its transition temperature is almost twice the maximum $T_c$ in $Nb_3Ge$ binary superconductors ($T_c$=23K). Comparing it with other types of superconductors (Figure 52), it can be seen that $MgB_2$ may be the "ultimate" low temperature superconductor with the highest critical temperature.

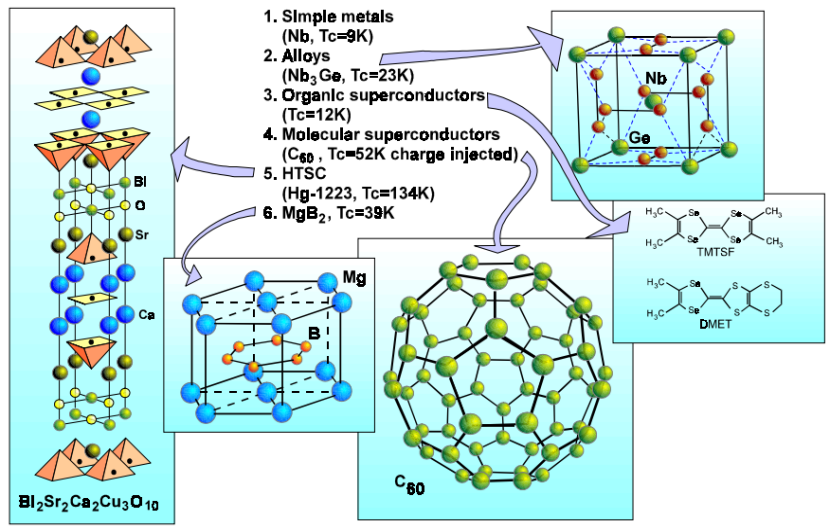

*Fig. 52.* Comparison between the structures of different classes of superconductors [123]. Reprinted from C. Buzea and T. Yamashita, *Supercond., Sci. & Technol.* 14, R115 (2001) with permission of IOP Publishing.

The properties of $MgB_2$ resemble those of conventional superconductors rather than the HTSCcuprates. These include the isotopic effect, the linear $T$-dependence of the upper critical field with positive curvature near $T_c$ (similar to borocarbides), and a shift in the temperature dependence of the resistivity $R(T)$ towards lower temperatures. On the other hand, the quadratic $T$-dependence of the penetration depth $\lambda(T)$, as well as the change in sign of the Hall coefficient near $T_c$, indicate that superconductivity, similar to cuprates, is unconventional. One should also pay attention to the layered structure of $MgB_2$, which may be the key to the higher critical temperature in both cuprates and borocarbides.

The response of the $MgB_2$ crystal structure to pressure is important for testing the predictions of competing theoretical models of the superconductor. For example, in simple metallic BSC superconductors such as aluminum, the critical temperature $T_c$ decreases under pressure due to a decrease in the electron-phonon interaction energy caused by an increase in the rigidity of the crystal lattice [124]. In addition, a large pressure derivative $dT_c/dP$ is a good indication that higher $T_c$ values can be obtained through chemical doping. In this regard, there is a prospect of using hydrogen to improve the characteristics of a superconductor. Doping is important from several points of view. First, it may increase the critical temperature of the selected compound. Secondly, this may indicate the existence of a related superconducting compound with a higher critical temperature $T_c$. Third, alloying elements that do not significantly reduce $T_c$ can act as pinning centers and increase the critical current density.

In [125], the critical current density of $MgB_2$ in a strong magnetic field was increased by implanting hydrogen nuclei—protons. Such hydrogen doping causes local damage to the crystal lattice structure and a local weakening of the order parameter in the superconductor, which increases the concentration of magnetic flux pinning centers. At the same time, the critical temperature $T_c$ decreases and the superconducting temperature transition broadens. The latter is probably due to the heterogeneity of damage caused by relatively high-energy protons. The influence of protons on the electronic structure of a superconductor was not considered in this work. The critical current density $J_c$ is obtained from irreversible magnetization measurements in a standard manner. Irradiation reduces $J_c$ in low fields (Fig. 53), partly due to a decrease in $T_c$. However, the decisive result is that the magnetic field dependence of $J_c$ is significantly weakened in irradiated samples. Even the lowest level of damage reduces the downward slope of the $\log(J_c)$ - H curve. Higher radiation doses further reduce this slope by about a factor of 2. The combination of low field degradation and high field enhancement produces an intersection such that the tested samples at 20 K have a higher $J_c$ than the original sample in magnetic fields (above about 2.5 T).

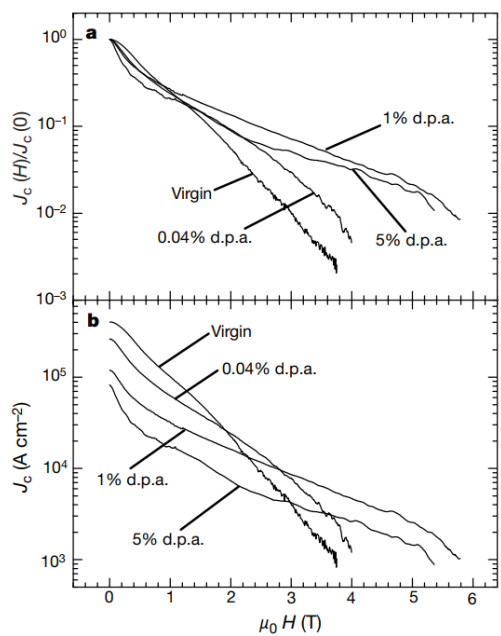

*Fig.53.* The magnetic field dependences of $J_c$ [125]. Reprinted from Y. Bugoslavsky, L.F. Cohen, G.K. Perkins *et al.*, *Nature* 411, 561 (2001).

Effect of irradiation on the field-dependence of $J_c$ at 20 K; the behaviour at other temperatures is similar. $J_c$ is obtained from the magnetization hysteresis width, using the Bean model. a, $J_c(H)$ of the virgin and irradiated samples at 20 K, with each sample normalized to its zero field value $J_c(0)$. Slower depression of $J_c$ by the field, as occurs even with the lowest dose, signifies more efficient vortex pinning in the irradiated samples. At higher doses, the improvement in the gradient of $\log(J_c)$ - H curves saturates, but $J_c(0)$ is suppressed; therefore there is an optimal level of disorder for enhancement of $J_c$ at high fields. b, Absolute values of $J_c$, which are somewhat uncertain, because the radiation damage is non-uniform.

Also, in these polycrystalline fragments, it is possible that defects have migrated to grain boundaries, interrupting the circulation of supercurrent [125]. Another way to quantify the dependence of $J_c$ on the magnetic field is to determine the irreversibility field $H^*$, above which $J_c$ falls below a certain level, the value of which is indicated in the caption to Figure 54. The position of the upper critical field $H_{c2}$ for an unirradiated $MgB_2$ sample is also shown there. Severely damaged (by neutron irradiation) $MgB_2$ was studied: disorder greatly reduces $T_c$, but leaves $dH_{c2}/dT$ practically unchanged ($J_c$ was not measured in this experiment). Therefore, it is likely that $dH_{c2}/dT$ does not change in the studied samples.

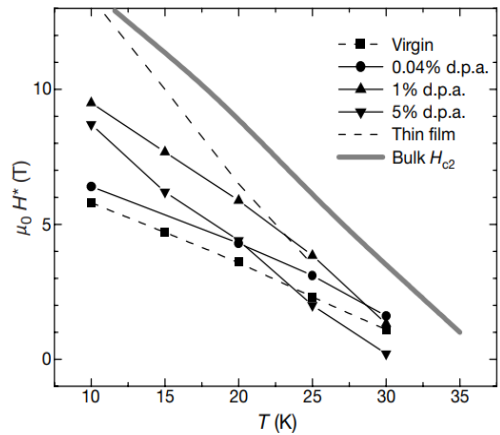

*Fig. 54.* Temperature dependence of the irreversibility field $H^*$ for different damage levels [125]. Reprinted from Y. Bugoslavsky, L.F. Cohen, G.K. Perkins *et al.*, *Nature* 411, 561 (2001).

The irreversibility field is that at which $J_c$ becomes immeasurably small; here we use a criterion of 1 kA cm$^{-2}$. There is a systematic increase of the slope of $H^*(T)$ curve with increasing dose. For the 5% doped sample (down triangles), this slope closely approaches the slope of $H_{c2}(T)$ in the virgin sample (squares). The irreversibility field of the nanocrystalline $MgB_2$ thin film (dotted line) is shown too [125]. Experimental observation of the effect of increasing $H^*$ in $MgB_2$ was also carried out in thin films deposited by laser ablation. A decrease in $T_c$ is also observed there, and the slope $dH^*/dT$ is even higher than in irradiated bulk samples (Fig. 54). These films are nanocrystalline, and, apparently, in such films, grain boundaries are responsible for the strong pinning of vortices [58, 123, 125].

At this stage, the detailed nature of the defects in irradiated $MgB_2$ has not been determined. The flow of protons can cause displacements of atoms in the lattice, and interstitial atoms can form loops between planes. At the end of their journey, the protons can diffuse to form $MgH_2$ or react to form boranes. In any case, the defects form stronger magnetic flux pinning sites than those that are natural for $MgB_2$, as evidenced by the sharp decrease in the slope of the plots of $J_c$ versus magnetic field strength $H$. Moderate disorder in the crystal lattice at a level that is achieved by chemical doping in volume of a superconductor, can significantly increase $J_c$ in *Fig.54.* Temperature dependence of the irreversibility field $H^*$ for different damage levels [125].

The performance of high-temperature superconductors has steadily improved over the years of their development. The best commercially produced long-length tapes, for example those containing the $Bi_2Sr_2Ca_2Cu_3O_x$ phase, achieve $J_c$ values of about $1.5 \times 10^5$ A $cm^{-2}$ at 20 K and zero applied field, falling by about a factor of three at 2 T. Both of these absolute values of $J_c$ and its (approximately exponential) field dependence are very close to those for irradiated $MgB_2$ (Fig. 54).

The superconducting properties of $MgB_2$ and the effect of hydrogen on them were also studied in [126]. It was found that hydrogenation at 100°C and $H_2$ pressure of 20 bar and 7 kbar leads to the formation of $MgB_2H_{0.03 \pm 0.01}$ without a noticeable change in lattice parameters and an insignificant change in $T_c$. AC magnetic field susceptibility measurements of a sample of $MgB_2$ hydrogenated at 20 bar suggested that the amount of superconducting phase was much less than exists after hydrogenation of $MgB_2$ at other pressures and temperatures [126], and the superconducting temperature transition was narrower. The hydrogen uptake was very low and independent of the applied pressure, and the volume of the $MgB_2$-type phase remained unchanged after hydrogenation.

In the studies presented, commercially available powdered $MgB_2$ (Johnson Matthey GmbH Alfa - 98% purity) was used. Hydrogen absorption was carried out at constant temperature and hydrogen gas pressure of 0.5–20 bar. The reaction temperature was set at 100°C and the pressure was set at up to 20 bar. The hydrogen concentration was determined by the Sievers method by observing the change in pressure in a calibrated sealed volume. Under the conditions described above, the hydrogen content did not exceed approximately $(3\pm1)\%$. In an attempt to increase hydrogen absorption, the hydrogen pressure was increased to 7 kbar, leaving the temperature at 100 °C. In this case, hydrogen absorption, as shown by mass spectrometry measurements, was the same and amounted to about 3%. X-ray diffraction analysis was carried out using the STOE diffraction system. The material was characterized by magnetic susceptibility measurements using alternating current. X-ray diffraction patterns of the original and hydrogenated samples are presented in Fig. 55. It is clearly seen that both the position of the diffraction peaks and their width are almost the same for all measured samples.

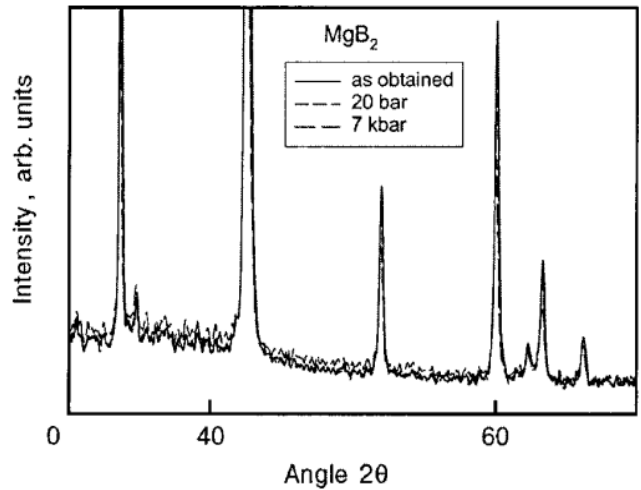

*Fig. 55.* X-ray diffractograms of as-obtained $MgB_2$ and hydrogenated under pressures of 20 bar and 7 kbar [126]. Reprinted from A.J. Zaleski, W. Iwasieczko, D. Kaczorowski *et al., FNT* 27, 1056 (2001) [*Low Temp Phys* 27, 780 (2001)].

It can be said that the structure and volume of the phase with a $MgB_2$ type structure did not change after hydrogenation. The results of measuring the susceptibility of $MgB_2$ hydrogenated at 20 bar are presented in Fig. 56. For comparison, the insets show the results for the initial $MgB_2$.

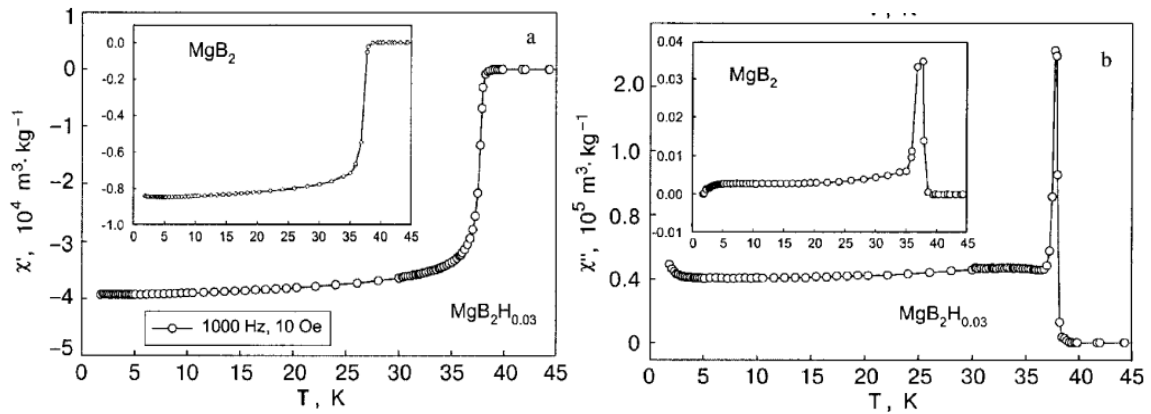

*Fig.56.* Real (a) and imaginary (b) parts of the ac susceptibility of $MgB_2$ hydrogenated under a pressure of 20 bar. The insets show the ac susceptibility for as-obtained $MgB_2$ [126]. Reprinted from A.J. Zaleski, W. Iwasieczko, D. Kaczorowski *et al., FNT* 27, 1056 (2001) [*Low Temp Phys* 27, 780 (2001)].

A comparison of the AC susceptibility of $MgB_2$ after hydrogenation at different hydrogen pressures is showing Fig.57. As noted above, the hydrogen absorption in both cases was very similar, despite the fact that the pressures during hydrogenation differ by a factor of 350. And, as in Fig. 56, it is clear that the critical temperature has remained unchanged and there is an additional decrease in the amount of the superconducting phase.

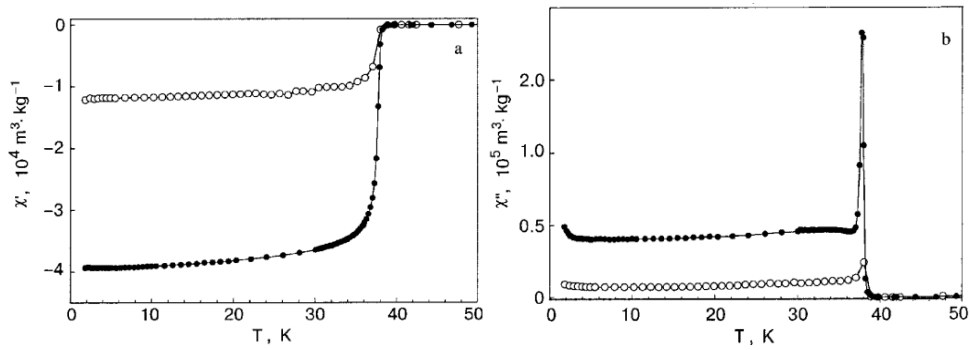

*Fig. 57.* The AC susceptibility of $MgB_2$ after hydrogenation at different hydrogen pressures. [126]. Reprinted from A.J. Zaleski, W. Iwasieczko, D. Kaczorowski *et al., FNT* 27, 1056 (2001) [*Low Temp Phys* 27, 780 (2001)].

For comparison, this hydrogenation procedure at low pressure and temperature was applied to the controversial compound $TaB_2$, studied in [127]. In this case, the hydrogen uptake was 36%, which is much higher than for $MgB_2$. And despite such a high hydrogen content, the critical temperature remained unchanged and amounted to about 9.5 K, but the diamagnetic signal was an order of magnitude lower than for the undoped material.

In [126], a significant decrease in the diamagnetic signal of hydrogenated industrial $MgB_2$ powder was observed, while its critical temperature remained practically unchanged. Hydrogen absorption was very low - about 4% and does not depend on the pressure used. This strange behavior can be explained by some opposing effects of hydrogen on the electron density of states (DOS), phonon frequency, electron-phonon coupling, and Coulomb screening.

At work. [96] managed to increase $T_c$ by approximately 1.2 K after holding $MgB_2$ in hydrogen under a pressure of 10 bar at 600°C.The $MgB_2$ powder was placed in a molybdenum boat that is pushed to the end of the finger. Once the finger was attached the system was evacuated and the reference volume filled to a predetermined pressure of hydrogen gas. This gas was then expanded into the sample volume and the end of the finger was heated in a small Heraeus split design tube furnace. After the sample had been heated to the appropriate temperature for the appropriate time, the furnace was switched off and the sample was allowed to cool.

Increasing the hydrogenation temperature to 700 C and 10 atmospheres resulted in a smaller increase in $T_c$, 0.4 K and 0.2 K, for soak times of 15 and 30 min, respectively. Each of these materials had similar ac field and frequency dependences as for the $MgB_2$ powder. Thus, lower temperature hydrogenation appears to result in a larger $T_c$ increase, but the optimum conditions have still to be determined. Hydrogenation of $MgB_2$ powder led to some increase in the superconducting transition temperature, determined by susceptibility. A constant magnetic field reduced the transition temperature in the same ratio as for pure powder.

Magnetic susceptibility was determined using a traditional inductively coupled bridge system. In this case, signal processing was performed automatically to obtain χ' and χ'' separately. The results of the susceptibility measurements are shown in Fig. 58. Powdered samples weighing ~40 mg were contained in cylindrical Teflon tubes sealed at one end with a small block of aluminum oxide. The tube was located on a copper pin in one half of the balanced secondary winding. The pin is connected directly to the cold head of the closed cycle helium refrigerator. An external constant magnetic field could be applied either parallel or perpendicular to the exciting alternating field. To measure χ($T$), a sample cooled in a zero field was heated at a rate of <1 K/min using a resistive heater. Between each magnetic field sweep, the samples were demagnetized by heating above the transition temperature.

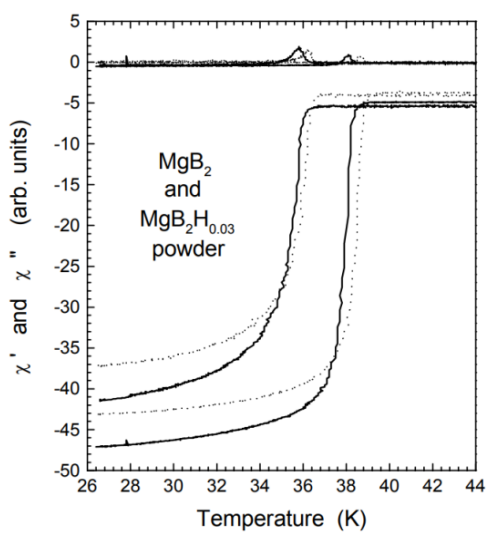

*Fig. 58.* AC susceptibility of pure $MgB_2$ (solid line) and hydrogenated $MgB_2H_{0.03}$ (dotted line) powder samples. Note the decrease in $T_c$ for each sample when a dc magnetic field of 0.5T is applied. The ac magnetic field is 0.35G, frequency 1kHz, the coils are slightly out of balance and the curves are displaced for clarity [96]. Reprinted from V.V. Flambaum, G.A. Stewart, G.J. Russell *et al.*, *Phys* C 382, 213 (2002) with permission of Elsevier.

Calculations using the obtained hydrogen absorption spectra showed that the chemical composition corresponded to $MgB_2H_{0.03}$. Treatment in Ar or He caused a smaller increase in $T_c$. A slight expansion of the lattice was observed in the direction $c$. These results once again confirm the theoretical assumption that $T_c$ depends on high-frequency modes for an electron-phonon superconductor.

A review article [123] shows that doping the $MgB_2$ compound with various chemical elements leads to a decrease in $T_c$, with the possible exception of zinc. The question of whether hydrogenation can significantly increase the $T_c$ of polycrystalline $MgB_2$ and be stable over time remains open.

In [128], the influence of the lattice parameter on the band structure of $MgB_2$ was calculated. It was established that the expansion of the unit cell increases the density of energy states of charge carriers at the Fermi level (DOS) $N_f(0)$ and thereby the critical temperature of the superconducting phase transition $T_c$. Thus, the effects of He and Ar and, to some extent, H in $MgB_2$ may be due to lattice expansion. Finding a superconductor with a higher $T_c$ by doping with $MgB_2$ will require a larger lattice constant $c$ and smaller $a$, $b$ values. A possible way would be to use a suitable substrate to force the $MgB_2$ film to have a larger size $c$ but smaller $a$, $b$. In [129], $MgB_2$ was synthesized in the presence of $H_2$ and Ar at 1173 K. An increase in the unit cell volume and an increase in $T_c$ by 0.6 K were obtained.

The authors of [130] hydrogenated $MgB_2$ with results similar to those of the studies described above. To confirm the $T_c$ dependence on the phonon frequency of the $E_{2g}$ mode and the unit cell volume, a pure $MgB_2$ sample was sintered at 800 °C for 10 h to eliminate the second sintering effects on its $T_c$. Then the sample was processed at 500 °C for 2 h in 30 atm $H_2$ atmosphere for hydrogenation. The unit cell volume calculated from the XRD pattern showed a small increase from 29.00 $Å^3$($a$= 3.0839 Å, $c$= 3.5208 Å) to 29.05 $Å^3$($a$= 3.0857 Å, $c$= 3.5232 Å), due to the hydrogen atoms, which are incorporated into the $MgB_2$ matrix and form $MgB_2H_x$ as interstitial atoms, but not substitutional atoms. The lattice parameter $a$ was nearly 0.1% larger in the sample sintered under hydrogen than in the sample sintered under argon. The $T_c$ for the hydrogenated sample was 0.6 K higher. The results of this work are in agreement with theirs. The phonon frequency of the $E_{2g}$ mode measured from the Raman spectra increases, as shown in Fig. 59. The $E_{2g}$ mode is dominant in the normal $MgB_2$, because the long sintering makes the material harmonic, while the PDOS (phonon density of states) peak becomes obvious for the hydrogenated sample, since the interstitial H atoms cause anharmonicity in the crystal. The Gaussian fitted results for the $E_{2g}$ mode and PDOS show that their frequencies shift from 587 and 754 $cm^{-1}$ to 612 and 773 $cm^{-1}$, respectively. Both the phonon frequency and the unit cell volume have increased for the $MgB_2$ sample after hydrogenation.

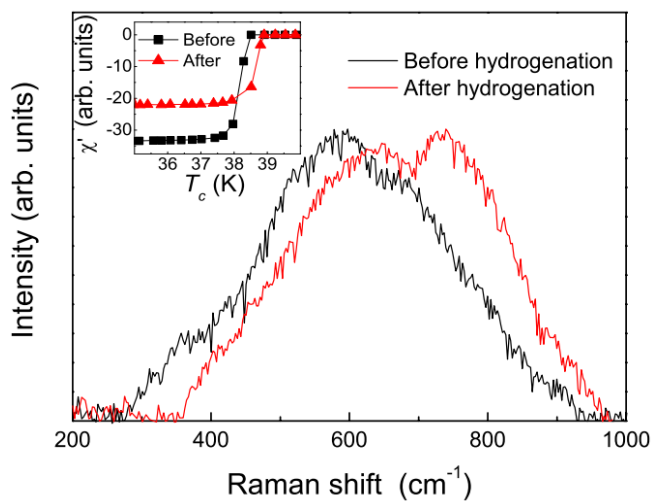

*Fig. 59.* The Ramanspectra of $MgB_2$ [130]. Reprinted from W.X. Li, Y. Li, R.H. Chen *et al.*, *Phys Rev B* 77, 094517 (2008) with permission of American Physical Society.

They concluded that the increase in $T_c$ due to hydrogen absorption was caused by interstitial H atoms, which did not affect the shape of the Fermi surface and the density of states, but was due to an increase in the phonon spectrum and in-plane Van der Waals interaction. However, other researchers note that this explanation contradicts the profound influence that the absorption of interstitial hydrogen has on the band structures of other metals.

In [131], tritium (T) was introduced into $MgB_2$ at room temperature and in a T + $N_2$ atmosphere. There was a small but noticeable increase in lattice parameter $c$ and a decrease in $a$, as well as an initial small increase in $T_c$ (by about 0.3 K), which slowly disappeared as tritium decayed to He.

Pure carbon does not exhibit superconductivity, with the exception of carbon nanotubes, although numerous carbon-based or carbon-containing materials have demonstrated superconductivity. The earliest discovered carbon-based superconductors were binary transition metal carbides with transition temperatures up to 11.1 K [132]. Carbon nanotubes exhibit superconductivity at 15 K. In [133], superconductivity was observed in single-wall carbon nanotubes with a diameter of 4 Å with a critical temperature of 15 K. It was noted that the smaller the diameter of the carbon nanotube, the higher the superconductivity temperature. A possible explanation for this improvement is that the greater curvature of the tube increases the interaction between electrons and lattice vibrations (phonons), which is an important property of superconductivity. The hydrogenation of single-walled nanotubes (SWNT) as a function of hydrogen concentration has been investigated with extensive first principles calculations [134].

Doping hydrogen into nanotubes gives rise to many properties that can become important in applications. One of the important effects of hydrogenation of single-walled carbon nanotubes (H-SWNTs) is the formation of metal nanotubes of square or rectangular cross-section with a high density of states at the Fermi level from zigzag nanotubes [135]. Uniform adsorption for zigzag nanotubes is metastable. Such a local minimum does not exist for all nanotubes, since uniformly adsorbed H atoms are rearranged during relaxation due to a coordinated exchange of C-H bonds with the formation of zigzag chains along the tube axis. Several snapshots of hydrogen dimerization on the tube are shown in Fig. 60.

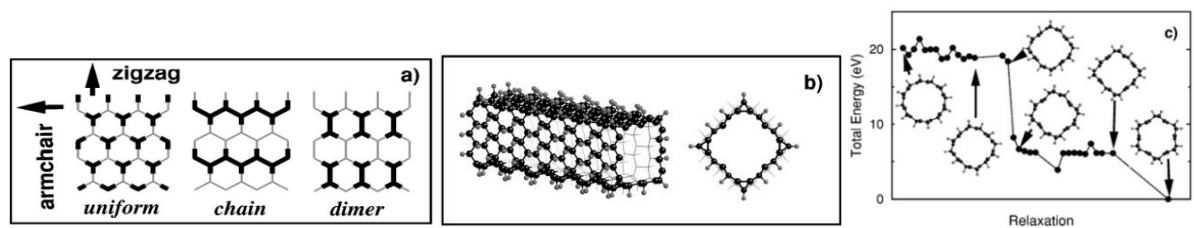

*Fig.60.* (a) A view of three different isomers of H-SWNT at half coverage. The left and up arrows indicate the tube axis for armchair and zigzag nanotubes, respectively. Carbon atoms which are bonded to hydrogens are indicated by dark color. (b) A side and top view of a (12,0) H-SWNT, indicating the square cross section of a uniformly exohydrogenated nanotube at half coverage. (c) Several snap shots during the relaxation steps of an armchair (6,6) H-SWNT, indicating that an uniform exohydrogenation at half coverage is not stable against forming a chain isomer [135]. Reprinted from O. Gülseren, T. Yildirim, S. Ciraci, *Phys Rev B* 66, 121401(R) (2002) with permission of American Physical Society.

In conclusion, it should be noted the attempts to influence hydrogen on the superconducting characteristics of various phases of aluminum [136] and [137]. Another unexplored area of possible beneficial effects of hydrogen on superconductors with a layered crystal structure (like cuprates, diborides, iron pnictides and iron chalcogenides) are superconductors based on nickel oxides – nickelites [138]. The recent experimental achievement in the detection of high-temperature superconductivity in La¬Ni$_2$O$_7$ (LNO) single crystals under pressure has attracted increased interest. Since the observed maximum SC transition temperature $T_c$ reaches 80 K at pressures exceeding 14 GPa, LNO represents a new platform for studying unconventional pairing mechanisms.

Since the discovery of high-temperature superconductivity (HTSC) in cuprates, commonly referred to as doped Moth- insulators, the quest to understand the relationship between the pairing mechanism in unconventional superconductors and strong electronic correlations [61] has continued to generate interest.

# 3. On the mechanisms of the influence of hydrogen on superconductivity

## 3.1. Metallic superconductors and hydrides.

The Bardeen-Cooper-Schrieffer microscopic theory of superconductivity (BCS - theory, 1957) explains the existence of superconductivity in metals by the exchange of two electrons (with opposite spins) with two phonons. As a result, electron pairing occurs and electron pairs can be transported through a metal conductor without resistance. According to the BCS theory, the critical temperature of a superconductor is given by the formula:

$$T_c \approx 1.14\theta_D \exp\left(-1/N(0)V\right)$$

(9)

where $V$ is the electron-phonon interaction potential, $N(0)$ is the density of electronic states (DOS - Density Of States) at the Fermi level, $\theta_D$ is the Debye temperature at which the maximum vibration frequency of the metal crystal lattice is excited. A further increase in temperature does not lead to the appearance of higher frequency oscillations, but only increases their amplitude. Density of states (DOS) is a quantity that determines the number of available energy levels of electrons in a unit energy range per unit volume of a conductor (sometimes the authors recalculate per single atom or per unit cell of a crystal lattice) and a superconductor in a three-dimensional case (per unit area in a two-dimensional case). From formula (9) it is clear that the transition temperature $T_c$ can be increased, in particular, by increasing $\theta_D$, if the electron-phonon interaction constant $\lambda = N(0)V$ changes little. The depth of penetration of a magnetic field into a superconductor and the coherence length (equal to the size of a pair of electrons), initially introduced through the phenomenological equations of London and Ginzburg-Landau, arise naturally in the BCS theory [138]. Considering that

$$\theta_D = hf_D / k_B = h\nu/(k_B a),$$

(10)

where $h$, $f_D$, $k_B$, $\nu$, $a$ are, respectively, Planck's constant, Debye oscillation frequency, Boltzmann constant, speed of sound in a metal, cell size of a metal crystal lattice, then from (10) it follows that the Debye temperature can be increased by decreasing the size ($a$) cells using external pressure on the metal. Indeed, experimental studies carried out at the beginning of the 21st century in laboratories in France, Japan and other countries showed that most chemical elements of the periodic table become superconductors with various low $T_c$ when they are compressed under high pressure. The only exception to the number of such low-temperature superconductors under pressure was hydrogen. Calculations showed that solid metallic hydrogen, which has a minimal crystal cell size, under high pressure (about 500 GPa) can have a $T_c$ of about 100 K. Attempts to obtain solid hydrogen have not yet yielded positive results. Therefore, the next step in increasing $T_c$ in this way was the use of compression of compounds of various chemical elements with hydrogen. This is how superconducting hydrides under high pressure appeared, described in Section 1 and Subsection 2.1, for which $T_c$ close to room temperature was successfully discovered. The results achieved were a remarkable confirmation of the BCS theory.

Simultaneously with the research into superconductors under pressure, work was carried out on the synthesis of hydrogen-containing superconductors at **normal** pressure. In these works, only in a few metal compounds it was possible to obtain a positive result explained by the BCS theory. Let us first consider the background information associated with classical metallic superconductors. Classical metallic superconductors can be divided into four classes (or groups) with four different superconductivity mechanisms [140, 141]. The first group includes *s-p*-electron metals, which belong to metals of groups 2 to 16 of the periodic table. These are the earliest known superconductors and are also called "soft" superconductors. This is the only group among superconducting elements where electron-phonon interaction is a known cause of the phenomenon of superconductivity. The second group includes transition elements or *s-d* electronic elements. This class also includes transition metal compounds, as well as compounds with *s-p* metals. In this class, the maximum transition temperature is found for numbers of valence electrons close to 5 and 7 per atom (even numbers are expected to correspond to filled bands and a small value of $N(0)$). The third class of superconductors includes non-magnetic metal compounds with a cubic crystal lattice structure with high Debye temperatures, including compounds of transition metals with light elements, in particular beryllides, nitrides, carbides and borides. Superconductors, which belong to the fourth and final class, consist of elements with a partially filled *f*-shell, such as lanthanides.

The desire to improve the superconducting characteristics of metallic and intermetallic superconductors has led to the use of various methods of doping them, including the smallest in size and most abundant element in the universe - hydrogen [139].

The good solubility of molecular hydrogen in metals is associated with its ability to diffuse through metals. Hydrogen is a light gas, so it has a high diffusion rate. Its diffusion ability is great at high pressure and temperature. If hydrogen atoms appear in a metal, they displace the metal atoms from equilibrium positions, causing expansion of the crystal lattice, which in some cases causes technical problems when using structural materials. Hydrogen interacts with almost all metals and can be present in them in the following states: adsorb on the surface, especially in the presence of microroughnesses, accumulate in micropores in molecular form, dissolve in the volume, form hydrides and interact with alloying elements. The state and distribution of atomic hydrogen, as well as its quantity in metals and compounds, depend on the characteristics and imperfections of the crystal structure. Schematically, the main processes of interaction of hydrogen with the surface of metals are presented in Fig. 61.

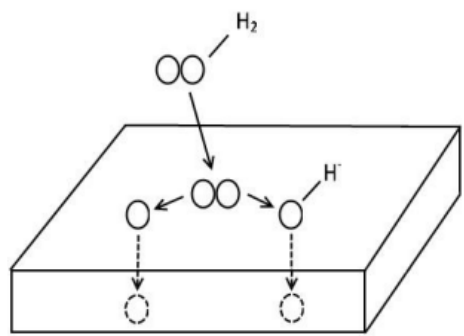

*Fig. 61.* Interaction of hydrogen with the surface of metals and intermetallic compounds [139]. Reprinted from Tatsuya Kawae, Yuji Inagaki, Si Wen *et al.*, *J. Phys. Soc. Jpn.* **89**, 051004 (2020).

The interaction of hydrogen with metals includes physical and chemical adsorption, diffusion, dissolution and the formation of chemical compounds. When a metal comes into contact with hydrogen, a layer of adsorbed gas forms on its surface. During physical adsorption, van der Waals forces arise between the surface and hydrogen molecules, while electron transfer does not occur and hydrogen molecules do not dissociate into atoms. During chemical adsorption, a covalent or some other bond occurs between surface atoms and hydrogen molecules due to the exchange of electrons. Adsorption accelerates with increasing temperature. In the process of chemical adsorption, hydrogen molecules disintegrate into atoms, then protons, which have high mobility and, having overcome the energy barrier, diffuse deep into the metal. Next, hydrogen dissolves in the metal or compound. Since hydrogen is diffusion-mobile, reverse processes can also occur simultaneously, such as the movement of protons from the bulk to the surface of the metal, the transformation of protons into hydrogen atoms, the appearance of molecular hydrogen due to recombination on the metal surface, and the desorption of molecular hydrogen into the gas phase. The absorption of hydrogen may be accompanied by the formation of solid solutions or chemical compounds. For example, at 300 °C, a solid solution first forms in titanium. When the hydrogen concentration reaches 0.15 wt.% (the hydrogen solution becomes saturated), the formation of the hydride phase begins. The penetration of hydrogen and its distribution in the volume of the metal is called absorption.

Depending on the sign of the thermal effect, it is divided into endothermic and exothermic. Endothermic absorption occurs with the absorption of heat, and therefore the amount of absorbed hydrogen increases with increasing temperature. In metals that absorb hydrogen by an endothermic reaction, the heat of hydrogen absorption is positive. With this absorption, no hydrides are formed. Exothermic absorption occurs with the release of heat, and therefore the hydrogen content in the metal decreases with increasing temperature. Metals that absorb hydrogen through an exothermic reaction form hydrides - these include, for example, titanium and zirconium.

Based on the nature of the bond between hydrogen atoms and metals, hydrides are divided into covalent, ionic and metallic. For example, metal hydrides are formed in titanium and zirconium, which are solid solutions. This type of hydride is characterized by an increase in volume, as a result of which internal stresses arise in the metal. Metal hydrides are brittle in many cases. Hydrogen atoms in transition metals are ionized under the influence of the potential field of the metal to form a positively charged ion - a proton, or a negatively charged ion. In the case of the formation of a proton, the lattice type and metallic properties of the metal are preserved, and in the case of the formation of a negatively charged hydrogen ion, a chemical compound with an ionic type of bond appears, which has a specific lattice. A proton is formed if the ionization potential of metal atoms is above 7 eV and the $s$-electrons of hydrogen move to the unfilled $d$-shell of the metal with a lower energy level. If the ionization potential is less than 7 eV, the metal donates electrons to hydrogen atoms and negatively charged hydrogen ions or hydrides are formed [142, 143].

The state of hydrogen in the form of protons in the lattice of transition metals is confirmed by experimental results, which indicate a change in the magnetization of these metals. The reason for the ionization of hydrogen atoms is considered to be the influence of the force electric field of a lattice, for example, nickel. After hydrogen chemisorption, the magnetization of nickel decreases, which means that the $d$-level is filled with electrons from hydrogen atoms. If complete ionization of hydrogen to a proton occurred in metals, then, due to its small size, the solubility and diffusion constants would not depend on the size of the gaps between the atoms in the metal lattice, between which the hydrogen atoms diffuse. In fact, the solubility and diffusion mobility of hydrogen atoms in metals are determined by the parameters of the crystal lattice, which indicates the large size of the hydrogen atom compared to the proton. Also, the relatively large size of hydrogen atoms in solid solutions is indicated by an increase in the resistance to plastic deformation of these solutions. Based on this, we can conclude that hydrogen in metals is not completely ionized.

Regions of the metal in which free atomic hydrogen is concentrated with low free energy compared to its energy in the lattice are called traps, and regions of accumulation of bound molecular hydrogen are called collectors. Traps are classified according to several criteria: open and closed, reversible and irreversible, saturated and unsaturated, movable and immobile. They are also divided into point ones, which include vacancies, substitution and interstitial atoms; linear – edge and screw dislocations, dislocation thresholds, intersections of three grain boundaries; two-dimensional – intergrain, interphase and twin boundaries; volumetric – areas of volumetric tension, accumulation of dislocations, pores and discontinuities. In superconductors, these defects must inevitably create effective pinning centers.

Experiments to study the processes of hydrogen sorption and desorption are carried out in the following sequence: the sample is weighed, the mass and density are entered into the software of the measuring complex used, the sample is placed in a chamber, the chamber is connected to a vacuum system and evacuated, linear heating (or cooling) is carried out with continuous pumping. After completion of the mode, hydrogen is pumped out of the chamber and cooled (or heated) [139, 143].

Interstitial hydrogen-enhanced superconductivity was first reported in 1970 in the thorium-$H_2$ system, where $Th_4H_{15}$ was found to be superconducting at temperatures of about 8 K, compared to $T_c$= 1.37 K for pure thorium. Within the BCS picture, the influence of hydrogen can be qualitatively understood as arising from the interaction of optical phonons with the electronic system, which leads to the expectation of an isotopic effect in which heavier interstitial sites reduce $T_c$ [94].

It is stated in [144, 145] that superconductivity has been repeatedly observed in electrolytically doped $PdH_x$ at higher temperatures (almost room temperature). Anomalies in the electrical and magnetic properties of $PdH_x$ at temperatures up to 70 K were reported in [146, 147], which wer were interpreted as indicating filamentary superconductivity. In 1950, Maxwell and Reynolds discovered that the critical temperature of a superconductor depends on the isotopic mass of its atoms [148, 149]. This important discovery pointed to electron-phonon interaction as the microscopic mechanism responsible for superconductivity.

In [150], a comparison of the calculated and experimentally obtained characteristics of the superconducting hydrides $PdH_x$ and $PdD_x$ was successfully carried out, which significantly expanded the understanding of the process of the effect of hydrogen on metallic superconductors. When calculating the concentration dependence of the superconducting transition temperatures of $PdH_x$ and $PdD_x$, including the inverse isotope effect, the first principles of electronic calculation of the band structure and measurement of phonon properties were used. Agreement with experimental results was obtained. The model used by the authors also provides a qualitative understanding of elevated transition temperatures in palladium hydrides with noble metals.

The inverse isotopic effect is explained by an effective increase in the force constant of the Pd-H interaction, $k_{Pd-H}$, compared to $k_{Pd-D}$ for Pd-$D_x$ interaction due to the increased anharmonicity of the H motion, as initially assumed from the analysis of neutron scattering measurements on $PdH_{0.63}$ and comparing them with data for $PdD_{0.63}$. It was found that the force constant for the case of protium is 1.2 times greater. And this leads to a 20% increase in the $(\omega^2)_D M_D / M_H$ ratio. The authors show that this increase is sufficient to quantitatively explain the observed inverse isotope effect. The dependence of $T_c$ in $PdH_x$ ($PdD_x$) on $x$ is due, according to the analysis, mainly to the rapid increase in the average electron-optophonon (i.e. H- or D-vibration) interaction $\lambda_{opt}$ as $x$ approaches 1. Although the total electron DOS decreases with increasing $x$, the H- or D-site components of DOS (see Fig. 62) increase such that $\lambda_{opt}$ increases with *x*. In fact, as the concentration of H or D increases, more electrons are found near the H or D sites. This effect continues for *x* > 1.0, and we argue that this explains the elevated $T_c$ values in palladium noble metal hydrides.

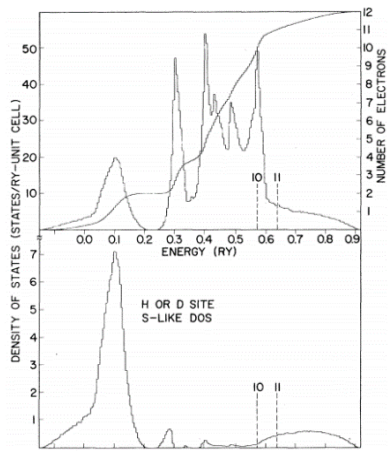

*Fig.62.* Calculated electronic densities of states for PdH or PdD vs energy. The vertical dashed lines show the location of the Fermi level, $E_F$, for ten or eleven electrons per unit cell (*x* = 0.0 and 1.0, respectively) [150]. Reprinted from B. M. Klein, E. N. Economou, and D. A. Papaconstantopoulos, *Phys. Rev. Lett.* 39, 574 (1977) with permission of American Physical Society.

Since the mechanism of the inverse isotope effect is related to the contribution of the optical mode, this effect tends to disappear with decreasing *x* (For example, at *x* = 1 for PdD$_x$ $T_c$ = 9.6 K, and for PdH$_x$ $T_c$ = 7.9 K). This is due to the fact that the effect of hydrogen isotopes on $T_c$ decreases as *x* decreases.

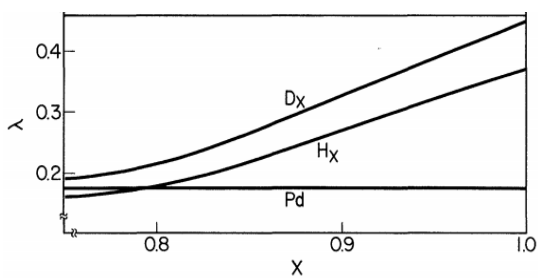

*Fig.63.* Calculated value of electron-phonon mass enhancement factors $\lambda_j = \eta_j / M_j \langle \omega^2 \rangle_j$ vs *x* in PdH$_x$ and PdD$_x$ [150]. Reprinted from B. M. Klein, E. N. Economou, and D. A. Papaconstantopoulos, *Phys. Rev. Lett.* 39, 574 (1977) with permission of American Physical Society.

It can also be seen from the results of [151] that increasing *x* to values greater than 1.0 will further increase $T_c$. We hypothesize that this is the reason for the higher $T_c$ in palladium noble metal hydrides.

In [143], the above described phenomena are explained by the ability of hydrogen atoms to create an additional high-frequency optical branch of vibrations of the solvent metal lattice. The existence of high transition temperatures to the superconducting state is largely due to the strong interaction of these optical phonons with the electronic system.

When doping, the binding energy of a hydrogen atom as a function of the electron density of the medium can be calculated using the effective medium theory [152, 153]. The binding energy of hydrogen atoms has a maximum at low electron densities and decreases with increasing electron density. When a hydrogen atom approaches the surface of a metal, the electron density with the maximum binding energy is always found. On their way through the metal lattice, the electron density is usually too high, except for vacancies.

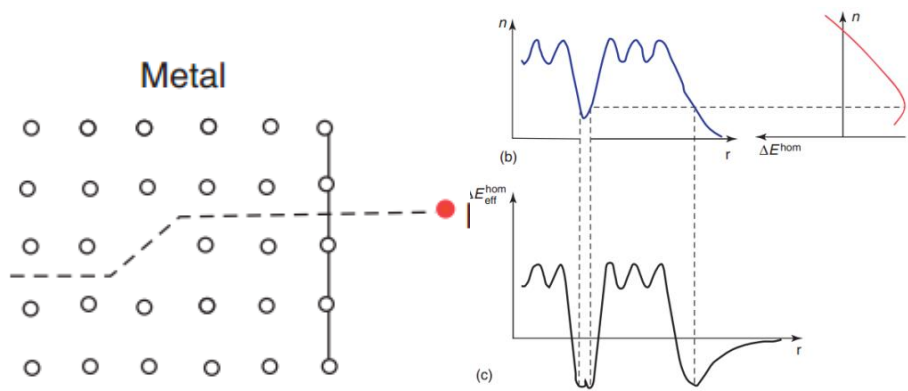

*Fig.64.* (a) A hydrogen atom on its way to the metal surface and then through the lattice. (c) Electron density n along the path of the *r* hydrogen atom and the change in its potential energy $\Delta E$ in this case [152]. Reprinted from Zuttel A., Hydrides. , *in Encyclopedia of Electrochemical Power Sources,* 2009 Elsevier B.V. with permission of Elsevier.

Hydrogenation is actually the introduction of electrons and protons into the electronic structure of the main lattice. For example, for transition metals, protons lower the energy of a portion of the *d* electrons and lead to new states about 4 eV below the Fermi energy $E_F$. On the other hand, electrons fill empty states at the Fermi energy $E_F$ and therefore increase the Fermi energy (Fig. 65). In this case, the *d*-zone of the base metal contains $N$ electrons, where $N$ is less than or equal to 10. The lowest zone hybridizes with the *s*–*p*-zone and forms a zone of predominantly *s*-character in the H sections, containing two electrons. Intercalating hydrogen typically lowers the *s*-type band by 3–4 eV. The remainder of the *d*-band can contain a maximum of eight electrons and is occupied by $N – 2$ electrons.

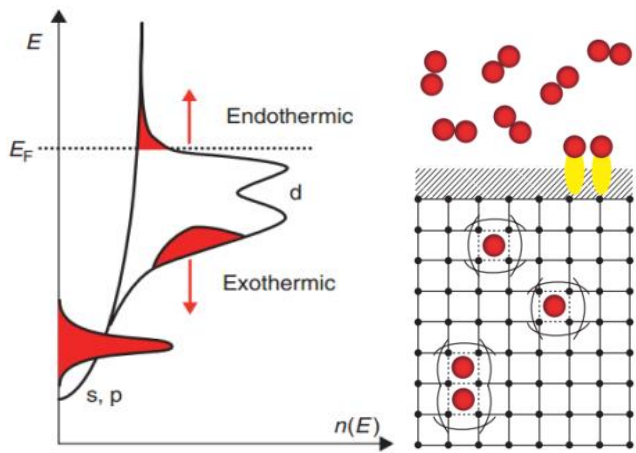

*Fig.65.* Schematic representation of hydrogen absorption (right) and the effect on the electronic structure, density of electronic states $n(E)$ (left) [152]. Reprinted from Zuttel A., Hydrides. *, in Encyclopedia of Electrochemical Power Sources,* 2009 Elsevier B.V. with permission of Elsevier.

The heat of formation of binary hydrides MHx is linearly related to the characteristic band energy parameter $\Delta E = E_F - E_s$, where $E_F$ is the Fermi energy, and $E_s$ is the center of the electronic band of the base metal with a strong $s$-character of bonding in interstices occupied by hydrogen.

It was shown in [153 – 158] that for most metals $E_s$ can be taken as the energy that corresponds to one electron per atom on the integral density of states (DOS) curve. The semi-empirical model mentioned above allows the stability of binary hydrides to be assessed if hard band theory can be applied. Attempts have been made to evaluate the possible influence of hydrogen on the crystal structure and current characteristics of aluminum [159].

Thus, in classical metallic superconductors at normal external pressure, a noticeable effect of hydrogen absorption on the critical temperature is observed. The BCS theory qualitatively explains the effect of hydrogen on $T_c$ through the electron-phonon interaction constant. The main factors are the effect of the added electron of interstitial H on the density of states at the Fermi level and the creation of optical phonon modes involving H, both of which change the electron-phonon coupling constant.

In conclusion of this section, we would like to note that despite criticism from Jorge Hirsch [160-165], both experimental and theoretical work is being successfully carried out and superconductivity under pressure is discovered in new hydrides every year. New estimates of the pressure required to obtain superconductivity in hydrides are also emerging [166, 167]. These compounds fit into the framework of the BCS theory, exhibit the properties of type II superconductors, and exhibit a positive isotope effect very close to α=0.5 of the BCS theory [87-89]. The critical current density of hydrides at low temperatures approaches the critical current density of commercial HTSC wires and the best LTSC, and the value of the second critical magnetic field of some hydrides can reach 200 T.

Measurements of the superconducting gap in hydrides by means infrared reflection spectroscopy; measurements of hydrides in pulsed magnetic fields up to 60-80 T and more to accurately construct the $H_{c2}(T)$ dependence; femtosecond reflection spectroscopy for direct measurement of the electron-phonon interaction constant of hydrides; Andreev reflection spectroscopy and microcontact spectroscopy to measure the energy gap of hydrides and its anisotropy; fabrication of microrings from hydrides to study magnetic flux trapping in them and fabrication of SNS structures; the use of microheaters and microthermometers to study the jump in heat capacity are promising. In high-temperature superconductors with a layered crystal structure, such as cuprate, iron-containing, diborides and nickelites, the BCS theory cannot describe the whole picture of the superconducting state, especially with hydrogen. A new microscopic theory for these superconductors needs to be developed.

## 3.2 Iron-based superconductors.

In some HTSCs with a layered crystal structure, such as cuprate, iron-based, diborides and nickelites, the BCS theory cannot describe the entire picture of the superconducting state. Therefore, it can be expected that the effect of hydrogen on this class of superconductors cannot be understood only in terms of changes in the electron-phonon interaction. As a study of the mechanisms of superconductivity, modification of superconductors with hydrogen (protium and deuterium) has proven to be very valuable. The role of hydrogen in creating Fe-based superconductors and improving their properties seems very promising today. Understanding superconductivity at high transition temperatures ($T_c$) remains one of the greatest challenges in condensed matter physics. It is believed that for cuprates the fundamental mechanism of $d$-wave pairing essentially lies in the $d_{x2-y2}$ orbital in the presence of strong Coulomb repulsion. It is usually called unconventional superconductivity, different from the more traditional type of Bardeen-Cooper-Schrieffer (BCS) superconductivity. Another striking example of unconventional superconductivity can be found in iron-based superconductors, where multiple $d$-orbitals are often involved in pairing. Similar conclusions were made in studies of diboride superconductors. Relatively recently, a new nickelate superconductor $La_3Ni_2O_7$ with $T_c \approx 80$ K at moderate pressures (up to 29.2 GPa) was discovered. It is important to note that it represents one of the rare examples of a superconductor whose critical temperature exceeds the boiling point of liquid nitrogen. Superconductivity, for example, in iron chalcogenides $Fe_{1+x}Te_{1-y}Se_y$ ($x = 0, 0.1$; $y = 0.1 - 0.4$), which have an anti-PbO crystal lattice structure, is strongly influenced by the number ($x$) of interstitial iron atoms located between the layers $FeTe_{1-y}Se_y$.

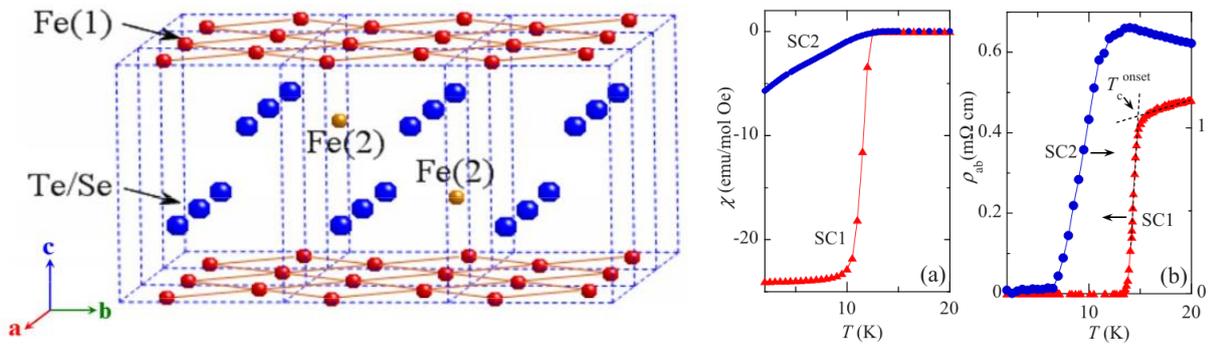

*Fig*. 66.: (left) schematic crystal structure of $Fe_{1+y}(Te,Se)$. The iron on the square-planar sheet is denoted by Fe(1); the iron partially occupying at the interstitial sites of the (Te, Se) layers is the excess Fe, denoted by Fe(2). (a) Magnetic susceptibility as a function of temperature $\chi(T)$ measured under a magnetic field of 30 Oe (applied along the *c* axis). (b) In-plane resistivity as a function of temperatures $\rho_{ab}(T)$. SC1 and SC2 represent two superconducting samples with 3% and 11% Fe(2). From [168]. Reprinted from T. J. Liu, X. Ke, B. Qian *et al.*, *Phys. Rev.* B 80, 174509 (2009) with permission of American Physical Society.

Iron chalcogenide (Fe*Ch,* where*Ch* =Te, Se, S) compounds consist of stacked layers of edge-sharing Fe*Ch*$_{4/4}$ tetrahedra, with up to 25% of the iron in the van der Waals gap (Figure 66). These interstitial iron atoms in $Fe_{1+x}Ch$ interfere with superconductivity by their magnetic moments and/or by their unfavorable contribution to the Fermi surface. The critical temperature of the FeTe$_{1-y}$Se$_y$ solid solution increases to 14 K at $y \approx 0.5$, depending on the amount of incorporated iron [169–172]. Thus, understanding and controlling excess iron in Fe*Ch* superconductors is also fundamentally important for possible applications, such as the production of superconducting wires.

Work [172] describes the use of additional annealing of the $Fe_{1+x}Te_{1-y}Se_y$ compound in a hydrogen atmosphere to stabilize its superconducting characteristics after annealing in oxygen to remove excess iron. The authors point out that non-superconducting samples of the nominal compound $Fe_{1.1}Te_{1-y}Se_y$, after annealing in an oxygen atmosphere, turn into superconductors with critical temperatures up to 14 K. The process becomes irreversible with subsequent hydrogen annealing.

Oxygen annealing was carried out by heating the samples to 300 °C for 2 hours in alumina crucibles inside sealed Duran© glass ampoules in an oxygen atmosphere ("O$_2$-annealed" samples). For hydrogen annealing, O$_2$-treated samples in alumina crucibles inside a Duran© tube connected to a bubble counter were heated to 200 °C for 2 hours under a continuous flow of hydrogen ("H$_2$-annealed samples").

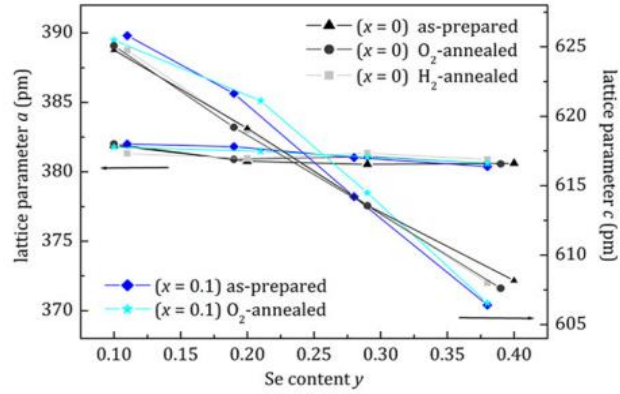

*Fig. 6*7. Lattice parameters of $Fe_{1+x}Te_{1-y}Se_y$ ($x = 0, 0.1; y = 0.1 –0.4$) [172]. Reprinted from Gina M. Friederichs, Matthias P. B. Wörsching and Dirk Johrend, *Supercond. Sci. Technol.* 28, 095005 (2015).

The lattice parameters $a$ with increasing Se content $y$ decrease slightly by 0.4%, and $c$ decrease more strongly, by 2.7%, as expected, due to the smaller radius of Se compared to Te. After annealing, only small changes were detected, not exceeding 0.1% (Fig. 67). Small changes in the amount of interstitial iron are not reliably detected by X-ray diffraction analysis due to the very weak scattering of only ≈10% $Fe^{2-}$. Moreover, it is likely that oxygen treatment occurs at the surface, leading to the formation of heterogeneous particles [170]. Data from electron transport measurements are consistent with these findings. The resistivity of freshly prepared, annealed in $O_2$ and post-annealed in $H_2$ $FeTe_{0.8}Se_{0.2}$ drops to zero at 14 K (Fig. 68). In contrast, $Fe_{1.1}Te_{0.8}Se_{0.2}$ becomes superconducting only after treatment in $O_2$.

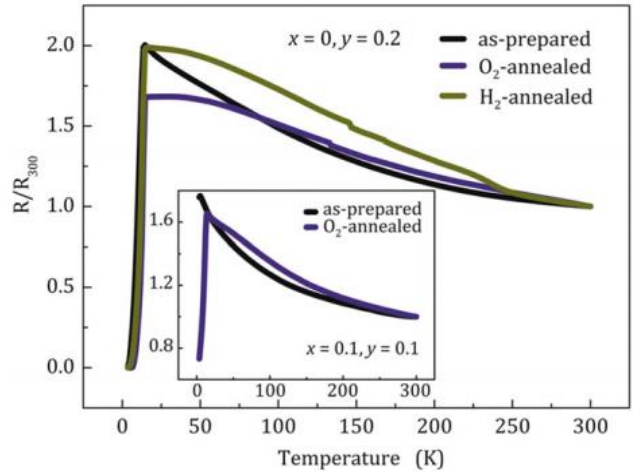

*Fig.* 68. Relative electrical resistivity of as-prepared, $O_2$-annealed, and $H_2$-annealed $FeTe_{0.8}Se_{0.2}$[172]. Reprinted from Gina M. Friederichs, Matthias P. B. Wörsching and Dirk Johrend, *Supercond. Sci. Technol.* 28, 095005 (2015).

The superconducting transition temperature ($T_c$), for example, in the FeSe compound can be significantly increased by several times through the application of pressure, electronic doping, intercalation of the intermediate layer and dimension reduction. Various ordered electronic phases such as nematicity and spin density waves have also been observed in high temperature superconductivity. In FeSe single crystals intercalated with $H^+$, using the ionic liquid implantation method [33], a series of discrete superconducting phases with a maximum temperature $T_c$ of up to 44 K was discovered. Simultaneously with an increase in $T_c$, suppression of the nematic phase and a transition from non-Fermiliquid to Fermiliquid behavior was observed. An abrupt change in the topology of the Fermi surface has been proposed to explain the discrete superconducting phases. In this case, a transformation of the band structure occurs, which favors the stability of the HTSC phase.

In [173], the effect of hydrogen doping on superconductors based on iron pnictides was studied using the first principles of density functional theory. It was found that the most stable arrangement of hydrogen atoms in LaFeAsO is near the Fe positions, which is consistent with NMR experiments [174]. The reason for the increase in $T_c$ is probably related to distortions of the original lattice, which promote superconductivity. Electronic doping, due to H atoms, can also be associated with an increase in the critical temperature. For illustration, figure 69 shows the crystal structure of the $(LaFeAsO)_8H_x$ supercell used in the calculations, as well as the most stable possible H positions in the lattice. To determine the position of hydrogen, the structure of $(LaFeAsO)_8H$ was optimized, where the H atom was initially distributed randomly as the initial configuration of the system. As a result of optimization, the most stable state was found, shown in figure 69 as an example for $x = 0.125$ and $x = 1$. In this configuration, the H atoms exist almost on the opposite side of the As atom relativeto the plane of the Fe atoms. The position is slightly shifted in the *ab* plane to avoid the influence of the upstream $La^{3+}$ cation.

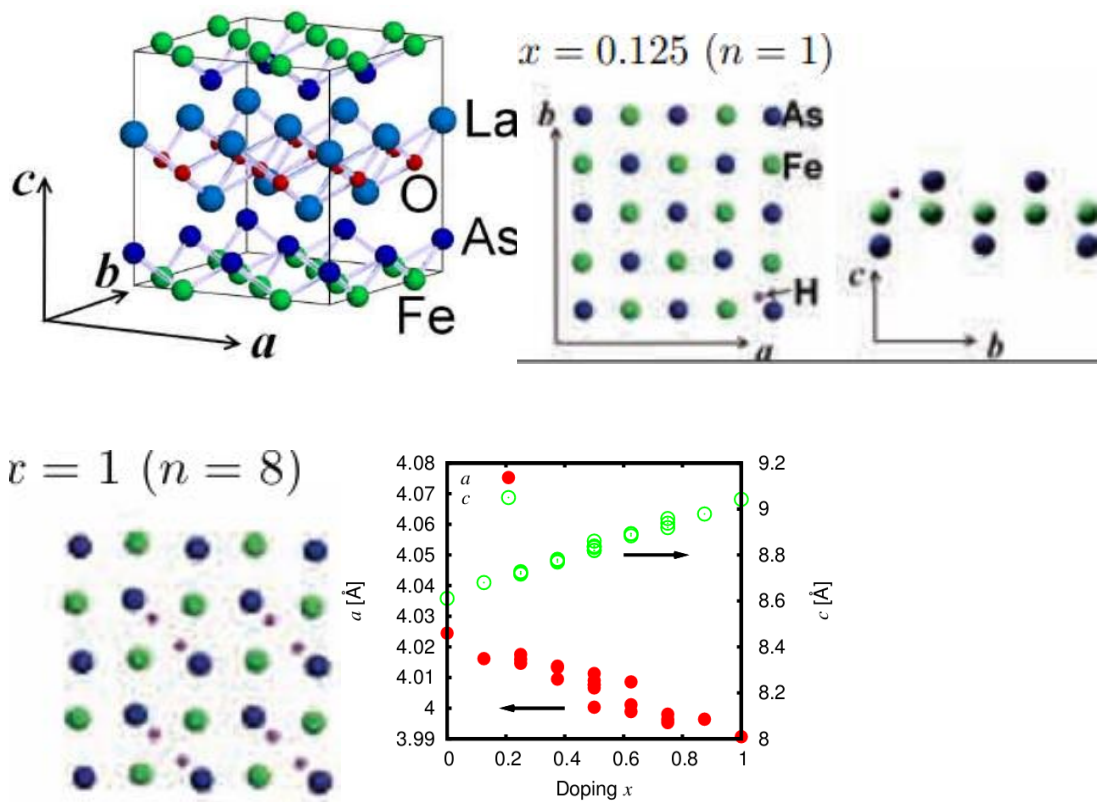

*Fig. 69.* Fragments of a drawings from [173]. Reprinted from Hiroki Nakamura and Masahiko Machida, *J. Phys. Soc. Jpn.* 80 SB009 (2011) with permission of The Physical Society of Japan.

As shown in Fig.69 (bottom right), the lattice constant *a* decreases with increasing *x*, which is consistent with experiment and explained by the authors of the work as follows. The FeAs layer isnegatively charged as (FeAs)⁻. When the H⁺ cation is introduced into this layer, the (FeAs)⁻ layer is compressed due to the Coulomb attraction of H⁺. The results show that the upper limit of hydrogen doping is about 0.25, above which the hole pockets on the Fermi surface disappear.

The iron-based bulk superconductor FeSe (type 11) has a superconducting transition temperature ($T_c$) of about 10 K at atmospheric pressure. Notably, the $T_c$ of bulk FeSe can be increased by applying compression [175]. Calculations in [176] show that the unperturbed electron-phonon coupling in bulk FeSe, assuming that momentum space is isotropic, is determined to be 1.8 meV in the absence of excess pressure (0 GPa). It is important to highlight that the strength of the initial electron-phonon interaction shows a tendency to decrease with increasing pressure, as shown in figure 70. This observation indicates that electron-phonon interaction alone is not sufficient to explain both the observed energy spectrum in the ARPES experiments and dependence of $T_c$ on pressure. However, the magnetic moment of Fe atoms and the exchange-correlation energy increase with pressure, which convinced the authors to combine antiferromagnetism with electron-phonon interaction to test whether electron-phonon interaction together with antiferromagnetic interaction can create such a strong response signal in the energy spectrum of the ARPES range.

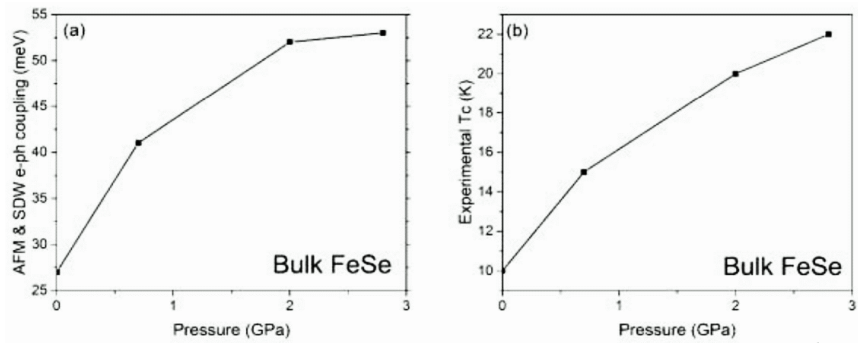

*Fig. 70.* Bulk FeSe; (a) AFM & SDW enhanced electron-phonon coupling. (b) The $T_c$ was measured in the experiments [176]. Reprinted from Chi Ho Wong, Rolf Lortz, *arXiv:*2401.02140.

The results of the above studies show that the significantly enhanced electron-phonon interaction arising from the presence of AFM (Anti-Ferro Magnetism) and SDW (Spin Density Waves) not only matches the experimental value observed in the ARPES range, but also shows a correlation with the superconducting transition temperature. These observations suggest that the instantaneous interaction between the AFM, SDW, and additional CDW (Charge Density Wave) states must play a critical role in shaping the electronic energy spectrum of the ARPES range. The effects of AFM and SDW-assisted CDW states were analyzed, in which SDW triggers the redistribution of AFM oscillations and then creates the CDW state. The presence of AFM, SDW, and CDW states in iron-based superconductors provides valuable information about the fundamental mechanisms that govern the unconventional patterns of ARPES spectra [177]. These discoveries could play a critical role in advancing the theoretical understanding of iron-based superconductors and the mechanisms by which hydrogen influences them.

An important characteristic of a superconductor that determines its magnetic and current-conducting abilities ($J_c$) is the ability to retain (pinning) magnetic fluxes induced in the surface layer or captured in the volume of the superconductor (Abrikosov and Josephson vortices, their bundles of varying densities). For study of the dynamics of magnetic fluxes in FeTe$_{0.65}$Se$_{0.35}$, measurements were carried out in the cooling mode of the sample in a given uniform magnetic field of the solenoid (FC - cooling mode in a DC magnetic field). As the temperature decreases during the experiment and the sample transitions to a superconducting state, most of the magnetic field is displaced beyond the boundaries of the sample, and part of it in the form of Abrikosov or Josephson vortices and their bundles is captured by various defects throughout the entire volume of the crystal. Thermally activated creep (slow hopping creep from one pinning center to another, having a lower energy level) of individual vortices and their bundles leads to redistribution and attenuation of bulk superconducting currents, the integral dipole moment $m(t)$ begins to decrease, and the average magnetization $M$ of the superconducting sample relaxes (Fig. 71).

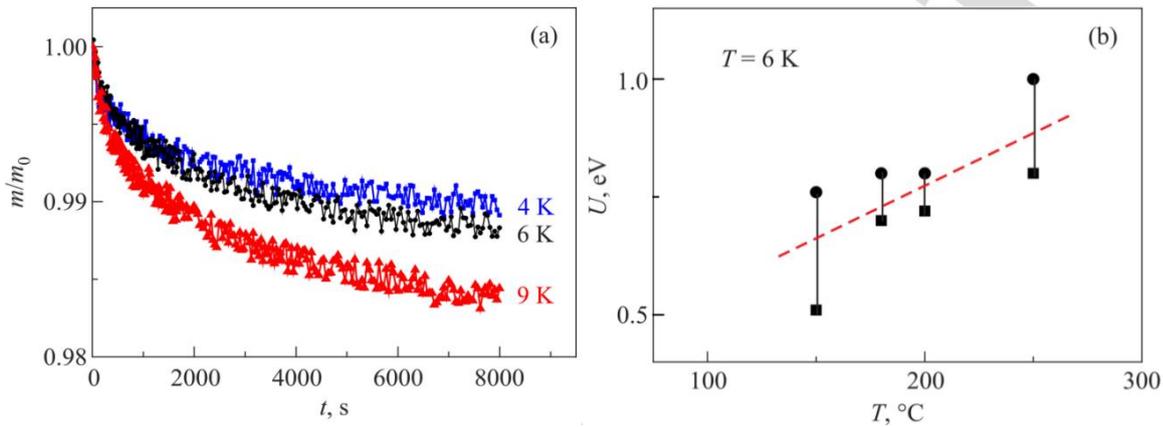

*Fig.* 71. (a) Curves of isothermal relaxation of the magnetic moment of the sample $m(t)$, normalized to its maximum, initial value $m_0$, for some temperatures of the tested single crystal. The sample was saturated with hydrogen at a temperature of 150 °C. (b) Change in the value of the effective pinning potential $U$ caused by the effect of hydrogen sorption procedures (as an example, the results of calculations of $U$ from measurements of $m(t)$ for a temperature $T = 6$ K are given). "Whiskers" (vertical lines) correspond to the scatter of the calculation results [178].

From the above figure 71(b) it follows that during the sorption of hydrogen in the studied temperature range of the FeTe$_{0.65}$Se$_{0.35}$ sample, it is possible to increase the effective pinning potential ($U$) with increasing temperature of this procedure, that is, with increasing intensity of hydrogen sorption from the gas phase. Therefore, according to the collective creep model of trapped magnetic fluxes, the critical current density ($j_c$) of superconductor should increase.

This result was confirmed in experiments with single-crystal samples of FeTe$_{0.65}$Se$_{0.35}$, which were also subjected to the procedure of hydrogen sorption from the gas phase [35].

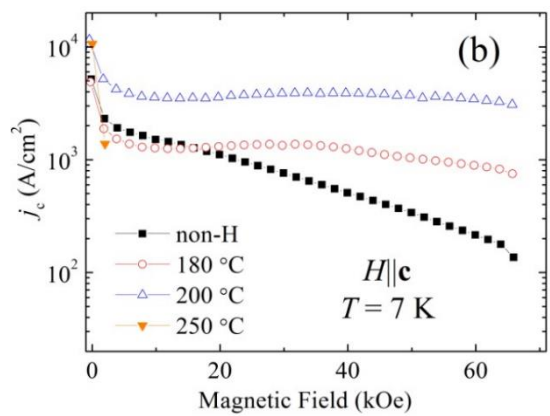

*Fig.*72. Field dependence of the critical current density, $j_c$, at 7 K for the FeTe$_{0.65}$Se$_{0.35}$ pristine single crystal hydrogenated at 180, 200, and 250 °C [35].

From the above experimental studies, the following conclusion can be drawn. Hydrogenation from the gas phase, carried out at a temperature of 180–200 °C, significantly improves the superconducting properties of the iron-chalcogenide compound $FeTe_{0.65}Se_{0.35}$. The bulk critical temperature increases by more than 1 K, and the transition to the superconducting state becomes more abrupt. In addition, new, very efficient pinning centers are introduced, which leads to a strong increase in the critical current density (4–30 times) compared to the original sample.

Iron-based superconductors have evolved into new high-$T_c$ content comparable to cuprates. The optimal critical temperature ($T_c$) of 56 K in electron-doped iron oxy-pnictides of type 1111 has attracted considerable attention from physicists and chemists. The doped element is not only necessary for inducing superconductivity, but is also a critical parameter that controls the electronic, magnetic and crystallographic properties of the ground states in high temperature superconductors. The hydride ion (H⁻), which is the anionic state of hydrogen, acts as an efficient electron donor in 1111 type IBSCs, leading to several important discoveries such as the double- dome structure of the superconducting phase, etc. Paper [179] summarizes the synthesis, physical properties and electronic structure of H⁻-containing iron-based superconductors of type 1111 together with corresponding phenomena of other superconductors. The work shows several general characteristics of iron-based superconductors with a high critical temperature (more than 50 K) and proposes a way to achieve a higher $T_c$. Why should H⁻ doping be used? To solve the serious problem of electron doping in 1111-type IBSC, we proposed hydride ion (H⁻) as an alternative electron dopant to fluoride ion. Hydrogen has a moderate electronegativity and then behaves like the simplest bipolar element in relation to its local environment. When highly electronegative elements such as oxygen and fluorine are bonded to hydrogen, a proton (H⁺) is produced by losing an electron from the $1s$ orbital. (Fig. 73(a)). On the other hand, the hydrogen anion H⁻ with the electron configuration $1s2$. stabilized with electropositive elements such as alkalis, alkaline earth and rare earth metals.

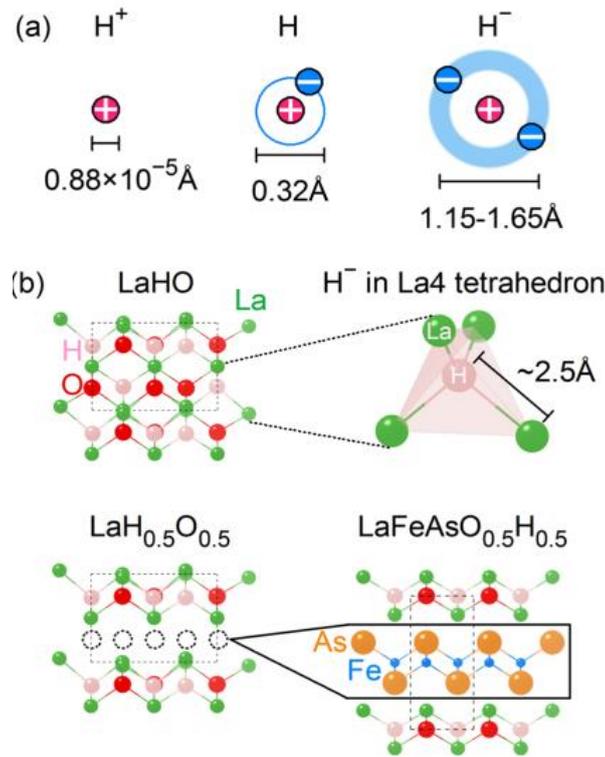

*Fig.* 73. (a) Schematic of electron configuration of $H^+$, $H^0$, and $H^-$. The rootmean-square charge radius of $H^+$ ($0.88 \times 10^{-5}$ Å), covalent radius of $H^0$ (0.32 Å) and the ionic radius of $H^-$ (1.15-1.65Å) are also shown. (b) Crystal structures of LaHO, LaHO with an ordered H and O vacancy, and $LaFeAsO_{0.5}H_{0.5}$. The ionic radius of $H^-$ ranging from 1.15 to 1.65Å is much larger than the size of $H^+$ ($0.88 \times 10^{-5}$ Å) and $H^0$ (0.32 Å), rather similar to those of $O^{2-}$ and $F^-$ (~1.3Å) [179]. Reprinted from Soshi Iimura and Hideo Hosono, *arXiv:* 2002.11218.

Due to the close ionic radii of $H^-$ and $O^{2-}$, both anions can occupy a crystallographically similar site. In Fig. 73(b) shows the crystal structure of the first oxyhydride LaHO. It crystallizes in a fluorite superstructure, in which both $O^{2-}$ and $H^-$ ions occupy the center of La4 tetrahedra, as does the $O^{2-}$ ion in the $[LaO]^+$ layer of La-1111. Therefore, if half of the anions in LaHO are replaced by [FeAs]-La-1111, the resulting structure corresponds to $H^-$ doped LaFeAsO ($LaFeAsO_{0.5}H_{0.5}$), where the replaced $H^-$ by oxygen site acts as an electron donor: $O^{2-} \rightarrow H^- + e^-$.

Taking into account the information presented above in the review, we can expect that the influence of hydrogen on superconducting characteristics will be especially noticeable in a monomolecular (1-ML) FeSe/STO film, where the role of electron-phonon interaction increases and the $T_c$ value can reach 109 K.

### 3.3 Cuprates and magnesium-boron compound.

In unconventional superconductors, which have a layered crystal structure and a strong two-dimensional nature of electronic correlations, a decrease in the strength of the electron-phonon interaction was observed due to the presence of effects associated with symmetry in momentum space. In particular, for example, in the presence of 4-fold symmetry in the superconducting gap, the electron-phonon coupling can be reduced by approximately 0.6-0.8 times [180]. Over a period of almost forty years, about 200 thousand experimental works have been published on the study of HTSC in cuprate compounds. A natural problem arises: which of these experiments can be considered decisive for understanding the HTSC mechanism? What does the critical temperature $T_c$ depend on and what does it correlate with? There seem to be no answers to these questions, just as there were with the initial discovery of HTSC. High-temperature cuprates have attracted attention precisely because their critical temperature can be high. But for each compound, the critical temperature $T_c$ depends on the doping or chemical potential. Let's recall the well-known parabolic approximation for close to optimal (corresponding to the maximum critical temperature) $\tilde{p}_{opt} = 0.16$ doping of holes per Cu ion in the $CuO_2$ plane

$$T_c/T_{c,\,max} = 1 - 82.6\,(\tilde{p} - 0.16)^2. \qquad (11)$$

This maximum is far from the metal-insulator transition, and high-temperature cuprates close to this maximum are, to a first approximation, ordinary metals for which the electron-band theory is well applicable.

In the review paper [181], among other problems of understanding the mechanisms of HTSC in cuprate compounds, the applicability of the BCS theory and the Hubbard model is analyzed. In particular, for doped hole cuprates, the coherent motion of quasiparticles in various versions of the (t–J) model is determined by the dynamic processes of spin waves in the antiferromagnetic (AF) lattice. That is, the hole can propagate only due to the local destruction of the long-range order of the AF observed in practice, which can be facilitated by the presence of hydrogen.

The role of hydrogen in superconducting $H_x YBa_2 Cu_3 O_7$ ($0.14 < x < 5.0$) was investigated using infrared (IR) and X-ray absorption spectroscopy (XAS) measurements [182]. The results indicate the formation of Cu-H bonds and the precipitation of a hybrid phase. X-ray measurements of one of the samples, $H_{1.0} YBa_2 Cu_3 O_7$, show partial restoration of the $Cu^{2+}$ state to the $Cu^{1+}$ state and deformation of both Cu(1) and Cu(2) sites. The XAS results also confirm that hydrogen is localized intrasite only near Cu sites.

To understand the mechanism and origin of superconductivity in high-temperature superconductors, a large number of studies have been carried out in which the constituent parts of HTSC are replaced by any others and/or the oxygen stoichiometry is controlled. Undoubtedly, a change in the composition of HTSCs will affect their structure, vibrational and physical properties, which can help in understanding the mechanism of superconductivity. For example, the superconducting transition temperature $T_c$ changes when Cu is replaced by other $3d$ metals (M) in $YBa_2Cu_{3-x}M_xO_{7-y}$, implying a crucial role of Cu in superconductivity.

The strong influence of the relative position of the Cu4$s$ level with respect to the Cu3$d_{x2-y2}$ level on the critical temperature $T_c$ shows [183] why $s$-$d$ hybridization of the conduction band is so important. This hybridization is proportional to the amplitude of $s$-$d$ exchange scattering between conduction electrons ($d$-pairing mechanism in the $CuO_2$ plane) and the influence of interstitial hydrogen atoms can play a significant role in this. The referenced paper describes how the $s$-$d$ Kondo interaction included in the BCS theory describes the well-known correlation between the critical temperature and the shape of the Fermi surface contour.

The fact that yttrium-barium cuprate YBCO is a "hospitable host" for hydrogen is beyond doubt, but the question is about the chemical state of the proton (hydrogen) included in yttrium-barium hydrocuprate ($H_2YBa_2Cu_3O_7$) and yttrium-barium oxyhydrocuprate ($H_2YBa_2Cu_3O_{7.8}$), open for discussion [184]. Opposing points of view are represented by the "hydride" and "hydroxide" hypotheses. In fact, these hypotheses can be formulated somewhat differently: (i) hydrogen can exist as a "free" monatomic particle ($H^-$, HO or $H^+$) in the crystal lattice and (ii) hydrogen can form a chemical bond with an oxygen ion ($OH^-$ or $H_2O$). The last statement seems more "natural" for oxides. However, historically, the hydride hypothesis appeared first and for a long time enjoyed greater popularity. For this reason, the issues of localization of the hydrogen ion in the YBCO lattice and its mobility (including self-diffusion, chemical diffusion and electromigration) still remain controversial or completely unresolved.

Another aspect of the "hydrogen-yttrium-barium cuprate" problem concerns the effect of intercalated hydrogen on the properties of the oxide. According to available data (for example, [185]), the proton turns out to be an extremely active guest, modifying the crystalline and especially the electronic structure of the host oxide. As emphasized in [186], soon after the discovery of HTSC in cuprate superconductors, it became clear that the carriers responsible for superconductivity in these materials are holes. It was found that when the chemical composition of the initial insulating compound is changed with the addition of hole carriers to the copper-oxygen planes, the critical temperature $T_c$ increases, passes through a maximum and then decreases to zero in the "overdoped" regime. It was then found that when the initial insulating material is doped with electrons instead of holes, superconductivity also arises, although with a lower maximum critical temperature. Experiments initially showed that the charge carriers in these electron-doped materials were indeed electrons.

After the discovery of electron-doped materials, the authors of [187] indicated that there is a natural explanation for why hole carriers of the same nature are induced in them as in hole-doped materials. In Fig. 74 schematically shows how holes on $O^=$ can arise as a result of electronic doping of $Cu^{++}$ using, for example, cesium ions.

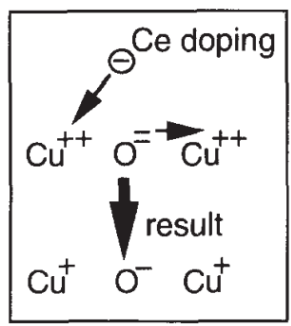

*Fig.*74. Schematic depiction of how holes are created by electron doping. The electron added to $Cu^{2+}$ repels an electron from $O^{2-}$ to the neighboring $Cu^{2+}$, leaving behind a hole in oxygen ($O^-$) [186]. Reprinted from J. E. Hirsch and F. Marsiglio, *Physica* C 564, 29 (1996) with permission of Elsevier.

Hole carriers, responsible for superconductivity in both hole- and electron-doped materials, are located in the zone resulting from the overlap of oxygen $p\pi$ orbitals in the Cu-O$_2$ plane, which are directed perpendicular to the Cu-O bonds, as shown in Fig.75. Electron carriers in electron-doped cuprates are located in the Cu-O zone formed by overlapping Cu $d_{x2-y2}$ and O $p\sigma$ orbitals directed along the Cu-O bond. Assumptions that hole carriers in hole-doped cuprates are located in O $p\pi$ orbitals have also been made previously in a number of studies (see references given in [186]).

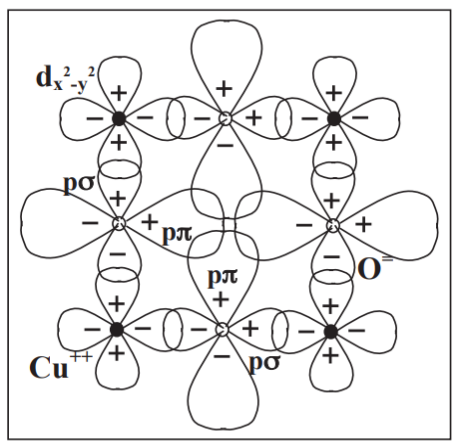

*Fig.* 75. Cu $d_{x2-y2}$ and oxygen orbitals in the Cu-O planes [186]. Reprinted from J. E. Hirsch and F. Marsiglio, *Physica* C 564, 29 (1996) with permission of Elsevier.

In the undoped parent compound the nominal valence is Cu$^{++}$ and O$^-$ and there is one hole in the filled Cu $d^{10}$ orbital. The O $p\pi$ orbitals point perpendicular to the Cu-O bonds, the $p\sigma$ orbitals parallel. We propose that doped holes reside in a band resulting principally from overlapping O $p\pi$ orbitals for both hole- and electron-doped cuprates [186].

As previously mentioned in this review, hydrogen is actively incorporated into the Cu–O bond, creating conditions for increasing the superconducting characteristics of HTSCs with a layered crystal structure. In contrast to cuprate HTSCs, the effect of hydrogen on the superconducting characteristics of magnesium diboride MgB$_2$, another promising compound in the HTSC field, has been studied relatively poorly. The effect of the hydration process on the superconducting characteristics and structure of MgB$_2$ was experimentally studied. Two facts—an almost unchanged critical temperature and a decreasing (during hydrogenation) amount of the superconducting phase—are quite surprising. The first fact means that the density of states, electron-phonon interaction and phonon spectrum remained unchanged after hydrogen absorption, while a decrease in the amount of superconducting phase may mean that the density of states changed significantly. Thus, to explain the above observations, in the picture of ordinary superconductivity due to electron-phonon interaction within the BCS model, the effect of hydrogenation on superconductivity must act in two opposite directions. This can be analyzed using the modified MacMillan equation:

$$k_B T_c = \frac{\hbar \omega_0}{1.2} \exp\left( -\frac{1.04(1+\lambda)}{\lambda - \mu^*(1+0.62\lambda)} \right),$$

(12)

where $k_B$ is the Boltzmann constant, $\omega_0$ is the average phonon frequency, $\lambda$ is the electron-phonon interaction constant, $\mu^*$ is the Coulomb pseudopotential.

Hydrogenation can affect the frequency of phonons. It was previously shown that the vibrations of B boron atoms are more strongly related to the electronic structure than the vibrations of Mg magnesium atoms. This can be judged by the fact that the isotope effect of boron is much stronger than that of magnesium. Both theoretical calculations and experimental work support the view that the in-plane stretching mode of boron $E_{2g}$ is the main source of strong electron-phonon coupling. This mode was found to be highly anharmonic. So, if hydrogen is adsorbed inside the planes of boron, where it can easily find a suitable place for it, and this can greatly influence the stretching mode of boron. But determining what impact this might have is not easy. This may be an increase or decrease in anharmonicity, a decrease or increase in the phonon frequency. Thus, hydrogen can also change $\lambda$ or $\mu^*$.

It is also impossible to exclude the absorption of hydrogen inside the Mg planes. This arrangement should not change important phonon frequency modes, but it may affect the ionicity of the magnesium and boron planes. This, in turn, should change the position of the Fermi level with respect to the boron and DOS s-bands. And along with it, Coulomb screening and electron-phonon interaction can also be changed. The authors believe that most likely a change in the ionicity of the layers should affect the value of the lattice constant c, perpendicular to the main planes of the crystal lattice. The authors of the cited work indicate that this unusual behavior of $MgB_2$ can be explained by some opposing effects of hydrogen on the DOS of charge carriers, phonon frequency, electron-phonon interaction and Coulomb screening.

The recent discovery of high pressure superconductivity of $La_3Ni_2O_{7-\delta}$ with a transition temperature around 80 K has generated extensive experimental and theoretical efforts. Several key questions regarding the pairing mechanism remain to be answered, such as the most important atomic orbitals, spin-orbit interactions, and the role of atomic defects. In $La_3Ni_2O_7$, direct electron counting shows that Ni $3d$ electrons form a half-filled $3d_{z2}$ band and a quarter-filled $3d_{x2-y2}$ band, and a strong correlation in the $3d_{z2}$ orbitals has been identified in a number of experiments. Consequently, a theoretical meaning of interlayer superexchange has been proposed that involves virtual hops of highly correlated Ni-$3d_{z2}$ electrons through the inner apical $2p_z$ oxygen orbitals connecting two adjacent $NiO_2$ planes in each unit cell. Following this picture, some studies argue that a slight oxygen deficiency at this site can suppress superconductivity.

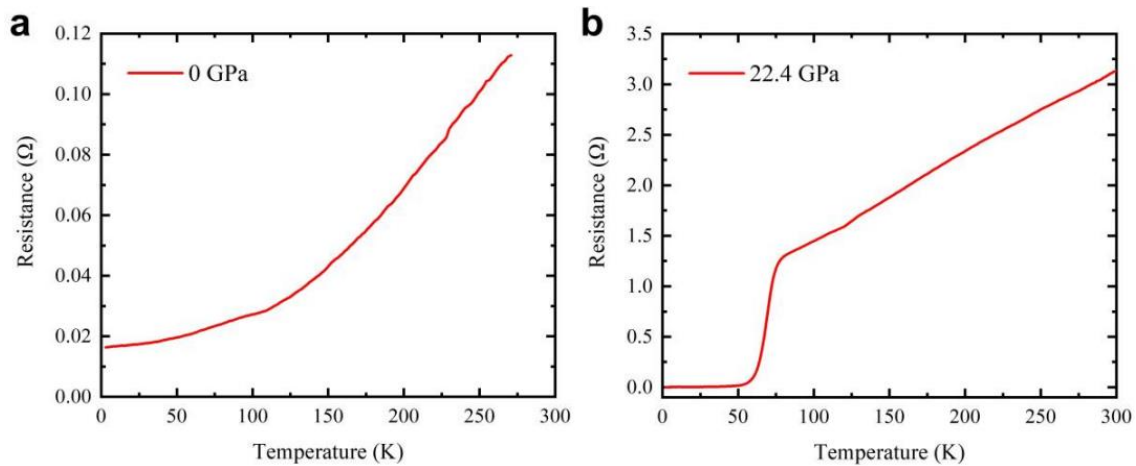

*Fig.* 76. Transport characterizations of the single crystal used in our measurements. a, Resistance of the $La_3Ni_2O_{7-\delta}$ single crystal versus temperature under ambient pressure. b, Resistance of the single crystal $La_3Ni_2O_{7-\delta}$ versus temperature under high pressure at 22.4 GPa, exhibiting a superconducting transition [187]. Reprinted from Zehao Dong, Mengwu Huo, Jie Li *et al.*, *arXiv:* 2312.15727.

## 4. Problems and prospects of hydrogen-containing superconductors.

The goal of modern superconductivity research is to achieve room temperature of the superconducting transition (about 291 K or 18 C) at normal air pressure (about 0.1 MPa). The large magnitude of the critical magnetic field and critical current density are also important. One of the ways to achieve this goal is to develop superconducting hydrogen-containing materials.

In Section 2 of the review it is shown that the superconducting transition temperature of about 291 K is currently obtained only under high external pressure on some hydrides of inorganic chemical elements. In particular, the transition temperature obtained was 287 K for the C-H-S compound, 294 K for the Lu-H-N compound, and 550 K for the La -H compound. It is important that the superconducting properties of these materials are well described by modern versions of the microscopic theory of superconductivity (BCS theory), and the structural and electron-phonon features of closely related metal hydrides have been previously studied quite well [188,189]. The discovery of superconductivity at room temperatures is of fundamental importance and is a powerful stimulus for the development of work on superconductivity and solid state physics in general. These circumstances contribute to further promising theoretical and experimental work on the search for high-temperature hydrides with room superconductivity at lower external pressures. The progress of work in this direction also depends on the number of necessary and quite complex experimental high-pressure installations in scientific laboratories. Unfortunately, there are few of them and this is one of the problems on the way to intensifying such research.

The second direction of research on hydrogen-containing superconductors is the search for ways to increase their critical parameters (superconducting transition temperature, value of the second critical magnetic field, critical current density) at normal external pressure. Unlike the first direction, this search was carried out using new high-temperature cuprates, pnictides and chalcogenides discovered after 1987. The features of this direction were the ability to conduct experiments at normal pressure and the relatively high initial critical temperature of these initial superconductors. The problem with this line of research is the lack of a microscopic theory of new superconductors and, accordingly, the uncertainty of the mechanism of interaction of hydrogen with them. At the same time, theoretical studies of the properties of hydrogen in metals and hydrogenated non-superconducting metals and alloys for solving strength problems were of certain interest. As a result, studies of hydrogenated superconductors at normal pressure are mainly carried out only by experimental methods. Let us dwell on the most noticeable changes in the properties of new superconductors under the influence of hydrogen. Section 3 presents the experimental results, from which it follows that the Fe-Te-Se chalcogenide crystal is compressed in volume by 15% under the influence of chemical pressure resulting from the thermal diffusion of hydrogen at a temperature of 200 K. In this case, the critical temperature of the crystal increases by 1 K. Generating large chemical pressures by introducing hydrogen may be a promising way to compress the crystal lattice to increase the Debye temperature, instead of the complex process of generating large external pressures using diamond pyramids. In addition, such crystals after treatment with hydrogen have a 30 times higher critical current density in a magnetic field of 7 T than before their treatment. The introduction of protons into the iron-based chalcogenide FeSe by an electrochemical method increases its critical temperature by 5 times. Chemical synthesis of chalcogenide with molecules of organic matter with a large number of hydrogen atoms increases the critical temperature by more than 5 times and reaches 50 K. These and other experimental results give there is reason to believe that hydrogenation of new materials is a promising direction.

This can ensure, in particular, after reaching a transition temperature above 77 K (i.e., the boiling point of liquid nitrogen), the practical use of hydrogen-containing superconductors at normal pressure. In addition, there are prerequisites for the creation of superconducting wires from Fe-Te-Se and Mg-B compounds, treated with hydrogen and having an increased critical current density in a high magnetic field. At the same time, in addition to the positive properties of such superconductors, there is an unexplored question about the temporary stability of maintaining the hydrogen concentration in some of them. This is especially true for the stability of the properties of chalcogenides chemically associated with organic matter. This problem is poorly represented in publications and requires additional study.

## Appendix: Physico-chemical properties of hydrogen.

The Appendix includes some background information on the atomic and molecular states of hydrogen. In Mendeleev's periodic system of chemical elements, the hydrogen atom ranks first with the minimum size and atomic weight, almost equal to the mass of the proton. Having one electron in a single electron shell, a hydrogen atom in chemical reactions can either give it away or accept it from an atom of another element. The electronic configuration of the hydrogen atom in the ground state is $1s^1$. In compounds, hydrogen is always monovalent. It is characterized by two oxidation states: +1 (donates an electron) and -1 (captures an electron). To separate an electron from the nucleus, i.e. the ionization of a hydrogen atom requires an energy of 13.6 eV. When an electron is captured, energy is released equal to 67 kJ/mol. About 100 years ago, it was the hydrogen atom that served as a natural model for Niels Bohr to create the quantum theory of the atom. This became the basis of modern quantum physics and chemistry. To describe the processes of interaction of the hydrogen atom with the crystal lattice of superconductors, the main dimensions of the hydrogen atom should be taken into account: atomic radius – 53 pm, covalent radius (half the distance between the nuclei of hydrogen atoms forming a covalent non-polar bond with each other) – 32 pm, as well as the radius of the negatively charged ion hydrogen – 54 pm. The value of the van der Waals radius ($\approx 110$ pm) is also important. This makes it possible to interpret the available crystallographic data and predict the structure of molecular crystals.

There are three isotopes of hydrogen: protium H (a nucleus of one proton), deuterium $2^1$H (or D, where the nucleus is one proton and one neutron) and tritium $3^1$H (or T, where the nucleus is one proton and two neutrons). Natural hydrogen contains 99.985% protium and 0.015% deuterium. Tritium is an unstable radioactive isotope that occurs in nature only in trace amounts. The chemical properties of hydrogen isotopes are almost the same, but the physical properties are different. This makes it possible to study the isotopic effect when studying the superconducting properties of compounds with the presence of hydrogen.

Hydrogen atoms tend to form a covalent bond by sharing two electrons belonging to different atoms.

The exchange integral for the hydrogen molecule $H_2$ in a certain range of distances between atoms is negative. The additional attraction causes the appearance of a minimum of electronic energy at a distance of approximately 1.5 Bohr radii (~ 75 pm). In this way, a covalent nonpolar bond is established between the atoms. The electron shells of two hydrogen atoms "pair".

At normal temperature and pressure, hydrogen is a diatomic gas with the chemical formula $H_2$, which forms spin isomers of molecules with antiparallel (para-form) and parallel (ortho-form) spin orientations of the atoms that make up the hydrogen molecule $H_2$. At room temperature or above, equilibrium hydrogen gas contains about 25% para form and 75% ortho form. The ortho form is an excited state having a higher energy than the para form by 1.455 kJ/mol, and it converts to the para form within minutes when cooled to a low temperature. In the liquid state ($T<20.28$ K), equilibrium hydrogen consists of 99.79% para-form (denoted as p-$H_2$ [190]. The magnetic properties of both forms of hydrogen were studied mainly to determine the magnetic moment of the proton. Experiments have shown that para-$H_2$ does not have a magnetic moment, since in this case the magnetic moments of the protons of the nucleus are antiparallel. For ortho-$H_2$ in the ground state, the observed magnetic moment consists of two parts: the magnetic rotational moment of the single-quantum state and the magnetic moment of parallel-directed protons. Assuming that the magnetic rotational moment in one quantum ortho state is equal to half the rotational moment of para-$H_2$, with $j = 2$ we obtain that the nuclear moment of ortho-$H_2$ is equal to approximately 5 nuclear magnetons (µN). It follows that the proton momentum is approximately 2.5 nuclear magnetons ($1\mu N = 5.05 \cdot 10^{-27}$ J•T$^{-1}$). Later experiments give a value for the proton's own magnetic moment of 2.79 nuclear magnetons, with the moment directed along the spin. The magnetic moment of the neutron is directed opposite to the spin and is equal to 1.91 nuclear magnetons. This circumstance can manifest itself when analyzing the effect of deuterium on metals and superconductors. The discovery of para- and orthohydrogen was one of the best experimental confirmations of the conclusions of quantum mechanics, which predicted in advance the existence of these two types of hydrogen.

Exposure to high temperatures and the associated increase in the kinetic energy of collisions of hydrogen molecules can lead to the breaking of covalent bonds between atoms. Dissociation of $H_2$ molecules occurs. The fraction of dissociated hydrogen molecules at atmospheric pressure as a function of temperature is presented below [191]. The dissociation energy of $H_2$ is estimated at 4.477 eV.

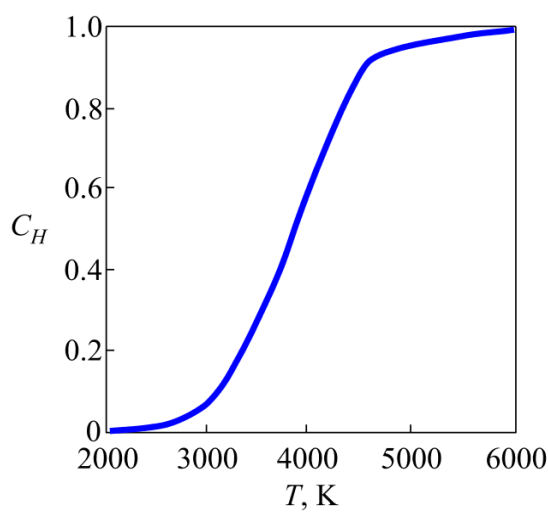

Fig. 77. Temperature dependence of the relative fraction of dissociated hydrogen atoms [191]. Reprinted from *Van Nostrand's Encyclopedia of Chemistry*. Wylie-Interscience. 2005

At significant temperatures or when exposed to other factors (for example, ultraviolet radiation), ionization of hydrogen atoms occurs, i.e. removal of an electron from an atom.

Hydrogen is highly soluble in many metals (Fe, Ni, Pd, Ti, Pt, Nb), especially in palladium (850 volumes of $H_2$ per 1 volume of Pd; its high diffusivity is associated with this property). Hydrogen molecules are strong enough; to enter into a chemical reaction, energy of 432 kJ must be expended for each mole of substance. Molecular hydrogen begins to diffuse into palladium at $24^0$C, into iron at $30^0$C, into nickel at $45^0$C, into platinum at $50^0$C, into copper at $64^0$C. The rate of diffusion increases with increasing temperature and pressure of hydrogen. At constant temperature, the rate of diffusion increases in proportion to the square root of the pressure. A.R. Ubbelohde [189] showed that hydrogen forms interstitial solid solutions with metals. Hydrogen dissolved in metals forms hydrides with reduced plasticity and increased electrical resistance. After dissolution, hydrogen is retained in the metal, forming numerous phases.

Hydrogen at elevated temperature and pressure exhibits oxidizing properties: the hydrogen atom adds an electron and turns into a negatively charged hydride ion $H^-$. At ordinary temperatures, atomic hydrogen interacts with active metals (for example, Na, Li, Mg). The interaction of hydrogen with some chemical elements is influenced by the value of its electronegativity (2,2) according to the Pauling scale. Electronegativity (EO) is the level of electron affinity of another atom. With elements having a higher EO coefficient, hydrogen establishes a non-valent hydrogen bond. As a result, an attraction of atoms (for example, O, N, C) occurs, which leads to the formation of a water molecule $H_2O$ and organic molecules with hydrogen.

This work was financial supported by the of leading Program of National Academy of Sciences of Ukraine "Fundamental research on the most important problems of natural sciences (section "Quantum nano-sized superconducting systems: theory, experiment, practical implementation"). State registration number of the work is 0122U001503.